\documentclass[useAMS,usenatbib]{mn2e}  
  
\usepackage{graphicx}  
\usepackage[fleqn]{amsmath}  
\usepackage{subfigure}  
\usepackage{lscape}  
\usepackage{bm}

\title  
[A super-Earth migrating in a disc with density waves excited by a gas giant]  
{ Outward migration of a super-Earth in a disc  
with outward propagating density  waves excited by a giant planet}  
\author[E. Podlewska-Gaca, J.C.B Papaloizou and E. Szuszkiewicz]  
{E. Podlewska-Gaca$^{1,3}$\thanks{E-mail: edytap@univ.szczecin.pl (EP)},  
J. C. B  
Papaloizou$^{2}$\thanks{E-mail:J.C.B.Papaloizou@dampt.cam.ac.uk (JP)} and   
E. Szuszkiewicz$^{1,3}$\thanks{E-mail: szusz@fermi.fiz.univ.szczecin.pl (ES)} \\  
$^{1}$Institute of Physics and CASA*, University of Szczecin,   
ul. Wielkopolska 15,  
70-451 Szczecin, Poland \\  
$^{2}$Department of Applied Mathematics and Theoretical Physics, University of  
Cambridge, Wilberforce Road, \\ \ Cambridge CB3 0WA, United Kingdom \\  
$^{3}$Kavli Institute for Theoretical Physics, University of 
  California, Santa Barbara, California 93106, USA} 
\begin{document}  
  
\date{Accepted; Received; in original form }  
  
\pagerange{\pageref{firstpage}--\pageref{lastpage}} \pubyear{2010}  
  
\maketitle  
  
\label{firstpage}  
  
\begin{abstract}  
In this paper we consider a new mechanism for stopping the  inward migration  
of a low-mass planet embedded in a gaseous protoplanetary disc. It  
operates when a low-mass planet (for example a super-Earth),  
 encounters   
outgoing  density waves excited by another  source in the disc. This source could be  
a gas giant  in an orbit interior to that of the   
low-mass planet.  As the super-Earth passes through the wave field,    
angular momentum is  transferred  
to the disc material and then   communicated to the planet through coorbital dynamics,   
with the consequence that its inward migration can be  halted or even reversed.

We illustrate how the mechanism we consider works in a variety of different physical  
conditions employing global two-dimensional hydrodynamical calculations.   
We confirm our results  by performing local shearing box simulations in which  the  
super-Earth interacts with density waves excited by an independent harmonically varying  
potential.  Finally, we discuss the constraints arising from the process considered here,  
on  formation scenarios  for  systems containing a giant planet and lower mass planet in an outer orbit   
 with a 2:1 commensurability such as GJ876.   
\end{abstract}  
  
  
\section{Introduction}  
\label{introduction}  
The orbital migration of low-mass planets in  protoplanetary discs may  
play an important role in shaping planetary systems.  
In our previous studies \citep{papszusz2005, paperI, paperII, papszusz2010}  
we have concentrated on the possibility of the occurrence of a resonant structure    
during early phases of the evolution of the  planetary system.  
 We have considered  
two-planet systems, containing at least one low-mass planet of several Earth masses,  embedded in a gaseous disc.   
In \cite{paperI} we  found that in a system with one low-mass planet (described here as a   
super-Earth) and a  gas giant,  
the formation of commensurabilities  readily occurs  when  the  migration of the two planets is convergent  
with  the gas giant being  on the larger orbit.  
However, as  shown in  
\cite{paperII},  this is no longer true if the migration is convergent but   
the gas giant   
is on the smaller  orbit.  
In this case  the super-Earth never gets close enough to the  
location of the 2:1 commensurability in order to be caught in it.   
The super-Earth  is initially found to  migrate towards the central star  but at some point   
stops or even starts to migrate slowly outward.  
However, no explanation of this phenomenon was given.   
The aim of our paper is to explore this behavior more fully  
using both global and local shearing box simulations with the aim of  
delineating the mechanism responsible for the outward migration,  
   showing  how it depends  
on model parameters such as the mass of the giant planet and the disc 
aspect ratio.  
  
Since it was realized that the   
fast migration of  low-mass planets \citep{ward97}  
could pose a serious  
challenge to scenarios for  planet formation,  
possibilities for slowing down such migration   
have been the subject of intensive  studies.  
A number of potential mechanisms have been found by focusing on a single  
low-mass planet embedded in a  protoplanetary  
disc. These mechanisms involve, among others,  entry into a magnetospheric cavity  
close to the star \citep{lin1996},  effects arising from the orbital eccentricity of a  
protoplanet \citep{paplar},  effects due to the possible  eccentricity of the protoplanetary disc  \citep{pap2002},   
magnetic fields \citep{terquem},  MHD turbulence \citep{laughlin2004,   
nelpap2004, johnson, adamsbloch},  sudden  jumps in disc state variables   
\citep{menougood,   
matsumura},  corotation torques  \citep{masset, ppa, ppb},   
disc thermodynamics \citep{PaarMel2006, BarMas2008a, KleyCrida2008, pp2008,   
kley2009, Paa2010, hasegawa, Paa2011, yamada} and instabilities of partial gap edges  
in low viscosity discs  \citep{li2009, li2010}.  
  
However, there may be mechanisms for affecting migration that depend  
on having a system of planets.  
For example, \cite{ogi} have suggested  the possibility of  ``an eccentricity trap''   
operating for a resonantly interacting convoy of planets in a way to halt type I   
migration  near the inner edge of a protoplanetary disc.   
 \cite{thommes05}  suggested  
that gas giants may be   efficient at capturing low-mass planets in their  
exterior mean-motion resonances, forming in this way a barrier in   
a protoplanetary disc ``a safety net for fast migrators''.  A follow up  
study based on numerical simulations of the disc planet interactions  has  been   
been presented by \cite{pierens}. They found that planets  
with masses in the range 3.5-20 $M_{\oplus}$ become trapped at the edge of the  
gap formed by the giant planet,  while  
 more massive planets are captured into resonance. 
 
  They  noted that the positive surface density 
gradient  resembled the configuration discussed by \citet{masset}  for their planet trap.
 Accordingly they raised the possibility that corotation torques
operating as in a tidally  undisturbed disc could be responsible.
However, the configuration differs significantly from that of \citet{masset}.
 The surface density profile in the  latter's   planet trap is maintained by a viscosity that
 increases rapidly inwards.  In the absence of a viscous angular momentum
 flux produced by the action of viscous  stresses at an inner boundary,  acceleration of the accretion flow
 would occur, that in turn causes a decrease in the surface density in a steady state.
 However, in a disc tidally disturbed  by a giant planet,  the profile is maintained by 
 the outward tidal transport of angular momentum produced by dissipating density waves
 rather than by either an accretion flow, or effects of comparable magnitude
 resulting from applied viscous stresses. The  wave transport also occurs in the coorbital
 region of a migrating planet and thus coorbital torques may be expected to behave
  differently  from those occurring  in an undisturbed disc.  Notably,  having an  external 
 source of angular momentum for the coorbital region means that  positive corotation torques
 can be sustained without  either the action of applied viscous stresses, or  the accretion  of exterior  fresh 
 material into the coorbital zone followed by its  loss 
 into the inner regions, which would otherwise be required.
 Put another way, an external source of angular momentum for the coorbital zone
 could  prevent corotation torques from saturating  as would be expected for a tidally undisturbed
 disc with small applied viscosity.
 Thus, as seen in our simulations,  long term 
  positive torques may be communicated to the planet even when the nominal
 viscosity is set to zero. In this way, even though there are some similarities,
  effects seen in a disc that is tidally disturbed
 by a giant planet   may differ from those   found in a disc 
 with a surface density profile maintained through the application of a 
 variable viscosity. 

\begin{figure*}  
\begin{minipage}[!htb]{160mm}  
\centering  
\includegraphics[width=70mm]{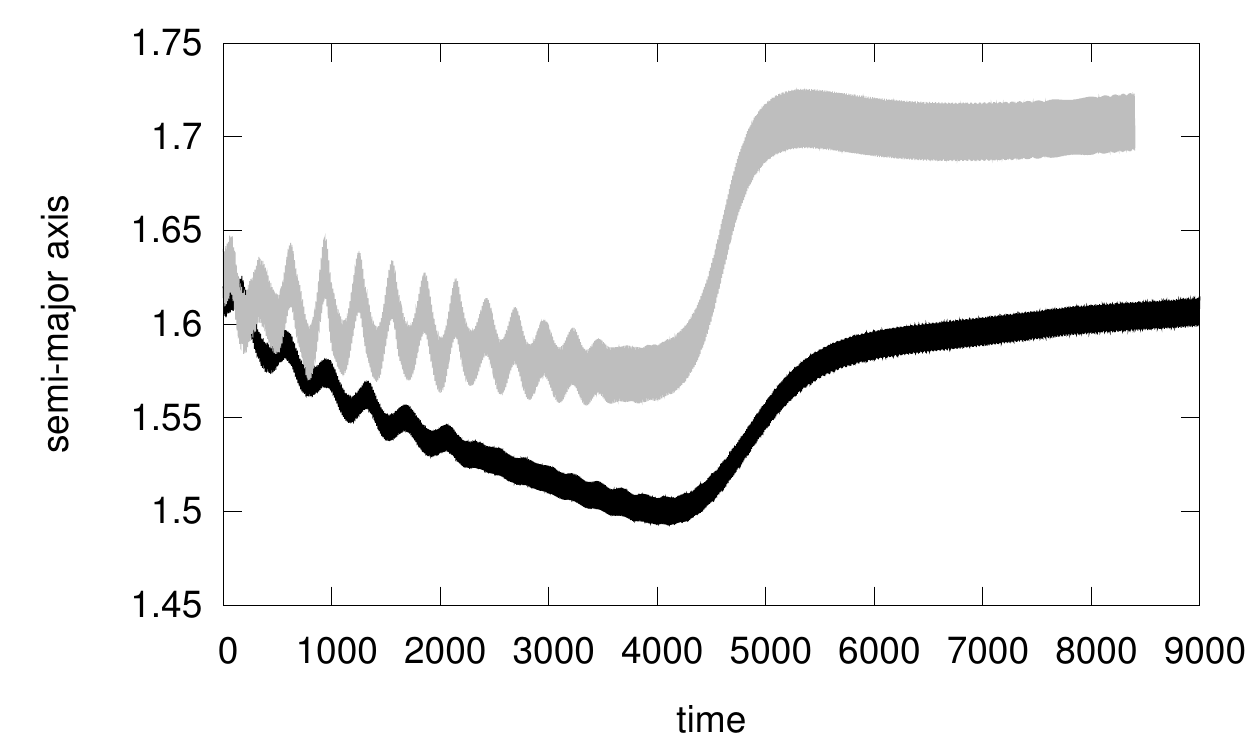}  
\includegraphics[width=70mm]{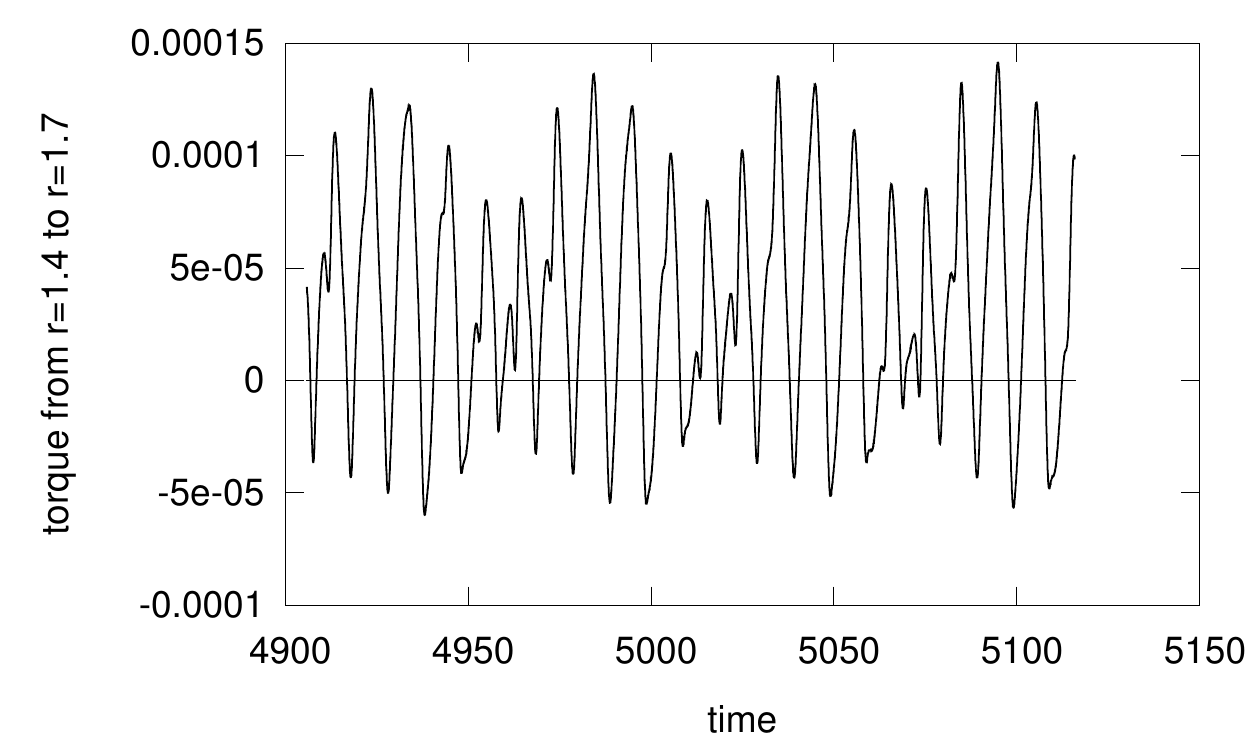}  
\caption{\label{2jow}{The left panel shows the evolution of the semi-major axis of a super-Earth  
in a presence of a gas giant with  one Jupiter mass (black curve) and two  
  Jupiter masses  (grey curve).  
The right panel shows the variation of the torque acting on the super-Earth: coming  
from the  disc  matter located  in the vicinity of the planet,   
in the region between $r=1.4$ and $r=1.7$.  
The timescale covers   
many periods of the super-Earth, which is located at  
$r \sim 1.545$  which is close to the azimuthally averaged surface density maximum that occurs
slightly exterior to the gap edge. Note during this time period, the giant planet is located at $r\sim 0.9,$ to which it has migrated.}}  
\end{minipage}  
\end{figure*}  

\begin{figure*}  
\begin{minipage}[!htb]{160mm}  
\centering  
\includegraphics[width=75mm]{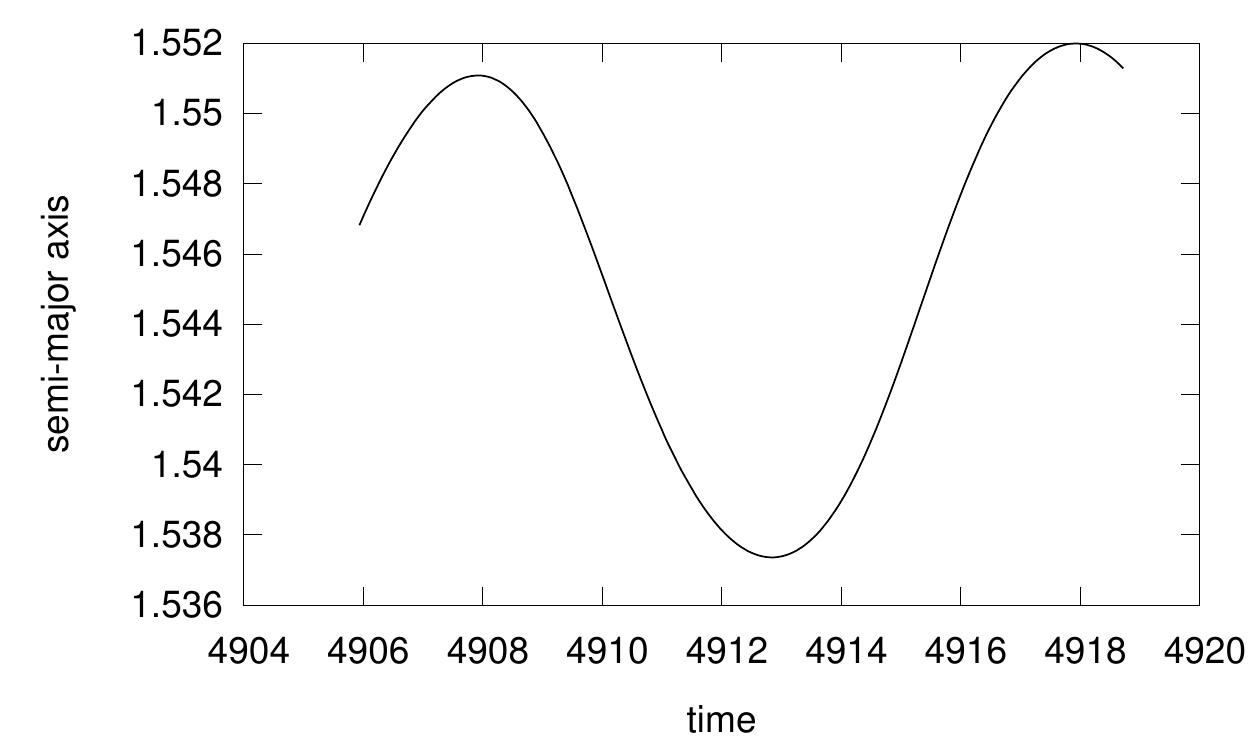}  
\includegraphics[width=75mm]{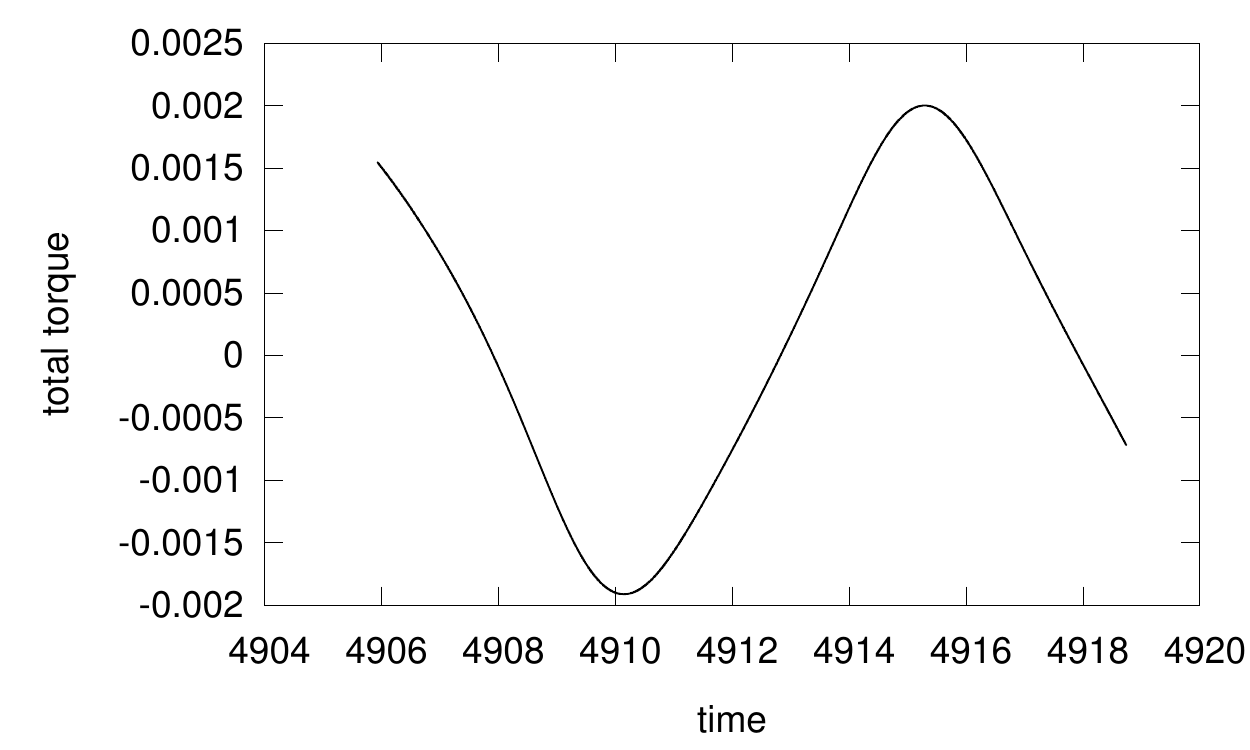}  
\end{minipage}  
\begin{minipage}[!htb]{160mm}  
\centering  
\includegraphics[width=75mm]{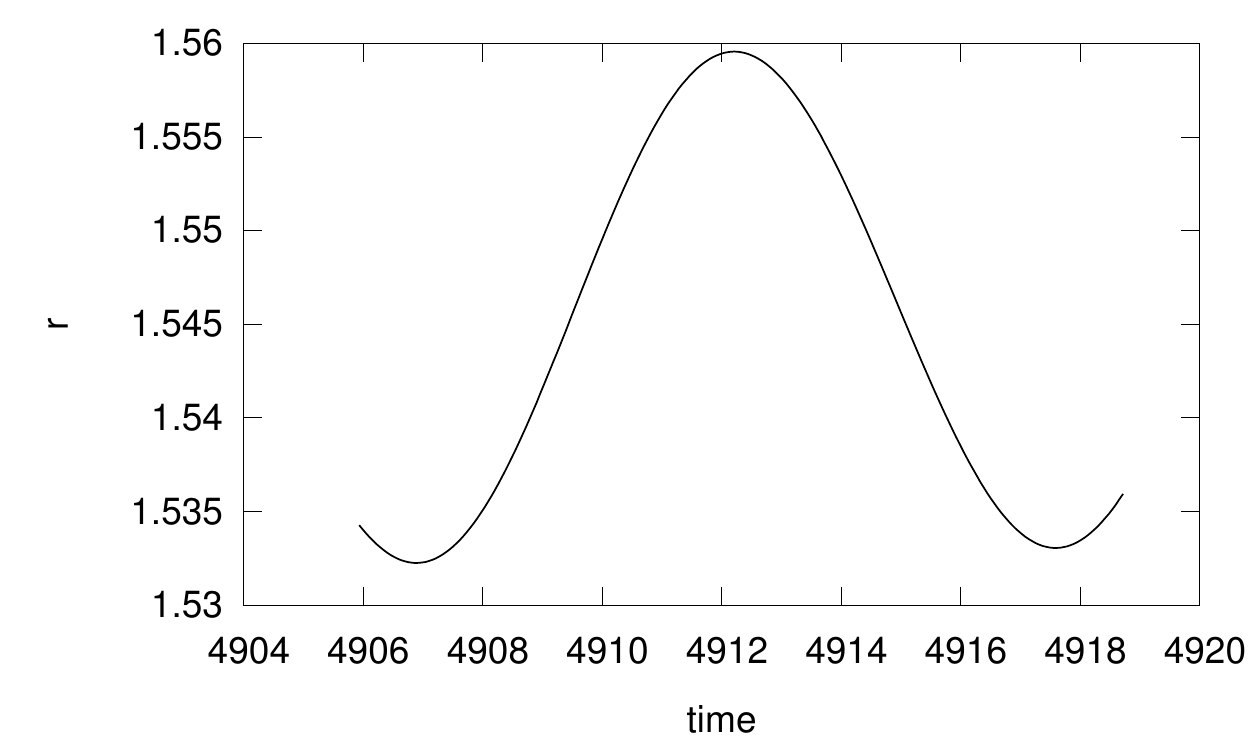}  
\includegraphics[width=75mm]{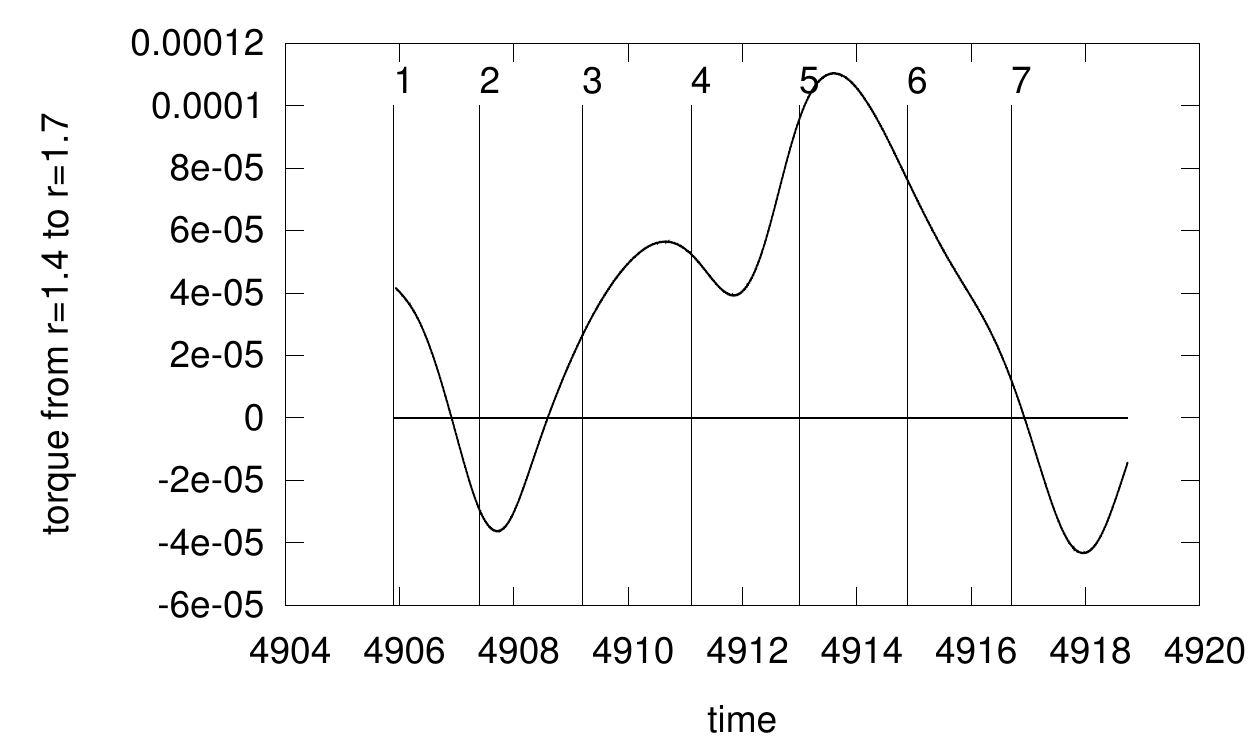}  
\end{minipage}  
\caption{\label{oneorbit}{({\it Left panel}) The semi-major axis of the   
super-Earth (top) and its distance from the central star (bottom)   
as a function of time. The time interval in each   
picture corresponds  roughly to one orbital period of the super-Earth. 
({\it Right panel}) The torque acting on the super-Earth: total torque (top), 
the one coming from the disc matter located in the close vicinity of 
the planet, namely between $r=1.4$ and $r=1.7$ (bottom). The vertical lines in  
the right bottom panel indicate particular moments of time for which the  
planet positions in the disc are shown in Fig. \ref{rysunek}.   
}}  
\end{figure*}  

The mechanism,  we  describe  in this paper, operates  in  
a similar physical situation where a low mass planet migrates in a disc  
which contains  trailing density waves  excited by another planet.  A  specific  
configuration of this type that we consider has  a gas   
giant in an interior orbit with a low mass planet in an exterior orbit  
around a central mass   
\citep[see][]{paperII}.   
However, local simulations that we perform indicate that the essential  
features are   
 a single planet  immersed in a field of propagating density  
 waves produced  by  an external source.  
 The waves are associated with an angular momentum flux which can   
 enter the coorbital region of the low mass planet and transfer angular momentum  
 to the disc material there through the dissipation of shocks.  
 This in turn can transfer angular momentum to the low mass planet  
 through coorbital dynamics. It is important to note that  
 this does not depend on vortensity gradients as do standard corotation  
 torques \citep[e.g.][]{masset, ppa, ppb}.  
In addition because the waves act as a source 
of angular momentum from outside the coorbital zone and the transfer 
involves viscosity independent shock wave dissipation, the usual saturation  
considerations relevant to standard corotation torques do not apply.

This paper is organized as follows.   
In Section \ref{modeleq} we describe the physical model and give the  
basic governing equations. Then  
in Section \ref{setup}  
we give a brief review of the  type I migration  
theory  applicable to an isolated  super-Earth embedded in a protoplanetary disc  
without propagating density waves.   
We go on to present simulations of  a super-Earth embedded  in a protoplanetary  disc  
 in which  density waves excited by an  
interior  giant planet propagate  
in Section   \ref{supergiant}.  We show that in the presence of   
outward propagating trailing density waves,  the inward type I migration of a super-Earth  
can be reversed  with the consequence that attainment of a 2:1 commensurability  
by migrating inwards from large radii is prevented.  
We go on to describe local shearing box simulations  of a  planet  embedded in a disc  
with density  waves excited by an imposed harmonically varying potential  
 in Section \ref {shearingbox}.  We find similar results to those for the global simulations.  
 The waves are found to induce angular momentum transport in the  disc material  
 that is then passed on to the planet. In  
Section \ref{theory} we consider this  mechanism, which  leads  to the  outward migration of the super-Earth  
in the global simulations, from  the point of view of the conservation of energy and angular momentum.  
 We then consider some consequences of the wave-planet interaction for the  
resonance capture  into an outer 2:1 commensurability of a super-Earth by a gas giant in  
Section \ref {observations}. In general density waves excited by the giant planet  
tend to prevent this, indicating that such commensurabilities , as seen for example  
in GJ876, may have formed through planetesimal migration after the gas disc dispersed.   
 Finally in  
Section \ref{conclusions} we give our conclusions.  
  
\section{Basic equations and model}\label{modeleq}  
We consider two planets of masses $M_1$ and $M_2$  orbiting a central star  
of mass $M_*.$  We adopt a cylindrical coordinate system $(r,\phi,z)$  
where $z$ is the vertical coordinate increasing in the direction normal to the disc plane  
for which the unit vector is ${\hat {\bf k}}.$  
 We integrate vertically to obtain a  
  two-dimensional flat disc model,  for which  
the governing hydrodynamic equations  
in an inertial  frame   
 are the continuity  
equation  
\begin{align}\label{continuity}  
\frac{\partial \Sigma}{\partial t}=-\nabla\cdot(\Sigma \mathbf{u}),  
\end{align}  
and the equation of motion  
\begin{align}\label{momentum}  
\frac{\partial \mathbf{u}}{\partial t} + \mathbf{u}\cdot\nabla \mathbf{u}  
 =  
-\frac{1}{\Sigma}\nabla P - \nabla\Phi + \mathbf{f}_{\nu},  
\end{align}  
 where  the vertically integrated pressure
 is  $P=c_s^2(r)\Sigma$ with  
$c_s(r) = h(GM_*/r)^{1/2},$    and  $h$ being  the assumed constant disc  aspect ratio. This
 is related to the
putative disc semi-thickness, $H(r),$  through $h = H(r)/r.$
Thus a  locally isothermal
equation of state with sound speed $c_s(r)$ is adopted.
The surface density is $\Sigma$  and ${\bf u}$ is the velocity.
The viscous force per unit
mass is  $\mathbf{f}_{\nu}.$  The detailed form of this  is given in \citet {nelson2000}. 
The effective  
gravitational   
 potential is  
given by  
\begin{align}  
\Phi = -\frac{GM_*}{r} - \hspace{8cm}  
 \notag\\  
\sum^{2}_{j=1}\left(\frac{GM_j}{\sqrt{r^2+r_j^2  
-2rr_j\cos(\phi-\phi_j)+\epsilon^2_p}} + \frac{GM_j r}{r_j^2}\cos{(\phi-\phi_j)}\right).\hspace{1cm}\notag\\  
\end{align}  
 The last term on  
the right hand side is the indirect term accounting for the  acceleration of the primary.  
In the above, the cylindrical coordinates of the planets  that are assumed to be   
 confined to the disc plane  
are $(r_i,\phi_i,0),  i = 1,2 $  
 and the  softening length  is $\epsilon_p. $  
   Note that  a non zero value of $\epsilon_p$ can be thought of as  
  accounting  very approximately  for the effects of vertical structure.  
 For the simulations presented here, we adopted $\epsilon_p=0.024r.$  
 Thus for  $h=0.03$ the softening length is  $0.8H.$  
 The corresponding quantities  for $h=0.05$ and  $h=0.07,$ for which we have also  
performed simulations  are $0.48H$ and $0.34H$ respectively.

 Setting the Keplerian angular velocity to be,  $\Omega(r)  = (GM_*/r^3)^{1/2},$
the unit of time is  taken to be $\Omega(r_{1,0})^{-1}.$
Here the gravitational
constant is $G,$  $M_*$ denotes the mass of the star and $r_{1,0}$ denotes the initial
radial position of the giant planet. The unit of time corresponds  to ($1/2\pi$)
times the orbital period of the initial orbit of the inner giant  planet.
The adopted unit of length is the initial
orbital radius of the giant
planet  which can be  assumed to  correspond  to 5.2 au.
However, it may be scaled from this to  arbitrary values by applying a scaling factor, $\lambda,$
provided that the disc surface density is scaled by $\lambda^{-2}$ preserving the mass scale.
The unit of time then scales  as $\lambda^{3/2}.$

\begin{figure*}  
\begin{minipage}[!htb]{160mm}  
\centering  
\includegraphics[width=100mm]{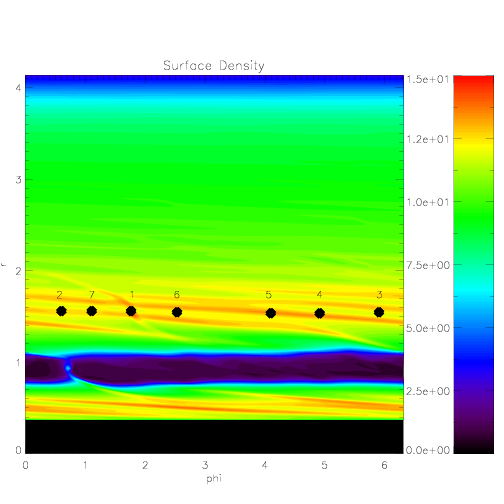}  
\caption{\label{rysunek}{  
The surface density contours in the disc at a time denoted by 1.  
The wake associated with the super-Earth can be clearly seen.  
The  planet positions at six additional times denoted $2-7$ are correspondingly marked  
to indicate their relation to the wave pattern excited by the Jupiter mass planet.  
The times correspond to those indicated in Fig. \ref{oneorbit} (right bottom panel),  
for which the values of torques have been given. The Jupiter mass planet  
is  at fixed azimuthal position as would be situation in a frame that corotates with it.  
}}  
\end{minipage}  
\end{figure*}  
\begin{figure*}  
\begin{minipage}[!htb]{180mm}  
\includegraphics[width=75mm]{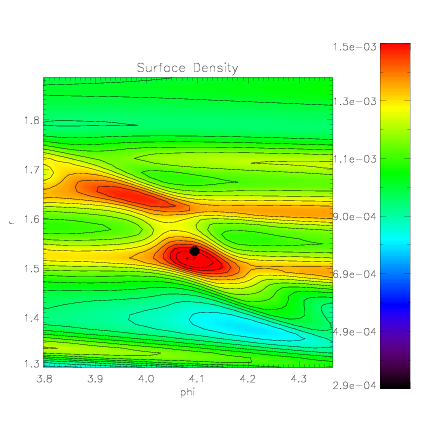}  
  
\vspace{-190pt}  
\hspace{240pt}  
\includegraphics[width=75mm]{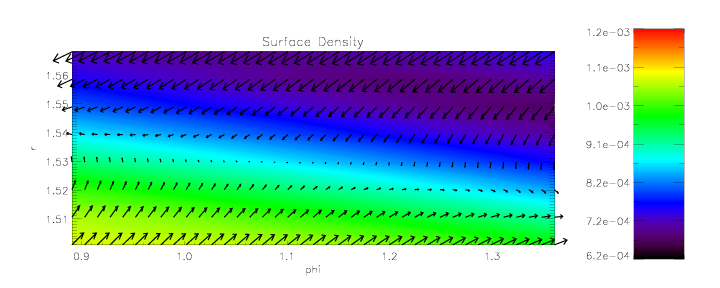}  
  
\hspace{240pt}  
\includegraphics[width=75mm]{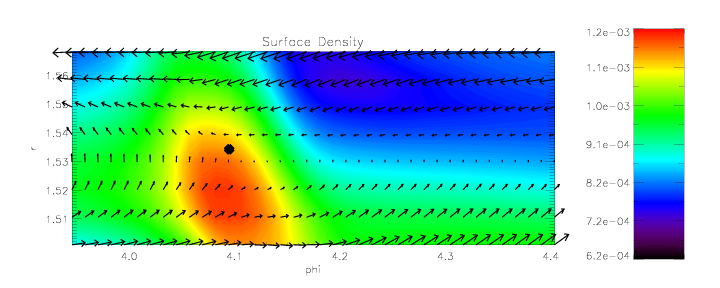}  
\end{minipage}  
\vspace{10pt}  
\begin{minipage}[!htb]{180mm}  
\includegraphics[width=75mm]{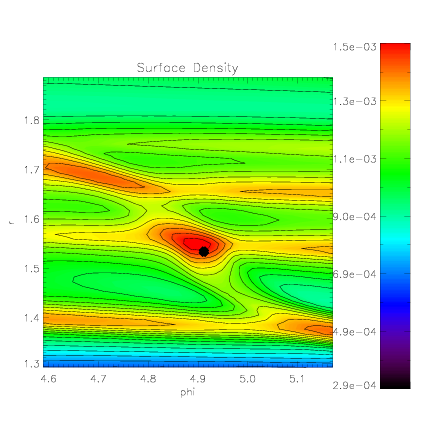}  
  
\vspace{-110pt}  
\hspace{240pt}  
\includegraphics[width=75mm]{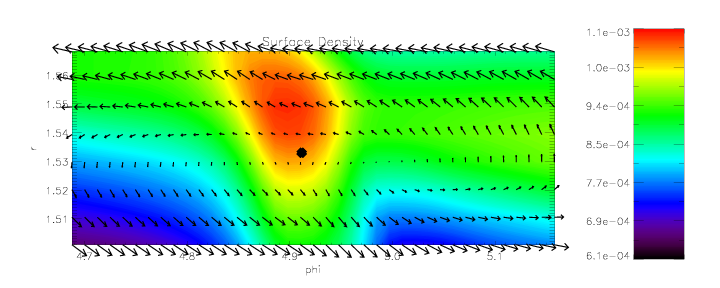}  
\end{minipage}  
\vspace{10pt}  
\begin{minipage}[!htb]{180mm}  
\includegraphics[width=75mm]{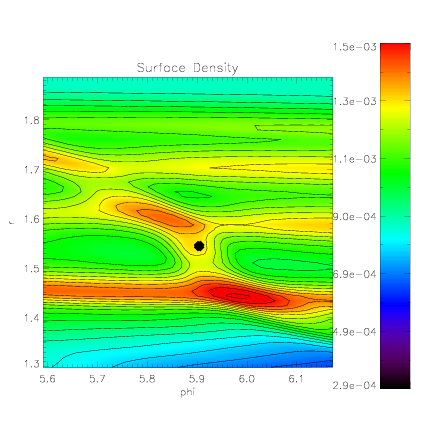}  
  
\vspace{-190pt}  
\hspace{240pt}  
\includegraphics[width=75mm]{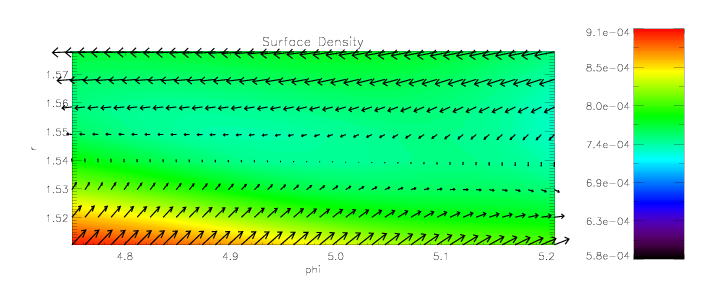}  
  
\hspace{240pt}  
\includegraphics[width=75mm]{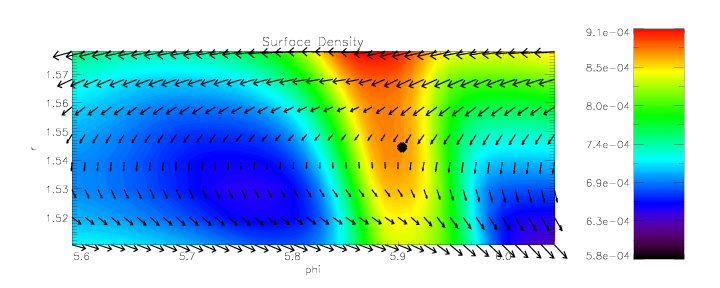}  
\caption{\label{fig3abc}{  
The  surface density  (left panels) and   streamlines   
in localized   disc segments  (right panels). The uppermost, middle and lowermost panels  correspond to the  first,  
second and third  moments of time  denoted in the lower  right  panel of  Fig. \ref{oneorbit}  
and Fig. \ref{rysunek}.  
}}  
\end{minipage}  
\end{figure*}  
\begin{figure*}  
\begin{minipage}[!htb]{180mm}  
\includegraphics[width=75mm]{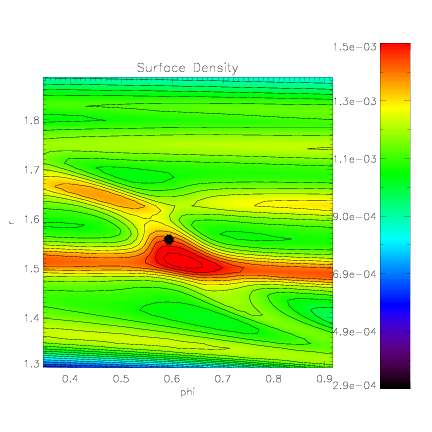}  
  
\vspace{-110pt}  
\hspace{240pt}  
\includegraphics[width=75mm]{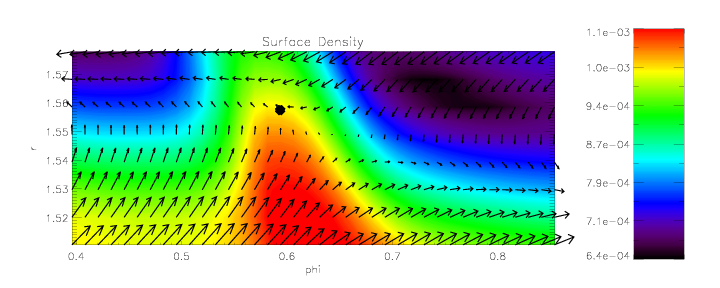}  
\end{minipage}  
\vspace{10pt}  
\begin{minipage}[!htb]{180mm}  
\includegraphics[width=75mm]{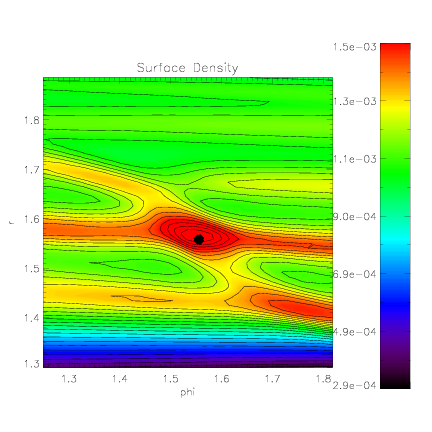}  
  
\vspace{-190pt}  
\hspace{240pt}  
\includegraphics[width=75mm]{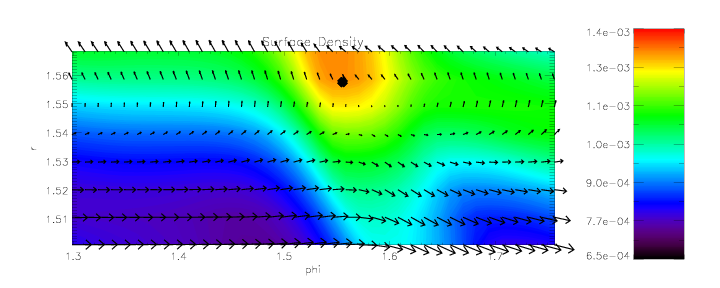}  
  
\hspace{240pt}  
\includegraphics[width=75mm]{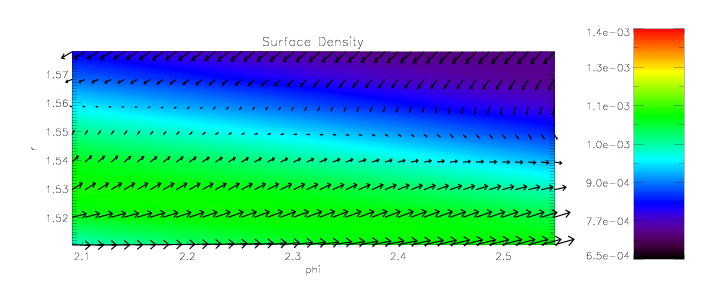}  
\end{minipage}  
\vspace{10pt}  
\begin{minipage}[htb]{180mm}  
\includegraphics[width=75mm]{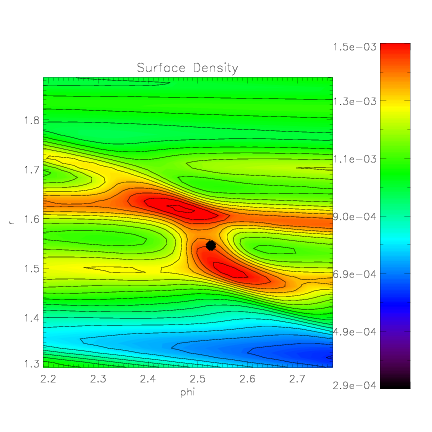}  
  
\vspace{-190pt}  
\hspace{240pt}  
\includegraphics[width=75mm]{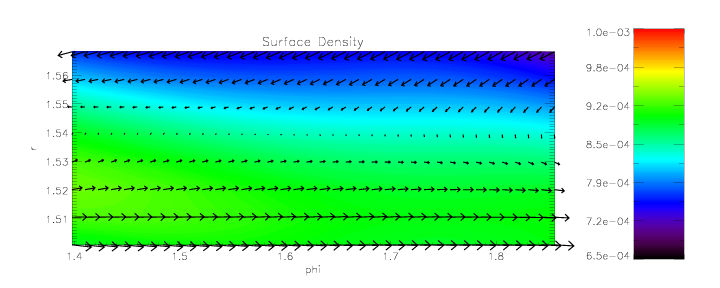}  
  
\hspace{240pt}  
\includegraphics[width=75mm]{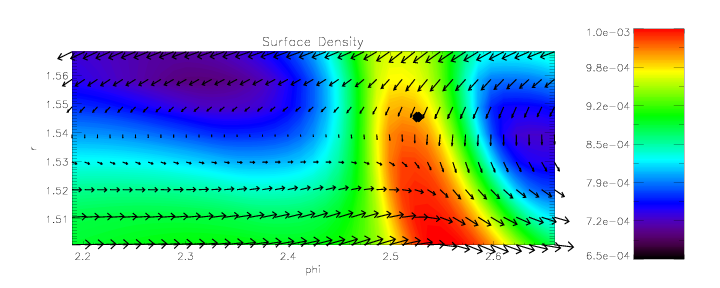}  
\caption{\label{fig3def}{  
The  surface density  (left panels) and   streamlines  
in localized   disc segments  (right panels). The uppermost, middle and lowermost panels  correspond to the  fourth,  
fifth  and sixth  moments of time  denoted in the lower  right  panel of  Fig. \ref{oneorbit}  
and Fig. \ref{rysunek}.  
}}  
\end{minipage}  
\end{figure*}  
  
We consider discs  which  undergo  
near-Keplerian rotation  with  $H$ being  small  
compared to  $r.$  
We have adopted two   
initial surface density profiles for  the disc.   
For the standard initial disc, which was adopted below unless stated otherwise, 
the profile was taken to be uniform with $\Sigma=\Sigma_0$ in the vicinity of the super-Earth  
with tapers at inner and outer radii prescribed  according to    
\begin{eqnarray}  
\label{shortdisc}  
\Sigma  = (0.1+4.5(r-r_{min}))\Sigma_0 \hspace{3mm}{\rm for} & r <r_{min}+0.2, \notag\\  
\Sigma  = \Sigma_0  \hspace{3mm} {\rm for} & r_{min}+0.2 \le r < 1.65, \notag\\  
\Sigma = \Sigma_0\times(r/1.65)^{-1.5}\hspace{3mm} {\rm for}& r > 1.65,    
\end{eqnarray}  
where $r_{min}$ being  the inner edge of the disc  
was taken to be $0.33.$  
  
Some of our simulations   
were carried out for an initial  disc  with a more extensive uniform surface density region,  
with as before, exterior and interior tapers but also a region of low surface density  
in the vicinity of the  orbital radius of the giant  planet,  corresponding to a gap.  
The surface density was given by  
\begin{eqnarray}  
\label{longdisc}  
\Sigma=(0.1+4.5(r-r_{min}))\Sigma_0 \hspace{3mm}{\rm for}& r < r_{min}+0.2, \notag \\  
\Sigma=\Sigma_0 \hspace{3mm}{\rm for} & r_{min}+0.2  \le  r < 0.7, \notag \\  
\Sigma=\Sigma_0\times(r/0.65)^{-11.25} \hspace{3mm} {\rm for} &  0.7 \le  r<0.75,\notag \\  
\Sigma=0.001\Sigma_0 \hspace{3mm}{\rm for} & 0.75 \le r <1.15,\notag \\  
\Sigma = \Sigma_0\times(r/1.43)^{11.25}\hspace{3mm} {\rm for} & 1.15 \le r <1.43,\notag \\  
 \Sigma=\Sigma_0\hspace{3mm}{\rm for} &1.43 \le r <4.3,\notag \\  
 \Sigma=\Sigma_0\times(r/4.3)^{-1.5}\hspace{3mm} {\rm for}&  r > 4.3.  
\end{eqnarray}  
This disc is referred to below as the one having a more extensive uniform
surface density region.
We remark that, due to nonlinear perturbations   
from the giant planet  the surface density profile, in the vicinity  
of and interior to the low mass  planet, tends to relax quickly to a form  
that does not depend on the initial profile used. 
However, differences at larger radii may remain throughout the simulations.   
  
\noindent In practice we  adopted  $\Sigma_0 = 6\times 10^{-4} M_{\odot}/ (5.2 au)^2 $  
 which, for $M_* = M_{\odot},$   
 corresponds to the minimum mass solar nebula   
(MMSN)  
with two Jupiter masses contained    
within a circular area of radius equal to $5.2 au.$  
Note that results may be scaled to different values of $M_*$  
by multiplying the planet masses by $M_*/M_{\odot}$ and surface densities  
and radii by $(M_*/M_{\odot})^{1/3}.$  
The mass $M_1$ is taken to be that of a giant planet  
and the mass $M_2$ is taken to be that of a super-Earth.  All planets are  
initialized on circular orbits.


We used the   
Eulerian hydrodynamic code NIRVANA \citep{Ziegler1998}  
to solve the governing equations  as  was done in our previous   
papers  
\citep{papszusz2005, paperI, paperII}. The details of the numerical  
scheme can be found in \cite{nelson2000}.     
The computational  
domain in the radial direction extends from $r_{min} = 0.33$ to $r_{max}$,   
which has been taken to be in the range $4 - 5.$  
The azimuthal angle $\varphi$ lies in  
the interval $[0, 2\pi]$. The disc  domain is partitioned into   grid cells   
in such a way that  a single grid cell  
is approximately square with  sides parallel to the radial  
and azimuthal directions being   approximately equal to  0.01 in  dimensionless units.   
The  radial boundaries  were taken to  
be open.   
  
When calculating  the gravitational  
forces on  the planets due to the disc, we include  the matter contained  
in the planet's Hill spheres.   
We remark that in practice only  the super-Earth is enveloped by the disc and for such  
a low-mass planet, the influence of the material inside Hill sphere  
can in any case  be neglected.  
The self-gravity of the disc is neglected.  
  
We have run our numerical experiments with disc aspect ratio,  
$h,$ ranging from 0.03 to 0.07 and constant kinematic viscosity $\nu.$    
Unless stated otherwise, we have adopted   
$\nu = 2 \times 10^{-6}$ in  dimensionless  units   
(this corresponds to the standard  
$\alpha$  
parameter  
being equal to $2.2 \times 10^{-3}$ for $h = 0.03$,  $8 \times 10^{-4}$   
for $h = 0.05$ and $4.1 \times 10^{-4}$ for $h = 0.07$).

\section{An isolated  super-Earth embedded in a protoplanetary disc }  
\label{setup}  
  
The migration of low-mass planets in a  gaseous disc has been  
calculated from the  
linear response of the disc to the perturbation caused by  
the presence of a planet \citep[e.g.][]{tanaka02}.  
This procedure is reasonable if coorbital torques do not play a role.  
The latter depend on the disc surface density and temperature  
profiles  and thermodynamics.  
When they are important, non-linear effects play a role  
\citep{Paa2011, yamada}. According to the recent  
studies, a single low-mass planet can migrate with a whole range  
of speeds, both  inwards and outwards, depending on the assumed  
 physical and structural  properties of the disc in which it is embedded.  
   
\begin{figure*}  
\begin{minipage}[!htb]{160mm}  
\centering  
\includegraphics[width=50mm]{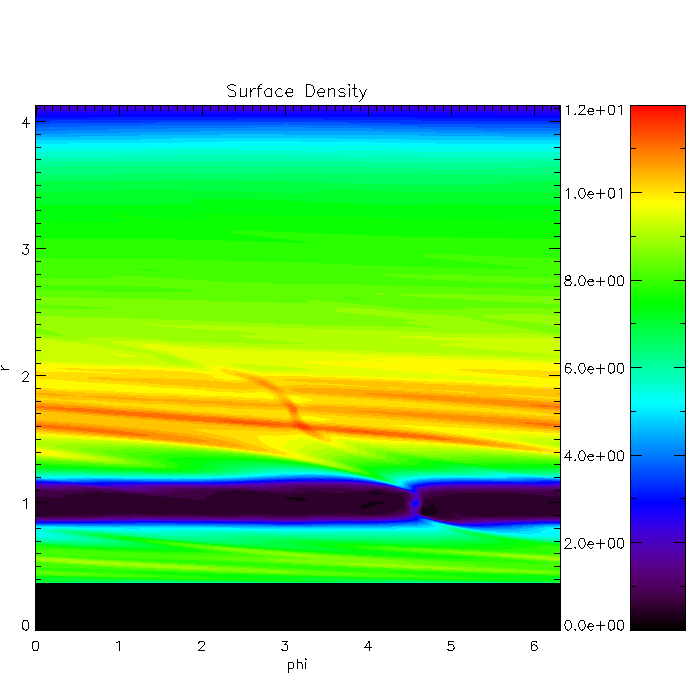}  
\includegraphics[width=50mm]{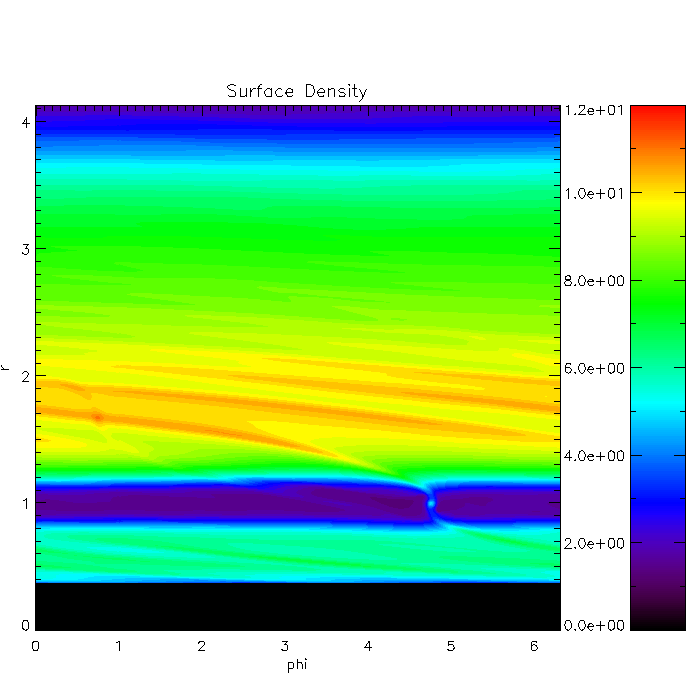}  
\includegraphics[width=50mm]{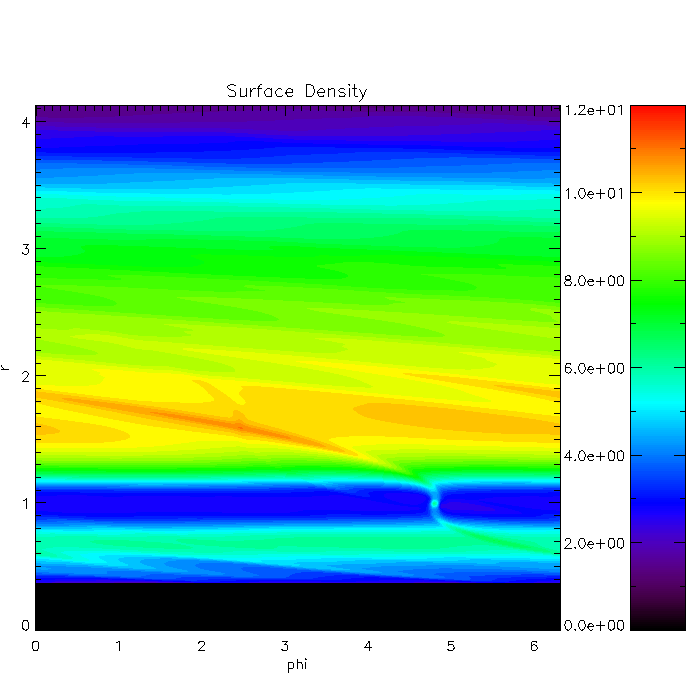}  
\end{minipage}  
\begin{minipage}{160mm}  
\centering  
\includegraphics[width=50mm]{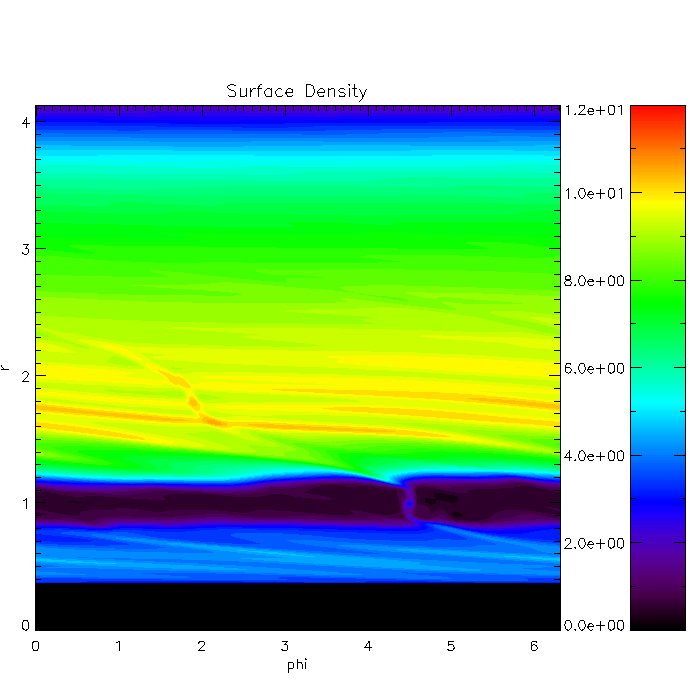}  
\includegraphics[width=50mm]{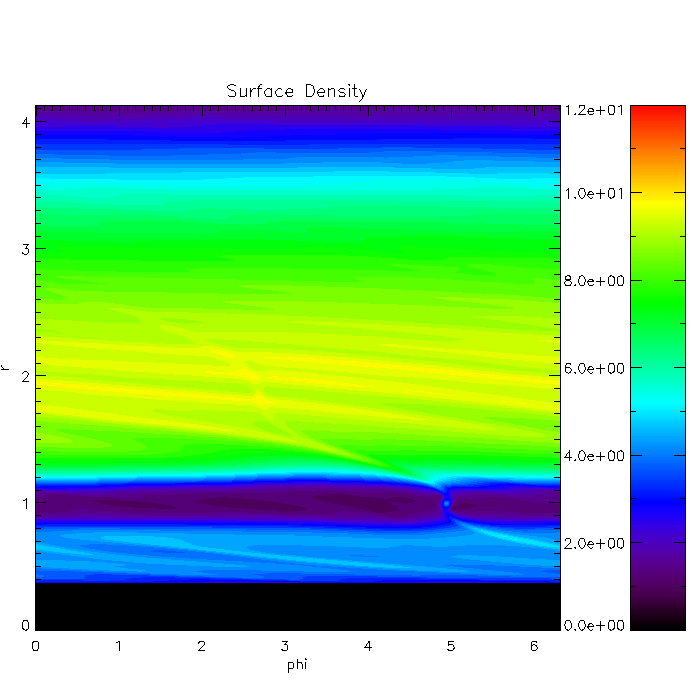}  
\includegraphics[width=50mm]{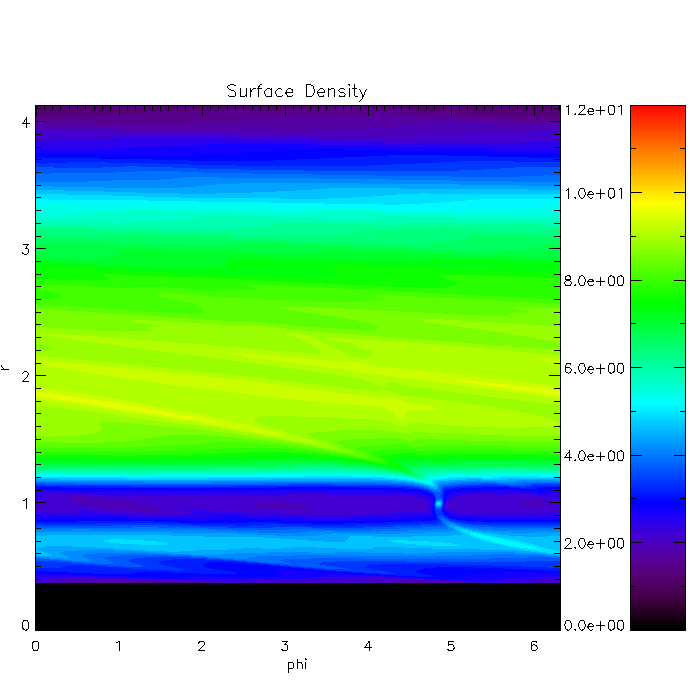}  
\end{minipage}  
\begin{minipage}[!htb]{160mm}  
\centering  
\includegraphics[angle=0,width=90mm]{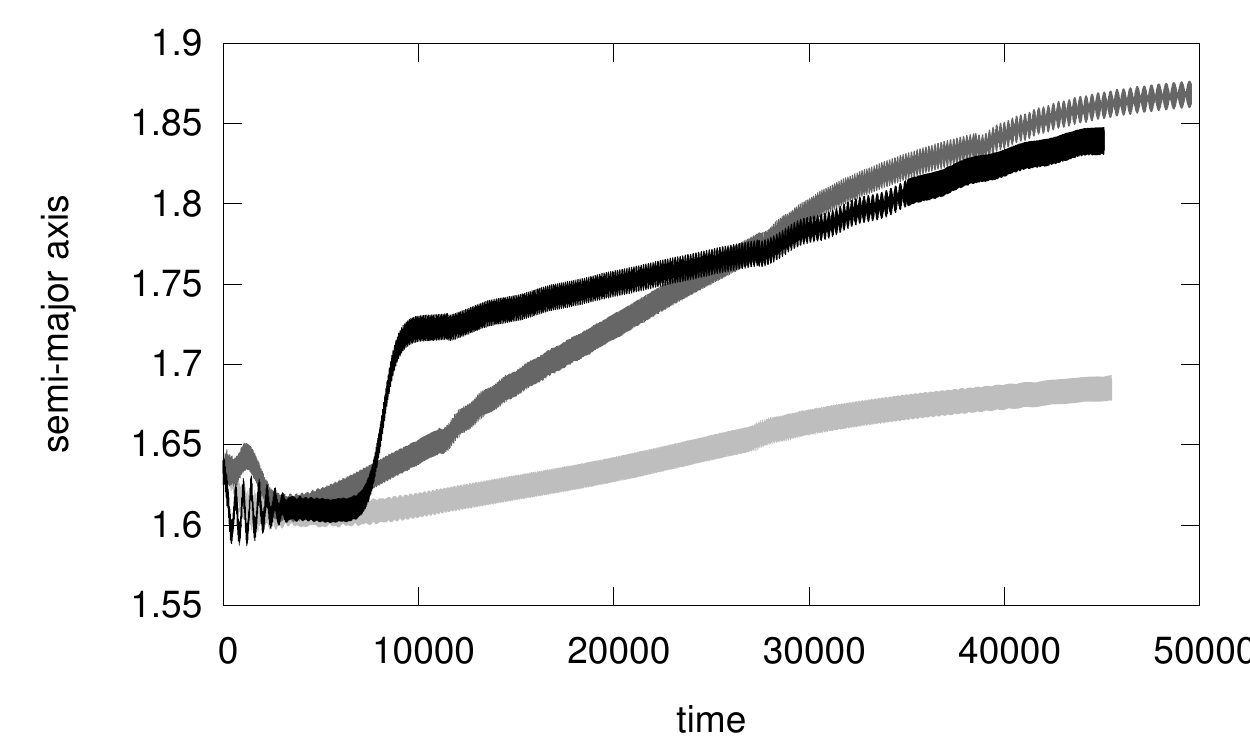}  
\caption{\label{h00}{The surface density contours for the disc with   
aspect ratio h=0.03 (left), h=0.05 (middle), and h=0.07 (right) at  
two different moments of the evolution, namely at about 10800  time   
units (three upper panels) and  45000  time units (three middle panels).  
The density waves weaken at later times because the gap widens and the disk  
tends to expand outwards along with the super-Earth.  
The evolution of the semi-major axis of a super-Earth  
in the disc with  
aspect ratio h=0.03 (black), h=0.05 (dark grey), and h=0.07 (light grey)   
(bottom panel).   
}}  
\end{minipage}  
\end{figure*}  
\begin{figure*}  
\begin{minipage}[!htb]{160mm}  
\centering  
\includegraphics[width=50mm]{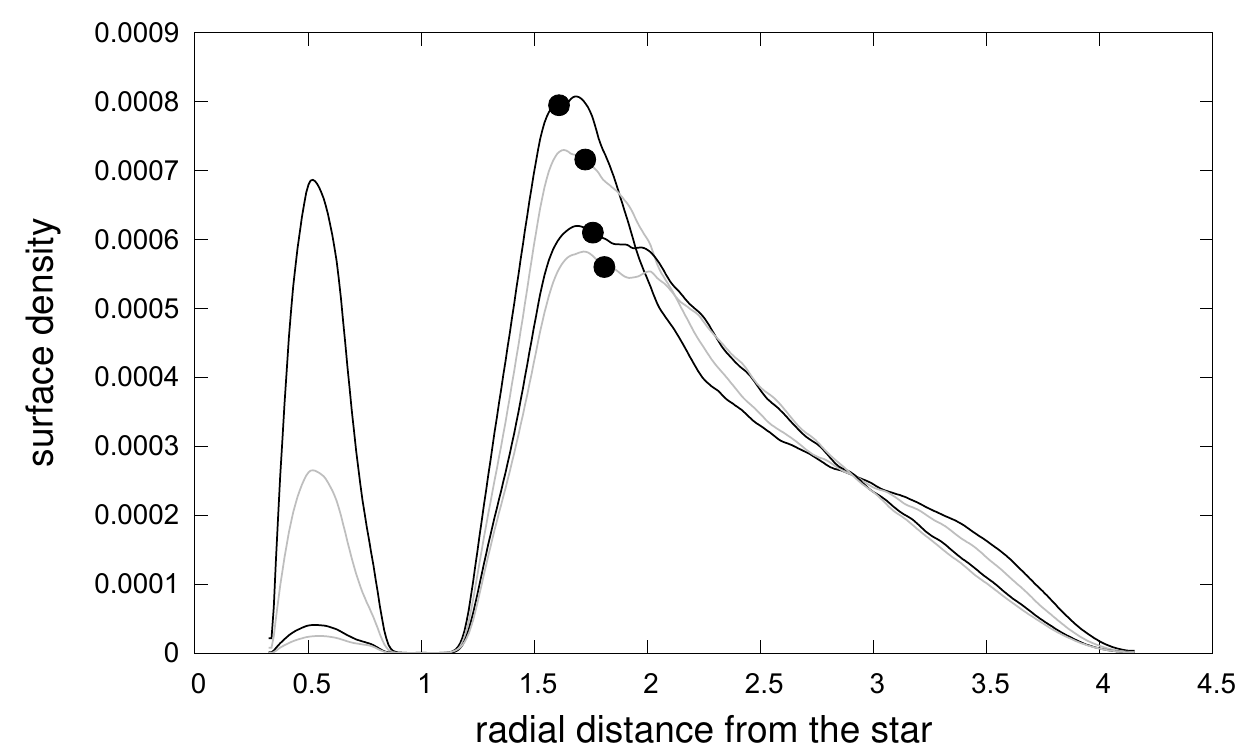}  
\includegraphics[width=50mm]{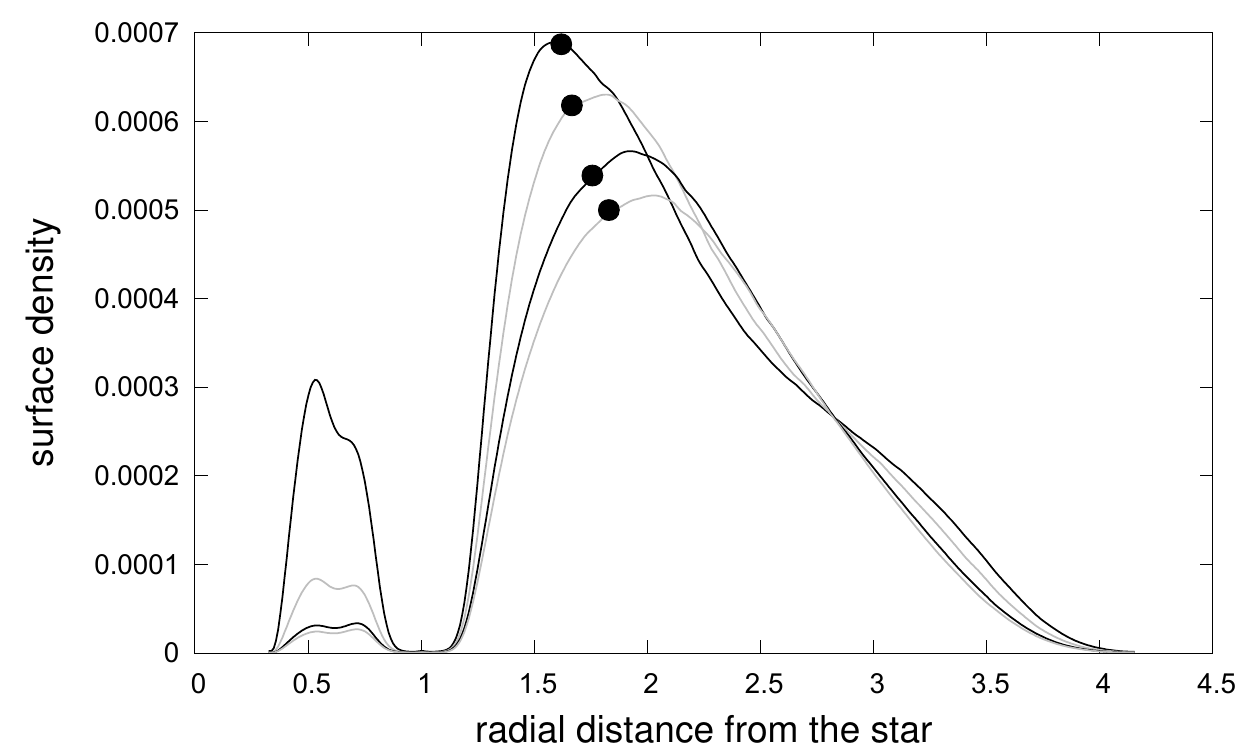}  
\includegraphics[width=50mm]{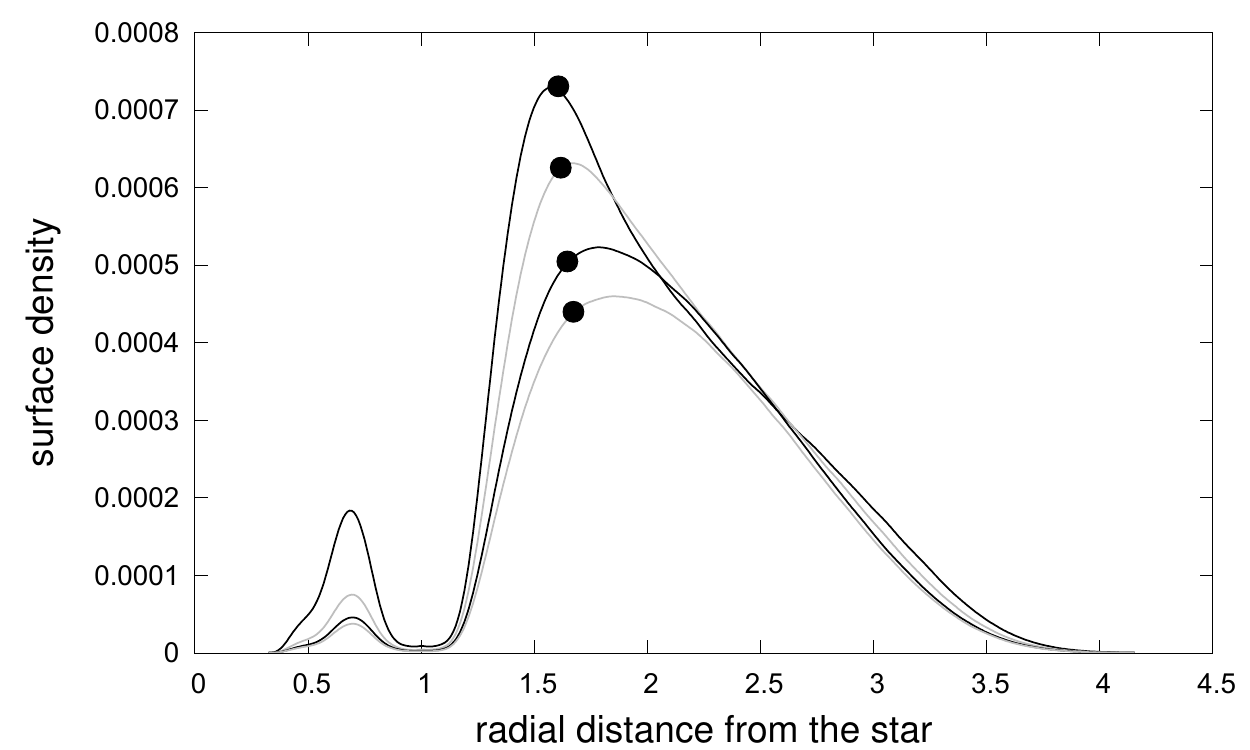}  
\caption{\label{sigma}{The azimuthally averaged surface density as a function of radius for the disc with   
aspect ratio h=0.03 (left), h=0.05 (middle), and h=0.07 (right).    
The position of the super-Earth is indicated by the black dots.   
The snapshots are taken after 6280, 12556,  
25132 and 37600  the time units.  
The surface density level at the planet location decreases  
monotonically with time.  
}}  
\end{minipage}  
\end{figure*}

For the disc models we consider,  
a  super-Earth with the mass   
of 5.5 $M_{\oplus}$ orbiting the Solar-mass star embedded in such a disc   
undergoes   inward  
migration with a speed similar to that   derived from  
the linear treatment of \cite{tanaka02}.  
The situation is very different if the super-Earth migrates in a disc 
in which density waves excited by a gas giant are present 
 and  have a significant 
interaction with it.  
We go on to consider  
this situation below.

\section{A super-Earth embedded  in a protoplanetary disc  in which  density waves excited by an  
interior  giant planet propagate}  
\label{supergiant}  
Before considering the migration of the super-Earth, we briefly review   
some well known relevant properties of the density waves excited  
in a gaseous disc by an orbiting giant planet.  
\subsection{Density waves excited by a giant planet}  
It is well known that an orbiting planet or satellite  excites density  
waves in a nearby gaseous disc that is orbiting the same central body  
\citep{gt1979}. The density waves are launched at Lindblad resonances  
and propagate away from the planet where they eventually shock and dissipate.  
Because of this they take the form of trailing waves propagating away from the planet which are always associated with outward angular momentum transport  
\citep[e.g.][]{lbk1972,balpap}.   
  
For a non migrating or slowly migrating protoplanet embedded in a disc  
without density waves viewed in a frame rotating with the mean angular  
velocity of the planet,   
there are closed streamlines that represent a coorbital flow.  
Each of these can be thought of as constituting a single coorbital cell  
that occupies almost the full $2\pi$ in azimuth, with the protoplanet  
located in the excluded region   
(see for example the 'librating streamlines' in Fig. 5 of \citet{maspap}).  
This of course is the same situation  that occurs in particle dynamics   
without pressure where the librating orbits correspond to horseshoe orbits.  
  
When density waves are present, the coorbital flows may be significantly  
affected by general pressure forces and shock waves such they depart   
from the normal horseshoe form.  
However, if fluid elements still remain trapped in the coorbital region,  
dissipative effects  acting on the coorbital  
flow   can lead to a net force or torque acting on the  
super-Earth, a phenomenon that will be discussed further in Sections   
 \ref{Interactionwaves} and  \ref{tor}  
below.   
  
This is analogous to the situation that occurs for particles  
that are subject to secular dissipative  forces  that   
cause them to slowly  migrate in the absence of the protoplanet.  
When a non migrating low mass protoplanet that induces horseshoe   
trajectories is present, particles on those trajectories can    
become trapped,  communicating the dissipative forces/torques that  
act on them to the protoplanet \citep[e.g.][]{Dermott}.  
  
In order for this phenomenon to result in a positive torque on the  
protoplanet, the local dissipation of the waves, in practice mostly  
through shocks, should produce a positive torque on the disc material,  
as is indeed expected for trailing density waves excited  
by an interior giant planet.  
Furthermore this transport should exceed any counteracting effects  
produced by a disc viscosity on average,  in order for a net positive torque to be  
communicated to the protoplanet.  Thus the torques  
associated with the  dissipation of the density  
waves should exceed viscous torques,  
implying that in a balanced state,  tidal effects exceed viscous effects   
with the balance communicated to the super-Earth.

\subsection{Interaction of an embedded super-Earth  
 with propagating density waves}\label{Interactionwaves}  
It  was  shown in \cite{paperII} that a super-Earth in  an orbit   
exterior to the giant, migrating in the disc    
 towards a Jupiter-mass planet,  never   
got close  
enough to the giant in order to  be caught in a  2:1 commensurability.   
Before it reached this location, it began to migrate   
outwards.   
Let us examine  the stage of the evolution when the super-Earth migrates outwards.

We consider a  disc with  $h=0.03$,  
$\nu=2 \times10^{-6}$ and $\Sigma_0=0.0006$ in dimensionless units.  
The giant planet was taken to be of one Jupiter mass and the initial  
orbital radius of the super-Earth was taken to be $1.62.$  
Tests have shown that when the super-Earth is located beyond the   
2:1 resonance with the interior giant its migration behaviour   
is not very much affected by the mutual  
gravitational  interaction  of the planets or its initial orbital radius.   
Accordingly,  to clarify  that it is the presence  
of the density waves that induces outward migration,   
we perform studies in which this interaction is   switched  off.  
The interaction of both planets   
with the central star and the disc remains intact.

The evolution of the semi-major axis of the super-Earth  
is shown as a function of time in Fig. \ref{2jow}.  
This shows that after an initial phase in which the  
semi-major axis decreases on average it eventually  
increases on average, the transition occurring around $t=4000.$  
At this point we emphasize that this ultimate  outward migration  
is not a consequence  of the   
particular disc model  or  initial conditions adopted (see below).

In order to see how the average torque  
acting on the super-Earth can then be  positive,  
 we consider the behaviour of the semi-major axis of the super-Earth and its   
distance from   
the central star. These are plotted for a time span of one typical orbital period   
of the super-Earth,  starting at $t \sim 4906,$ where the mean migration  
rate is outwards,  in  Fig. \ref{oneorbit}.   
In Fig. \ref{oneorbit} we also show the torques acting on the   
super-Earth arising from  both  the whole disc  
and only  the region located  in the domain $1.4  < r < 1.7$  
that is  close to  the planet.   
The average positive torque  exhibited  over the  one orbital period of the super-Earth  
occurs  for subsequent orbital periods at approximately the same level  
 (see Fig. \ref{2jow}) although the  torque itself shows   
  a  long period modulation implying some unsteadiness.    
  
 We have  
calculated the values of the torques acting on the planet  
at seven times  which are indicated  
by the vertical lines in Fig. \ref{oneorbit}.   
To illustrate the location of the super-Earth with respect to the density  
waves excited  in the disc by the giant at these times,  
we  indicate the  planet positions as seen in a frame that corotates with the giant   
 on the top of the surface density contours  at  a single moment of time in  Fig. \ref{rysunek}.   
Here we remark  that the density wave pattern is approximately fixed as seen in a frame that  
corotates with the giant.  Note too that at this time  Fig. \ref{rysunek} indicates that the giant planet has migrated
inwards to $r \sim 0.9.$ 
   
In Figs. \ref{fig3abc} - \ref{fig3def} we show the surface density  
contours  in a close vicinity of the planet at the first six times  
 indicated in the lower right panel of Fig. \ref{oneorbit}. These fall within one  
orbital period of the super-Earth.  
We also show  the  streamlines   
 both in the vicinity of the planet  
and for  some cases  also a coorbital azimuthal domain   
separated from the planet. This is to confirm that the streamlines  do close.  
For all but the second time, the torque exerted in the vicinity of the   
super-Earth is positive.  Although it is sometimes difficult to discern   
because of cancellation effects, this is in general  seen to be expected from the  
form of the density contours. For example at the first time  
there is a surface  density excess to the right of the planet that contributes  
a positive torque, while at the second time there is a surface density excess  
to the left of the super-Earth that contributes a negative torque.  
  
The coorbital streamlines  are closed. At the fourth time they resemble  
the form associated with steady state horseshoe orbits with an X point near the planet's  
location. Thus there are oppositely directed turns on either side of the planet.  
However,  because of the presence of density waves the pattern is different in other cases, for  example at the first and   
the sixth times,  
the planet is inside the circulating region with an X point significantly shifted  
from the planet at $\phi \sim 1.1$ and  $1.5$ respectively.  
Because the situation is a dynamic one,  it cannot be said that fluid elements  
follow these streamlines but they do indicate that they are turned close to the planet  
at least in some cases. Just as in the standard horseshoe case this is associated  
with angular momentum exchange with the planet even though the situation is  
more complex because of the time dependence and the deflection of the streamlines  
by density waves altering their form.

We have performed a series of  
simulations in order to investigate  the dependence of the outward torque   
on the physical set up.      
The evolution  
of the semi-major axis of a super-Earth for  giant planet  
masses of  one  Jupiter mass   
(black curve) and two Jupiter masses (grey curve) is shown in  
Fig. \ref{2jow}.    
In these calculations, the gas giant is allowed to migrate and  
the super-Earth does not interact gravitationally  with a gas giant.  
This ensures that the torque acting on the super-Earth arises entirely from the  
disc material.  
  
We remark that  density  
 wave excitation is expected to be   
 enhanced by placing a  more massive giant in the disc.  
The initial  inward migration of the super-Earth is slower in the case with the   
two Jupiter mass giant.   
Also in this  case, outward migration causes the super-Earth   
 to reach larger radii at equal large times.    
This is consistent with the idea that the outward torque is  
produced by  
some fraction of the outward angular momentum transfer rate  
associated with density waves excited by the giant. This torque  
competes with the torques associated with standard type I migration  
to determine how far the super-Earth can migrate outwards.

We have also  performed  
 calculations in which the super-Earth  
is  embedded in  discs of different  aspect ratio
and for which the giant planet was  not allowed to migrate.
  
The pitch  of the spiral arms excited by the giant planet  
depends on the inverse Mach number, or equivalently the aspect ratio  
$h = H/r$ of the disc. The smaller the aspect ratio the more tightly the  
spiral arms are wound. In this case the larger Mach number results in the waves more readily becoming nonlinear and forming shocks.   
  
In  Fig. \ref{h00}  we show the disc surface density contours  
for discs  with aspect ratio $h=0.03$ (left panel), $h=0.05$ (middle panel),  
 and $h=0.07$ (right panel)  
for a one  Jupiter mass giant planet.  
The snapshots are taken after 10800  and 45000 time units.   
It can be seen that as the   
aspect ratio increases,   
 the density waves launched by the giant planet become  weaker and less  
tightly wound.  
  
The migration of the super-Earth  
in these three different environments is  
shown in Fig. \ref{h00} (bottom panel).   
The initial orbital radius was 1.64 in each case.  
The black curve shows the evolution of the semi-major axis  
of the super-Earth embedded in the disc with aspect ratio 0.03,  
the dark grey curve  in the disc with $h=0.05$, and the light grey  
curve in the disc with aspect ratio 0.07.

 For the disc with $h=0.03$ the interaction of the  
 nonlinear  density waves with the super-Earth  
causes it to increase  its  
radius  from $r=1.6$ to $r=1.72$ in approximately  
4000 time units.  For the disc with $h=0.05$ the density waves are weaker  
and interact with the planet less strongly  than in the previous case.  
Indeed, the outward migration is not so fast and proceeds almost at the same  
rate during the whole evolution. For the $h=0.07$ the density waves are even  
weaker as is their   
 interaction with the planet with the result that it  
 migrates outward very slowly.  
  
To complete our   discussion of these simulations we present   
the azimuthally averaged surface density at different times during  the  
dynamical evolution of the super-Earth for the   
discs with the different aspect ratios. They are shown in  
Fig. \ref{sigma}  
 after 6280, 12556,  
25132 and 37600  time units.  
The azimuthally averaged surface density  at the planet location decreases  
monotonically with time.  
The planets  continue to migrate outwards  
but  they  remain   in the neighbourhood of the azimuthally averaged surface  
 density maximum in all these  
runs. This is an indication that the angular momentum transfer  
due to the density waves continues to widen the gap as well as move the planet  
outwards. This might be expected if a positive torque acting on the   
planet requires a non zero outward angular momentum flux carried in density  
waves.  The  presence of such a flux might be expected to be associated with gap  
production interior to the planet.

In Section \ref{shearingbox} we study   
the effect of  density waves on planet migration  
by doing a local calculation of  a planet in  the presence of forced waves.  
This approach has the advantage that  a very deep gap  does not necessarily   
result.  However, the forcing of the planet  
is  similar to what we see in discs where the  
waves are excited by  a  giant planet.  
Thus we  gain  a strong indication in favour of the hypothesis that the  
outward migration studied here is a consequence of the  
direct interaction of a low-mass planet with the spiral waves.

\begin{figure*}  
\begin{minipage}[!htb]{160mm}  
\centering  
\includegraphics[width=70mm]{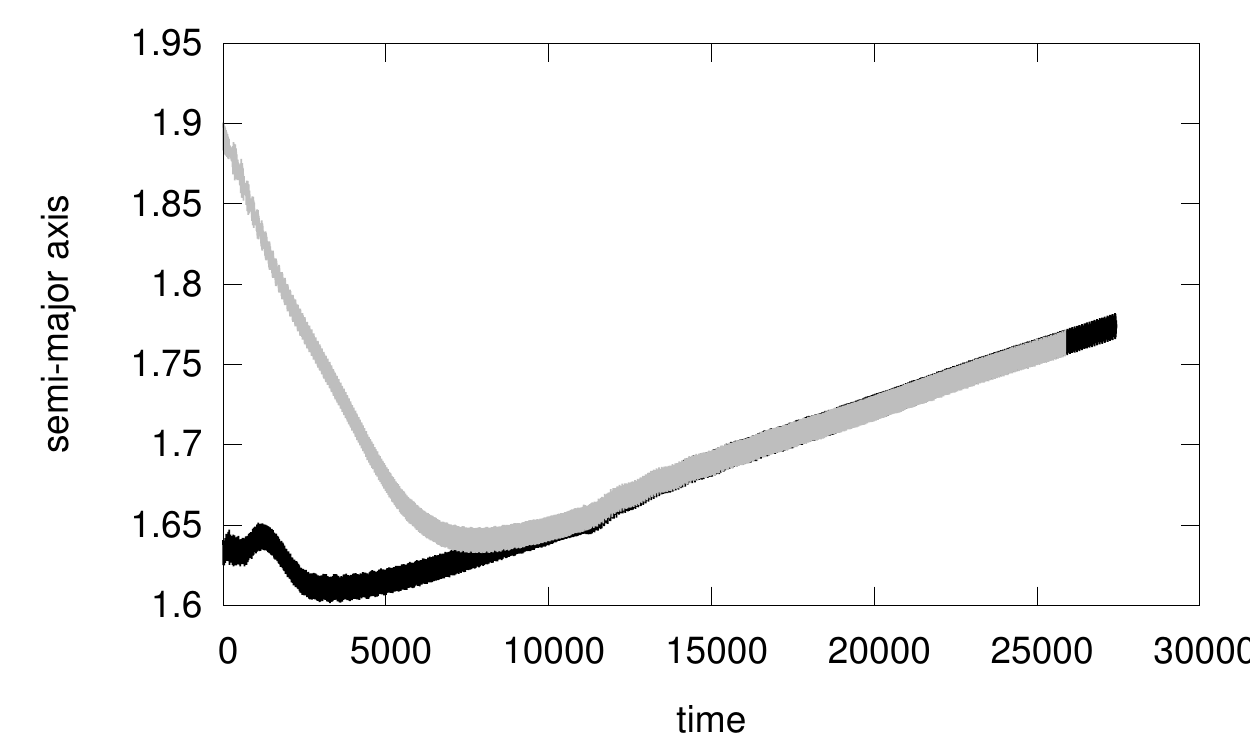}  
\includegraphics[angle=0,width=70mm]{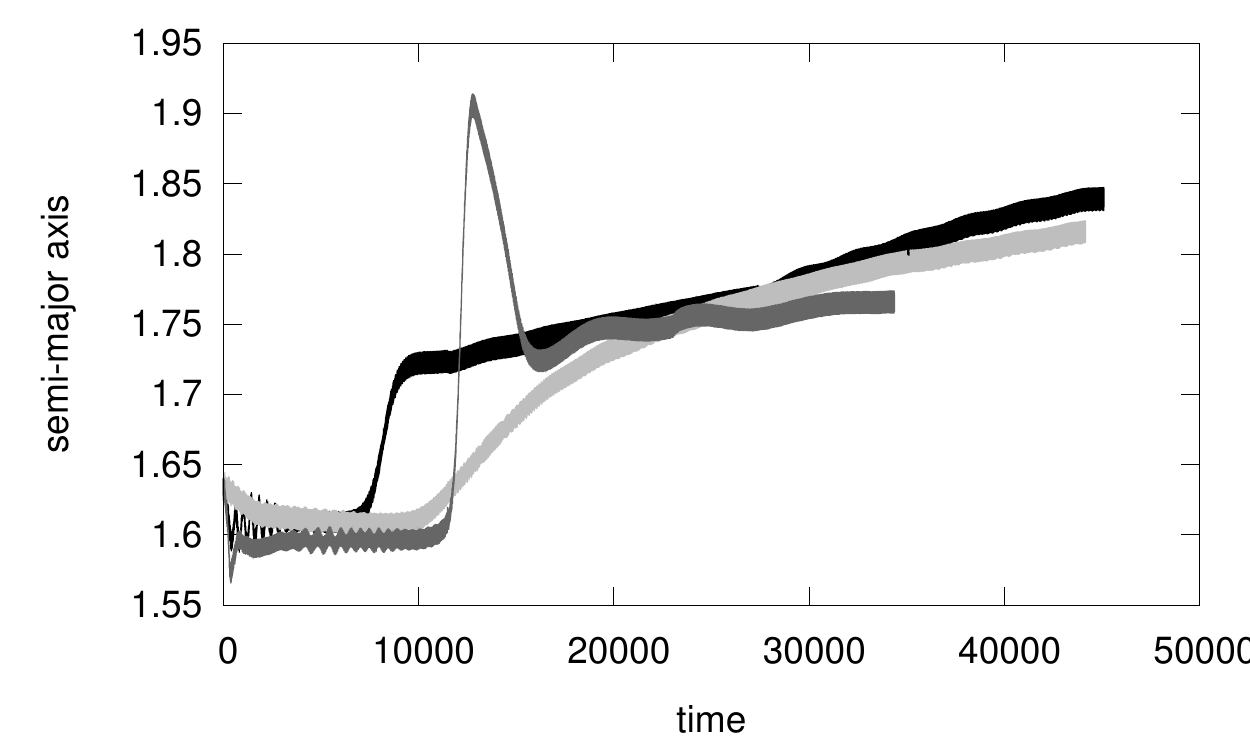}  
\caption{\label{1empor}{The left panel shows the evolution of the semi-major axis of a 5.5 Earth  
 mass planet for which the  initial distance from the central mass  was 1.64  
(black line) and 1.9 (grey line).  
The aspect ratio of the disc is h=0.05. The right panel shows the evolution of the semi-major axis of a low mass  
planet with a mass of 1 Earth mass (light grey curve),  5.5 Earth masses  
(black curve) and 10 Earth masses  
(dark grey curve).  The disc   
aspect ratio $h=0.03.$   
}}  
\end{minipage}  
\end{figure*}

We have  also investigated  how the  evolution of the super-Earth  
depends on its initial location  in the disc.  
In  the left panel of Fig. \ref{1empor} we show the results of   simulations for which the  
initial orbital radii were $1.64$ and $1.9.$  
 These  show  that regardless of the initial  orbital radius,    
the radial position of the planet    
always  approaches  the same   curve.  
  
Another important issue is  the dependence of  
the  migration behaviour  of  a low-mass  
planet  on its mass.   
 For this purpose we have performed simulations  
with  embedded planet   masses  of one Earth mass,   
5.5 Earth masses,  and   10 Earth masses embedded in a  disc   
with aspect ratio  $h=0.03.$ Their initial orbital radii were $1.64.$    
The results are also  shown in Fig. \ref{1empor}  (right panel).  
   The giant planet, of one Jupiter mass,  is not allowed to migrate and, as previously,  the  planets  
don't interact with each other gravitationally.  
The results  differ from  
what  would be  expected  assuming that the outward torque    
occurs as a direct  result of   
interaction with the spiral wave.  
In that case it would be proportional to the planet mass.  
 As the torques arising  from disc planet interactions are proportional to the square  
 of the mass in the linear regime,  the planet with the lowest mass  
should be pushed out  the most rapidly.  
This is not the case in our calculations, which indicate the opposite in the early stages.  
  
\begin{figure*}  
\begin{minipage}[!htb]{160mm}  
\centering  
\includegraphics[width=70mm]{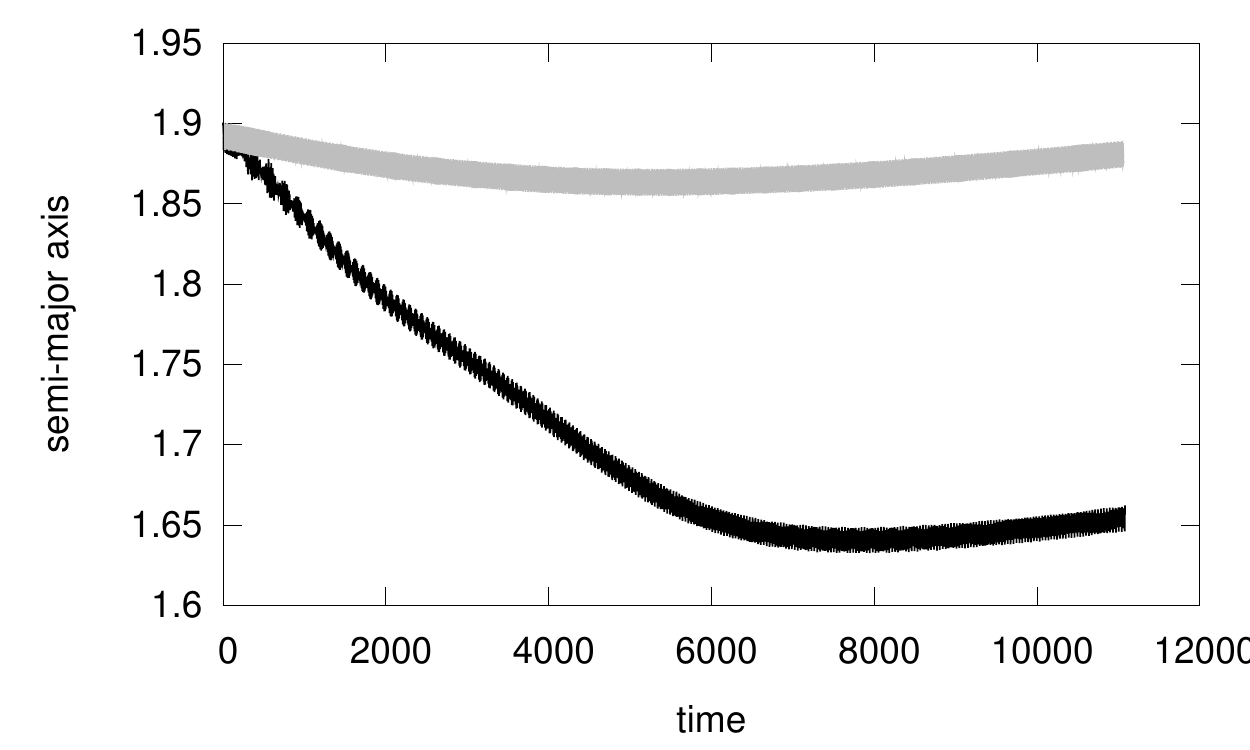}  
\caption{\label{1empor1}{ The   
 evolution of the semi-major axis of a  
super-Earth  
in a disc with  standard initial  surface  density distribution (black curve)  
and in a  disc with the more extensive  initial  surface density distribution  
(grey curve).  
}}  
\end{minipage}  
\end{figure*}

\begin{figure*}  
\begin{minipage}[!htb]{160mm}  
\centering  
\includegraphics[width=70mm]{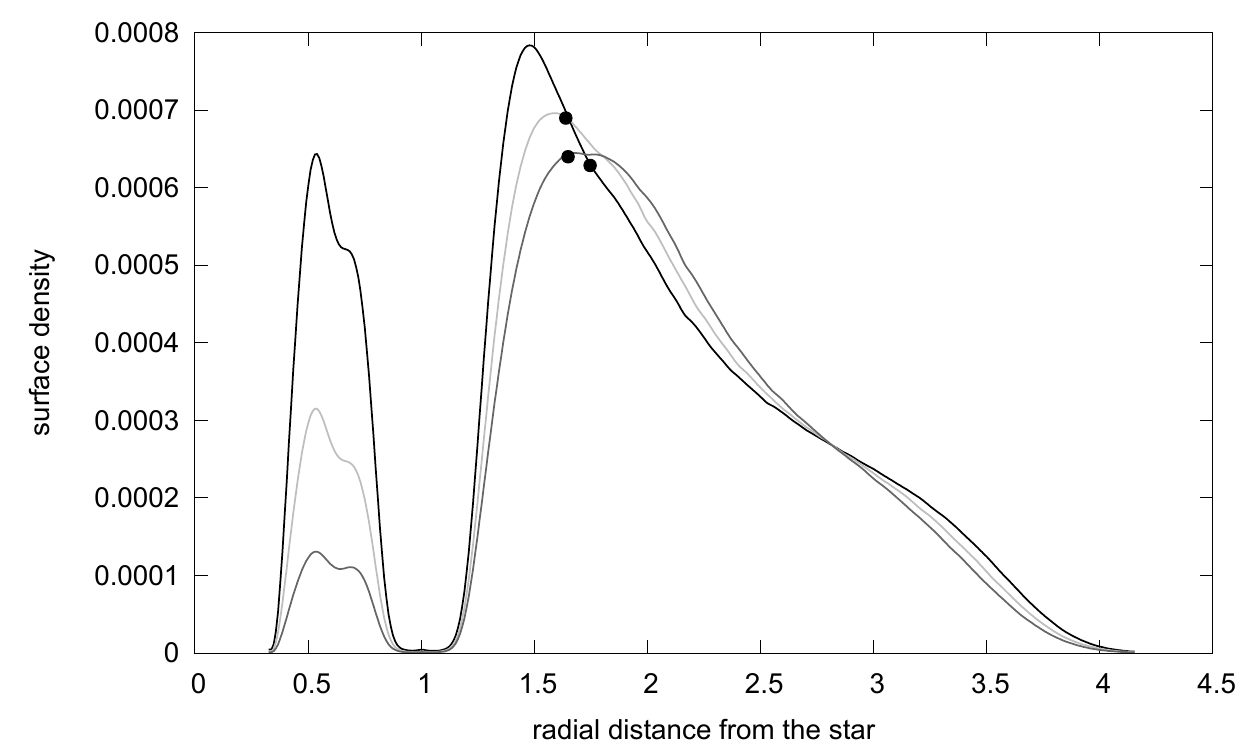}  
\includegraphics[width=70mm]{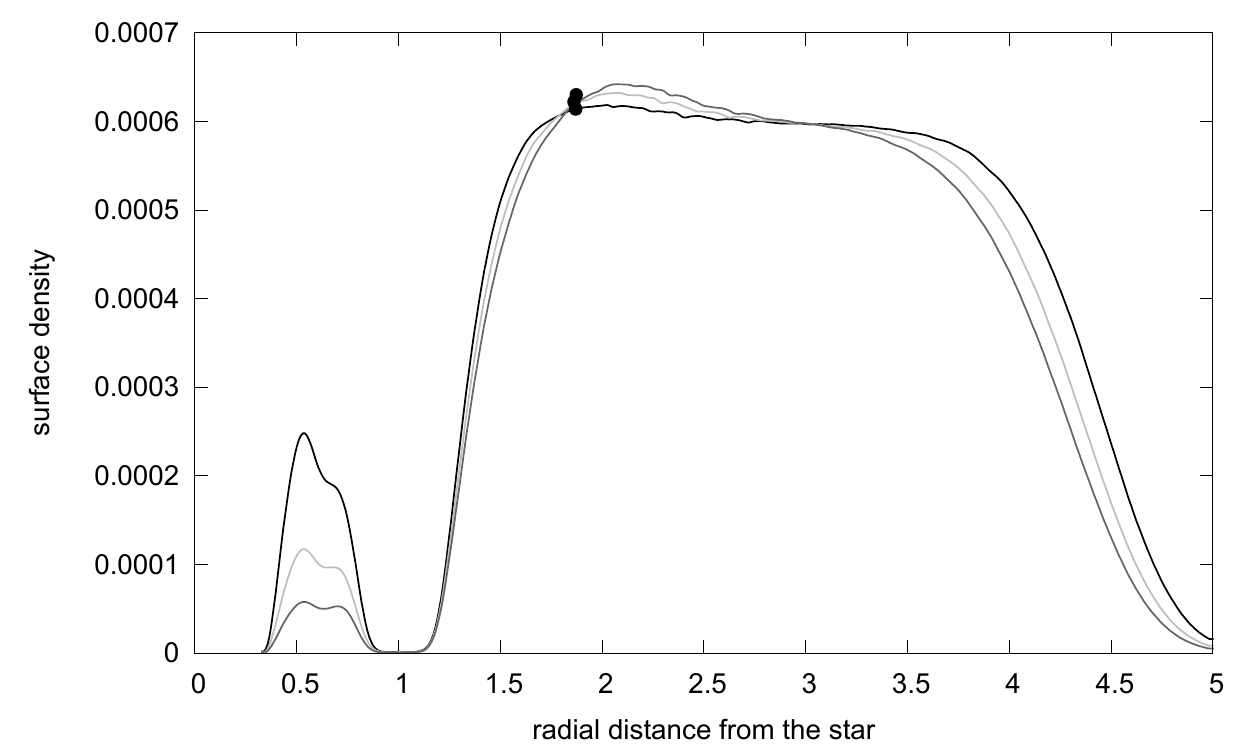}  
\caption{\label{sigmalong}{The azimuthally averaged surface density profiles  
 for an initial surface density distribution corresponding to the standard disc (left)  
  and  for an initial surface density distribution corresponding to the more extensive  disc (right).   
The  snapshots are taken  
after 3100 (black), 6200 (light grey)  and 10000 (dark grey)
 the time units. 
}}  
\end{minipage}  
\end{figure*}

All the numerical experiments discussed so far indicate that  low-mass   
planets    
reverse their  inward migration  at  
a location, which is very  
close to  a strongly peaked azimuthally averaged  surface density maximum.  
This feature of our calculations appears due to the particular choice of  
the initial standard  surface density profile  Eq.(\ref{shortdisc}). To illustrate this,  
we replace this profile by the more extensive one given by  Eq.(\ref{longdisc}).  
The outer edge of this disc is now far   
enough  away so that it does not  affect the surface density distribution in the vicinity   
of the planet.  The initial  profile incorporates a gap   
 so that we eliminate the surface density bumps that appear at    
gap edges as a result of  the gap opening process.  
In  Fig. \ref{1empor1}  we show the evolution of the semi-major axis of a   
planet that  starts from the initial value of $1.9,$ in  both  
   the  standard disc  and the more extensive disc  
given by  Eq.(\ref{longdisc}).   
The aspect ratio is taken to be $h=0.05$ in both cases.   
The azimuthally averaged  surface density profiles for these two cases   
are shown in Fig. \ref{sigmalong}. The  snapshots    
are  taken     
after 3100, 6200   and  10000   time units.  At the last time  
there is slow outward migration in both cases.   
There is a significant difference in the behaviour of  the super-Earth migration. The  
super-Earth  initially  migrates much more slowly  in the case starting with   
surface density distribution  given by Eq. (\ref{longdisc}),  which
is in line with the behaviour of standard corotation  torques which are expected to be stronger
for the case with more rapidly decreasing azimuthally averaged surface density \citep[eg.][]{masset, ppa}.
However, it could also be a consequence of the tidal effects of the giant being less effective
at larger radii in that case.

\begin{figure*}  
\begin{minipage}[!htb]{160mm}  
\centering  
\includegraphics[angle=0,width=70mm]{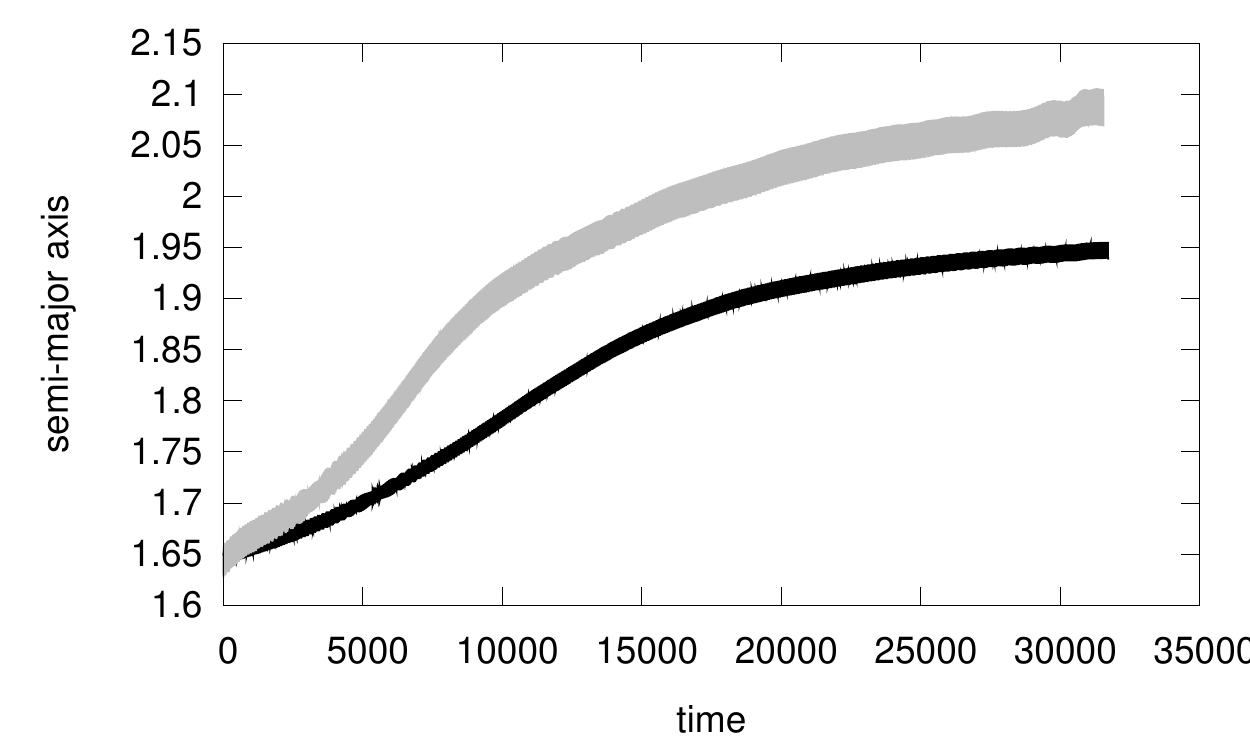}  
\includegraphics[angle=0,width=70mm]{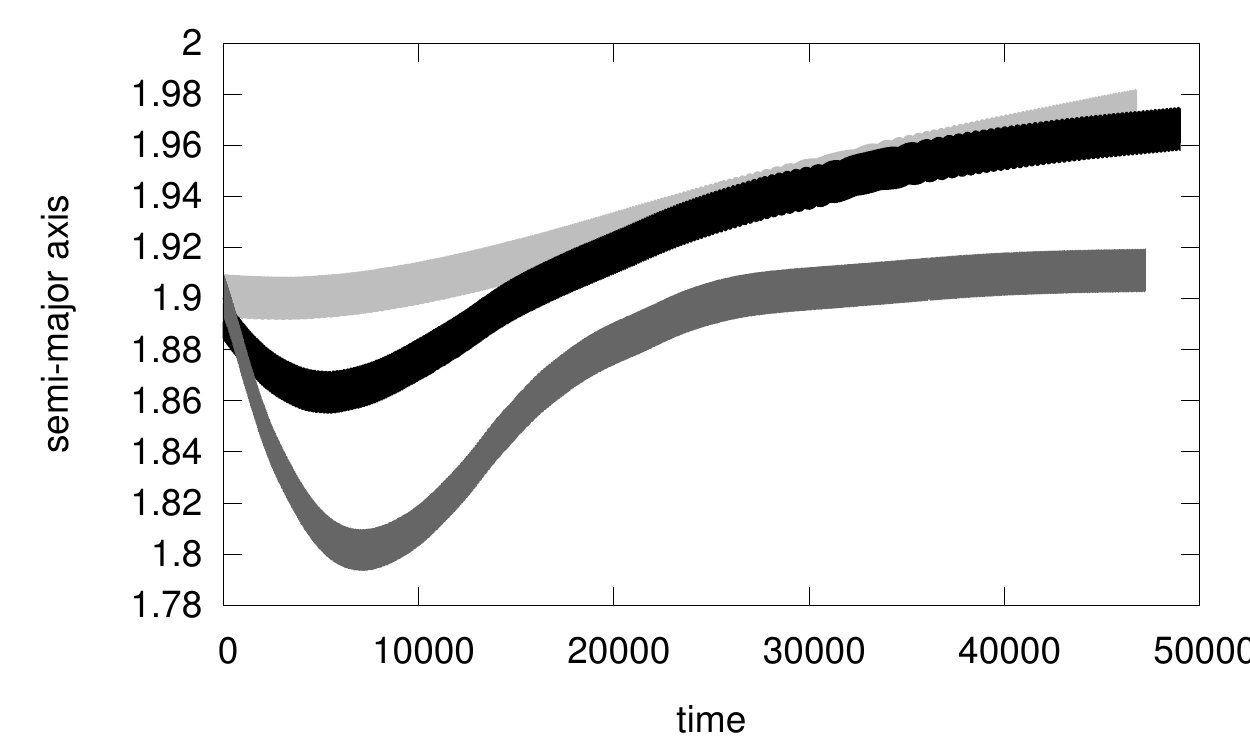}  
\caption{\label{1i2jowisze}{ Orbital evolution   
for simulations starting with the more extensive  
surface density profile given by equation (\ref{longdisc})  is  illustrated.  
The left panel shows the evolution of the semi-major axis   
of a super-Earth in a presence of a gas giant with one  Jupiter-mass   
(black curve) and two Jupiter  masses  (grey curve).  
The initial value is $1.65$ in each case.  
The right panel shows the evolution of the semi-major axis of a   
planet of 1 Earth mass (light grey curve), 5.5 Earth masses  
(black curve) and 10 Earth masses (dark grey curve)  
in  a disc with  
aspect ratio $h=0.05.$ The initial value is $1.9$ in each case.  
}}  
\end{minipage}  
\end{figure*}

 Let us see how the results of our previous  
experiments depend on the adopted surface density profile of the disc.  
 In  the left panel of Fig. \ref{1i2jowisze}  we show the evolution of the  
semi-major  
axis of the super-Earth for  gas  giant  masses of  one Jupiter mass and   
 two Jupiter  
masses. As before, other things being equal,  the super-Earth outward  migration rate  is  larger  
if the gas giant has a larger mass.  
  
Now, we come back to the simulations carried out for  three different masses of the  
low-mass planet (Fig. \ref{1empor}) and redo them starting  
with the more extensive surface density profile  Eq.(\ref{longdisc}).  
We present a comparison of the evolution of planets  
with masses 1, 5.5 and 10 Earth masses in  a disc with aspect ratio  
$ h=0.05.$  
In Fig. \ref{1i2jowisze} (right panel) we show the evolution of their semi-major axes.  
We see that  
the larger is the mass of the planet, the greater is the extent  of  the initial  inward   
migration and the larger the  mass of the planet,  the faster is the subsequent outward   
migration.    
The second of these conclusions  could be inferred from  the previously  
discussed low-mass planet evolution in the disc with the standard  initial surface  
density  profile (Fig. \ref{1empor}) immediately after the reversal of the  
direction of the migration.   
  
\subsection{The effect of varying viscosity}  
\begin{figure*} 
\begin{minipage}[!htb]{160mm} 
\centering 
\includegraphics[width=75mm]{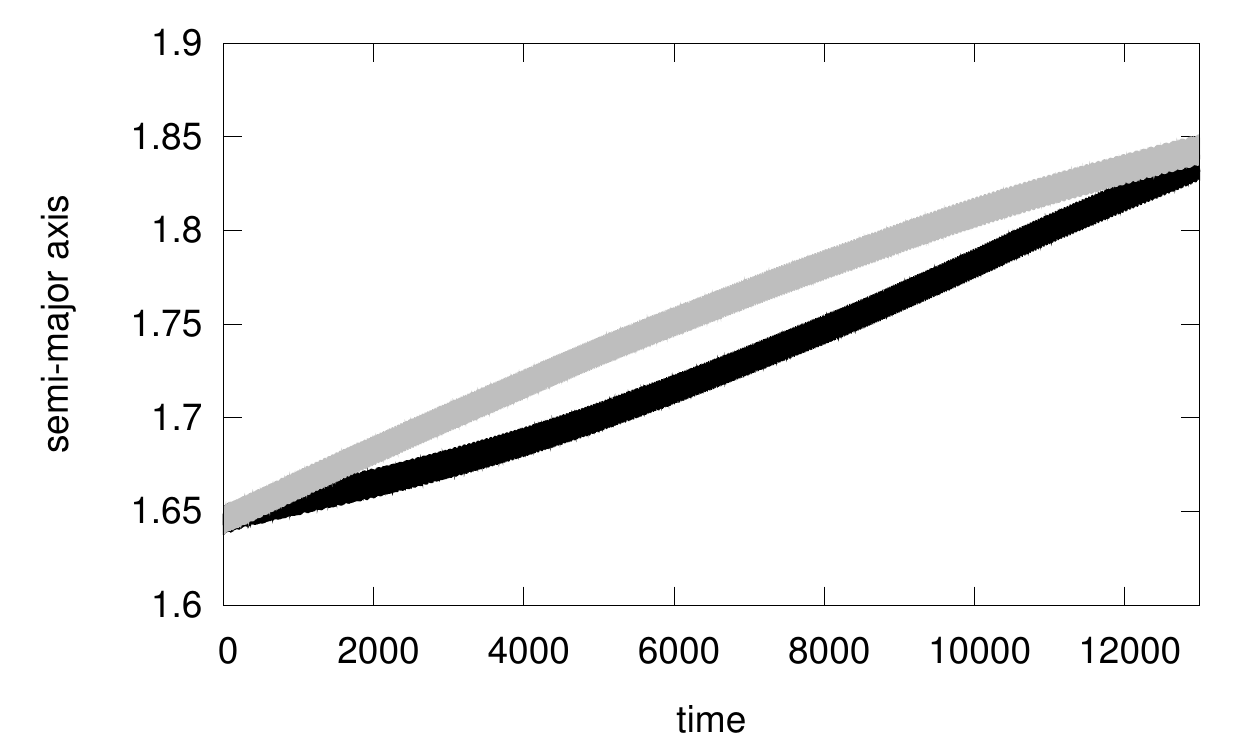}
\includegraphics[width=75mm]{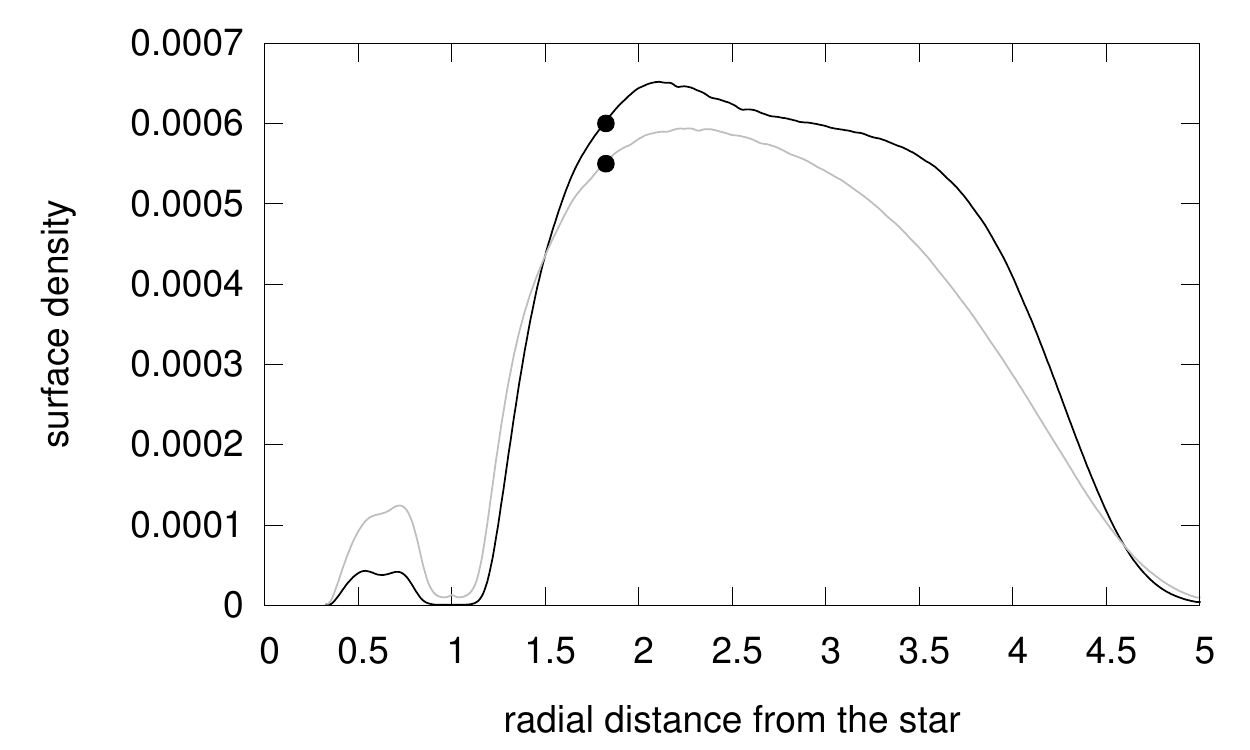}
\end{minipage}
\begin{minipage}[!htb]{160mm}
\centering
\includegraphics[width=75mm]{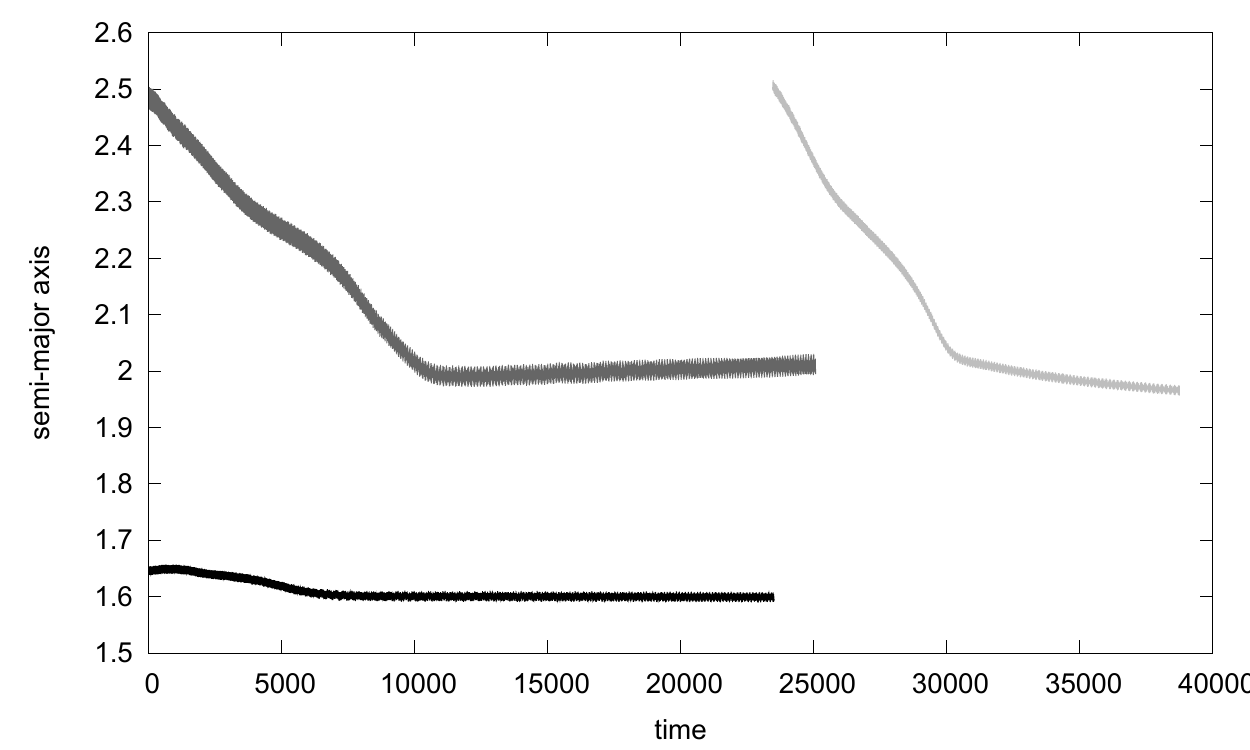}
\includegraphics[width=75mm]{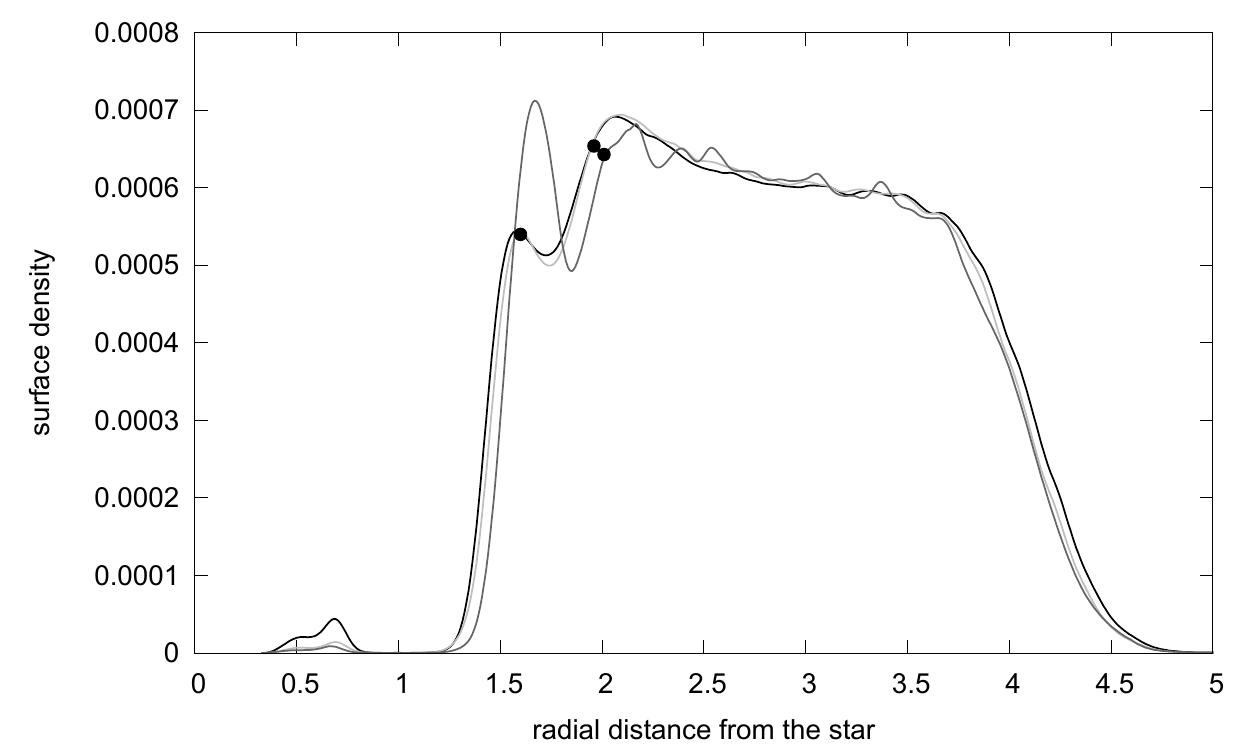}
\end{minipage}
\caption{{\it Upper panels}:
The  semi-major axis of the super-Earth as a function of time
for $\nu=2\times 10^{-6}$ (lower  black curve) and $\nu=10^{-5}$(upper
 gray curve) is plotted
in the left panel.
The azimuthally averaged surface density  profiles, with the planet position indicated,  for
$\nu=2\times 10^{-6}$ (upper  black curve) and $\nu=10^{-5}$(lower
grey curve)
at $t=12000$  are  plotted in the right panel.
   {\it Lower panels}: 
The left panel shows the semi-major axis of the super-Earth as a function of
time for the simulations with $\nu=0.$   Results are presented for a  one Jupiter mass  
giant with a super-Earth starting
at $r=1.65$ (black), $r= 2.5$ in a continuation run (light grey) and a  two Jupiter mass giant 
with a super-Earth starting at $r=2.5$
(dark grey).
The right panel shows  the azimuthally averaged  surface density
as a function of radius with the
final positions of the planets indicated for the three cases.
\label{rysuneklep1}}
\end{figure*}

We have checked that the similar results as found above occur  
for viscosity $\nu$ in the normally considered range $\nu  \le 10^{-5}.$  
For example Fig. \ref{rysuneklep1} compares the migration behaviour  
and azimuthally averaged surface density profile at $t=12000$  
for $\nu=2\times 10^{-6}$ and $\nu= 10^{-5}.$  
For these simulations the giant planet was one Jupiter mass  
and they were initiated  with the  extended surface density profile.  
  

\begin{figure*}
\begin{minipage}[!htb]{160mm}
\centering
\includegraphics[width=70mm]{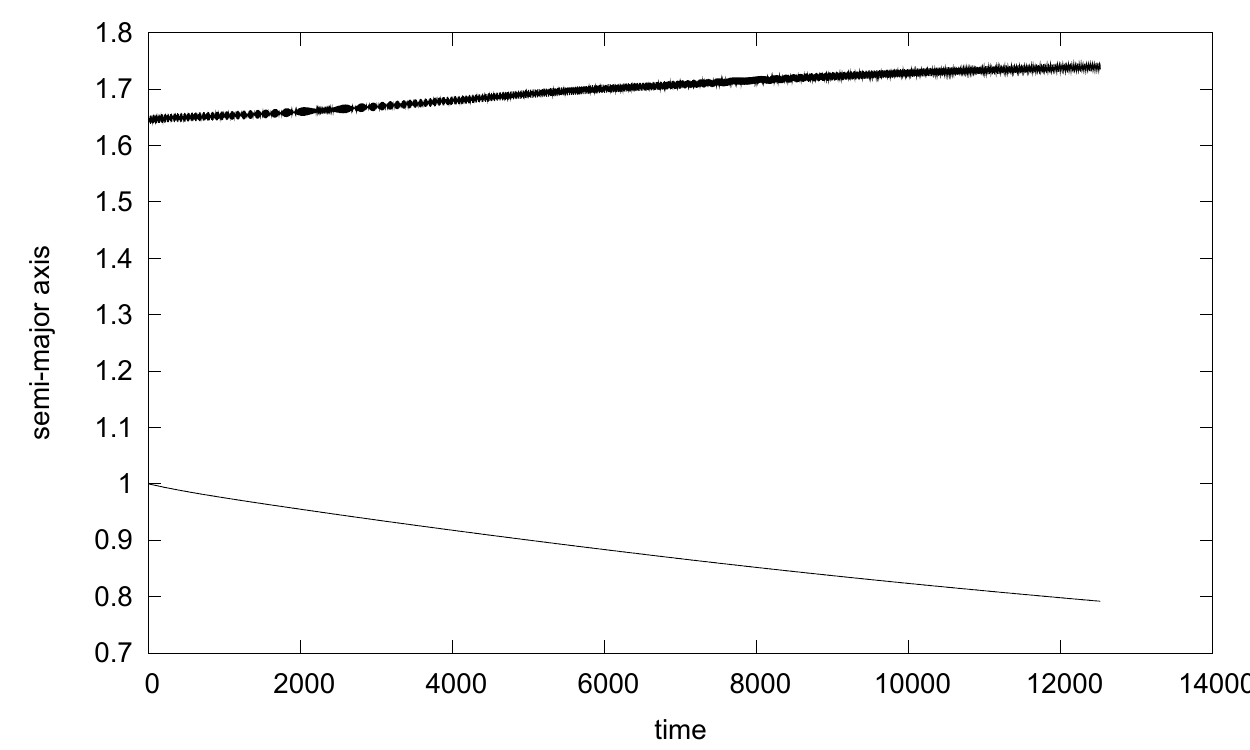}
\includegraphics[width=70mm]{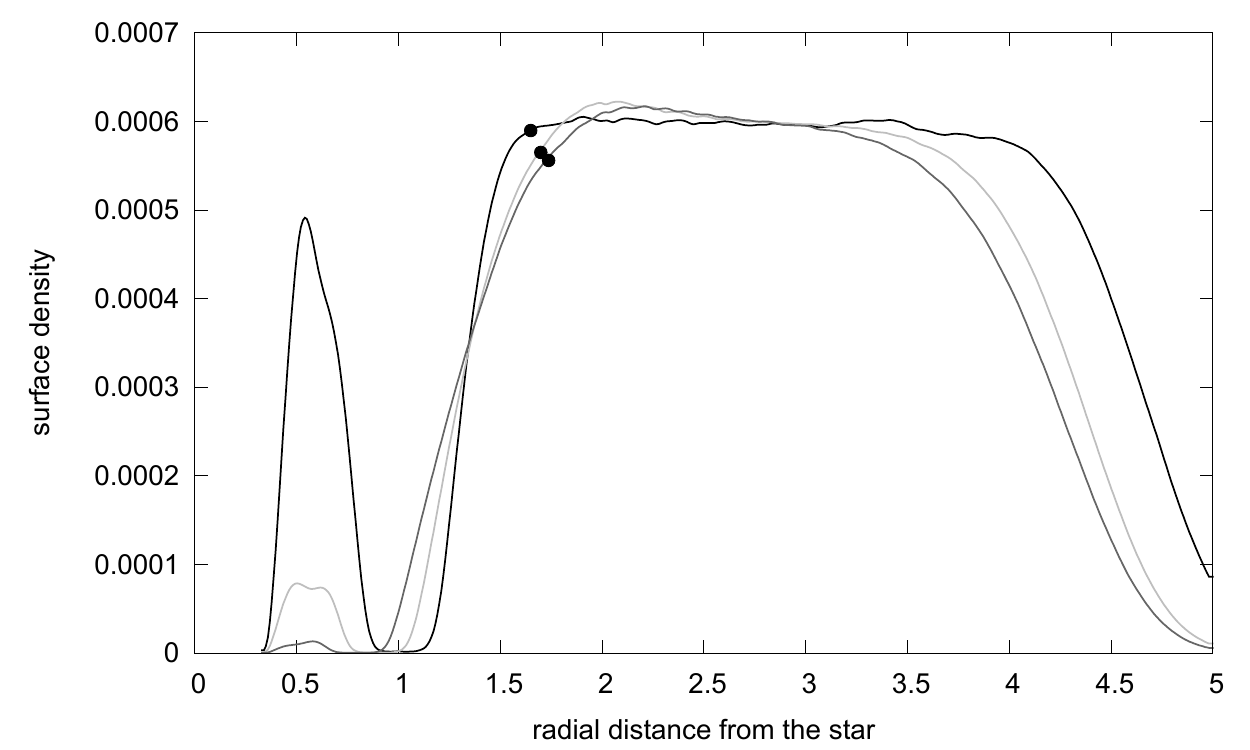}
\caption{\label{rysunekmig}{
  Left: The evolution of the semi-major axes of the super-Earth (upper curve)
and the Jupiter mass  planet (lower curve) when the latter is allowed to migrate. 
Right: The azimuthally averaged  surface density profiles of the disc
 after 400 (black), 6000 (light grey) and 12000 (dark grey)  time units with the  positions of the super-Earth
 indicated.
}}
\end{minipage}
\end{figure*}

The surface density profile is smoother in the case with larger  
viscosity and as might be expected, more material leaks interior to the giant planet  
orbit in this case.  

 We have also confirmed that similar stalled or outward migration occurs
when the applied viscosity is set to zero.
We remark that \citet{devalborro7} find that numerical diffusion 
in NIRVANA under conditions like those adopted here 
corresponds to at most $\nu \sim 10^{-7}$ at a particular grid location
, while usually amounting to much less.   Thus 
 the lack of the smoothing effect
provided by even a small  applied viscosity results in an initial condition dependent
oscillatory form
for the azimuthally averaged surface density profile beyond the gap  that comes about
from the angular momentum  transport induced by the tidal effect of the giant planet
planet acting on the initial disc.
As there is no long term evolution of the outer disc due to the action of an internal viscosity,
 these features persist.
This in turn leads to some differences in the  migrational behaviour  of the low mass
planet when compared to cases with a  non zero $\nu.$
In these simulations  we  adopted an initial 
disc with the extended surface density profile and   aspect ratio $h=0.05.$ 
 The mass of the super-Earth was $5.5$ Earth masses.

We have considered the case of a non migrating Jupiter mass  giant planet
with an initial radius for the super-Earth, $r=1.65.$
The results of this simulation
as well as the others undertaken with zero applied viscosity
 are illustrated in the lower panels of Fig. \ref{rysuneklep1}.
The low mass planet migrated inwards initially but then subsequently halted
with radius  slightly exceeding $1.6$ dimensionless units.
This situation could be followed for about $1500$ orbits of this planet.
Because the azimuthally averaged surface density in these cases
has an oscillatory structure
as a function of radius, that   may result 
from the effect of density waves being launched at different Lindblad resonances 
 \citep[see eg.][] {gt1979}, we restarted this run 
after moving the super-Earth to  a new radius $r=2.5.$ It was  eventually found to migrate inwards
but, when compared to the original rate, at a very small and decreasing rate. This indicates the possibility of either reaching $\sim r=1.6$  or stalling
at some point further out. 
At $r\sim 2,$ where inward migration starts to slow,   the surface density profile is 
affected by tidal effects due to the giant planet and thus the migration of the
super-Earth can be affected.
Finally we ran the same simulation but with the giant planet mass taken to be two Jupiter masses, starting with the same initial surface density profile
but with the initial radius for the super-Earth taken to be $r=2.5.$
In this case the initial inward migration stalls at $r\sim 2$ after which the
planet undergoes slow outward migration which could be followed for $\sim 1000$
planet orbits.

We comment that in all of these cases that have sustained
interrupted   inward migration over long periods of time,
we do not see any removal of the effect  on
an estimated libration time scale for the horseshoe region, as might be
anticipated if corotation torques operated in an 
 otherwise quiescent coorbital region.
In this context we remark that parameters are comparable to those adopted 
in  \citet{pp2008}  (see eg. their Fig. 21) with the libration period being 
$100-200$ planet orbits, which is much less than the duration of our simulations.  A
 similar situation applies to the inviscid local simulations  we have carried out 
 (see  the discussion at the end of section \ref{LOCALSIM}).
\subsection{ The effect of  migration of the giant planet on the migration of the super-Earth}\label{NOTMIGG}
 To illustrate  the  effect on the migration of the super-Earth of allowing  the giant planet to migrate,
we present the results of a simulation starting with the extended surface density profile, 
aspect ratio h=0.05 and $\nu =2\times 10^{-6}.$
 The mass of the super-Earth was $5.5$ Earth masses. Thus this run has the same parameters  as  for a corresponding one
presented in
the upper panels of  Fig. \ref{rysuneklep1}, except that  the Jupiter mass planet is allowed to migrate in response to the disc
torques acting on it. Results from this simulation are illustrated in
Fig. \ref{rysunekmig}. 

If effects that interrupt inward migration are associated with the  neighbourhood
of the giant planet, the super-Earth should naturally tend to  move with it.
 In this case,  although the giant moves from $r=1$ to $r=0.8,$
the super-Earth still moves slowly outwards. However, the ratio of its orbital radius to that
of the giant reaches $\sim 2$ where  calculations  presented above indicate that
tides still operate. Thus the above simulation is approximately consistent with the
super-Earth maintaining the same relationship with the giant planet as when 
it does not migrate.

\section{ Local shearing box simulations  of a  planet  embedded in a disc  
with  independently excited density  waves }  
\label{shearingbox}  
  
The results obtained in Section \ref{supergiant} for the migration of  a super-Earth   
may be complicated by the   
presence of a  deep gap opened by the gas giant and the particular way that  
forces density waves in the disc.

To establish the fact that the essential features are a planet embedded    
in  a disc hosting outward propagating density waves,  
 we decided to investigate the phenomenon by  performing   
local simulations with a planet  embedded in a disc  with   density waves  
independently forced by an independent mechanism  
that did not involve additional planets producing a very deep gap.  
 These simulations may also be performed at much  higher resolution  
of the coorbital region than the global ones and could be used to show that the qualitative nature of  
process was robust to changes in parameters such as gravitational softening  
density wave amplitude and applied viscosity.

\subsection{Local Model}\label{modelboxeq}

  
\begin{figure*}  
\begin{minipage}[!htb]{170mm}  
\centering  
\vspace{-0cm}  
$\begin{array}{cc}  
\hspace{-2cm}\includegraphics[width=90mm,height=65mm]{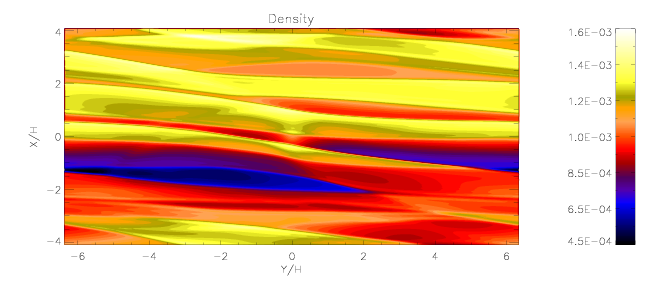}&  
\includegraphics[width=70mm,height=65mm]{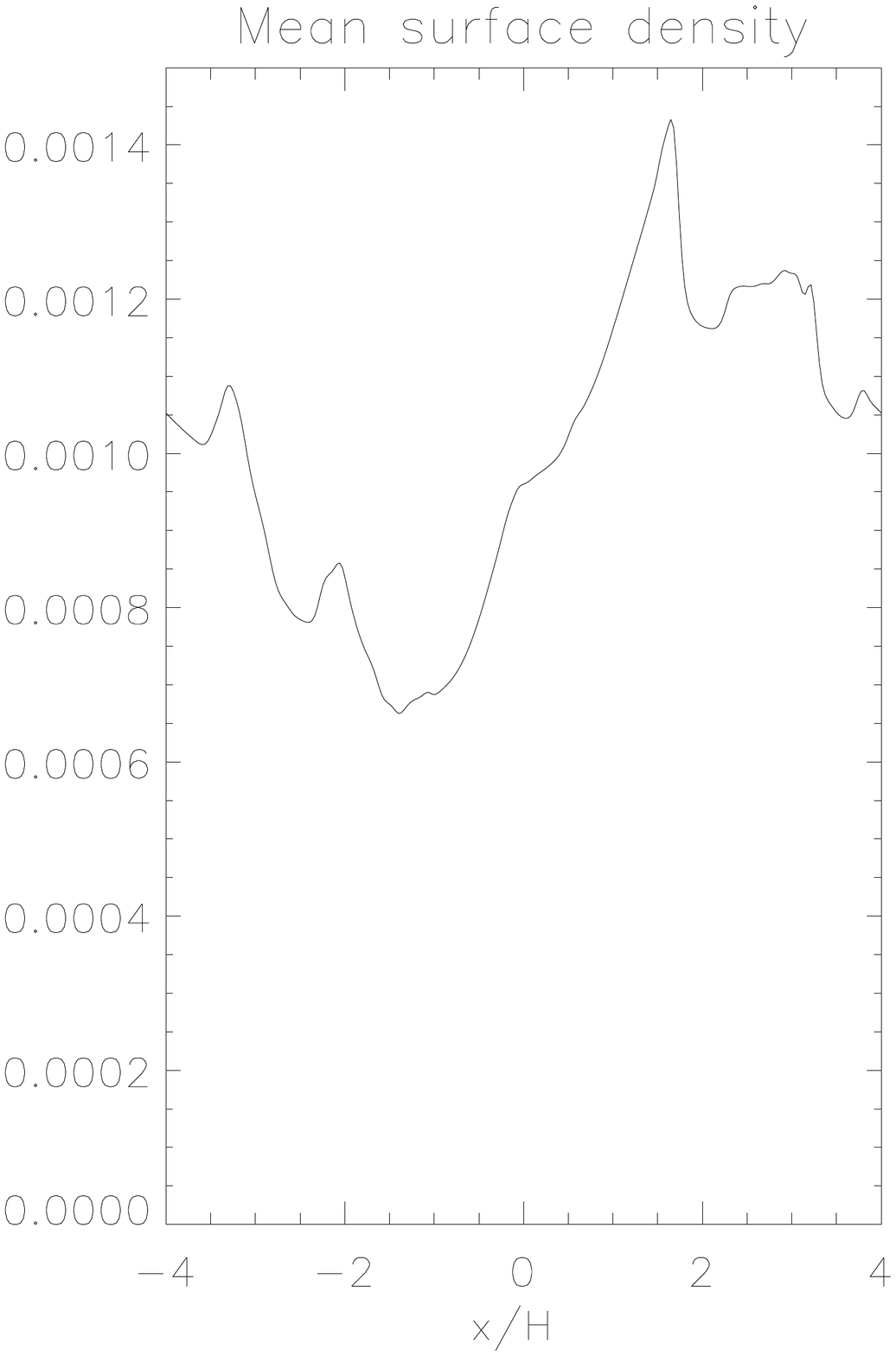}\\  
\end{array}$  
\end{minipage}  
\begin{minipage}{170mm}  
\centering  
\vspace{-0cm}  
$\begin{array}{cc}  
\hspace{-1cm}\includegraphics[width=80mm,height=65mm]{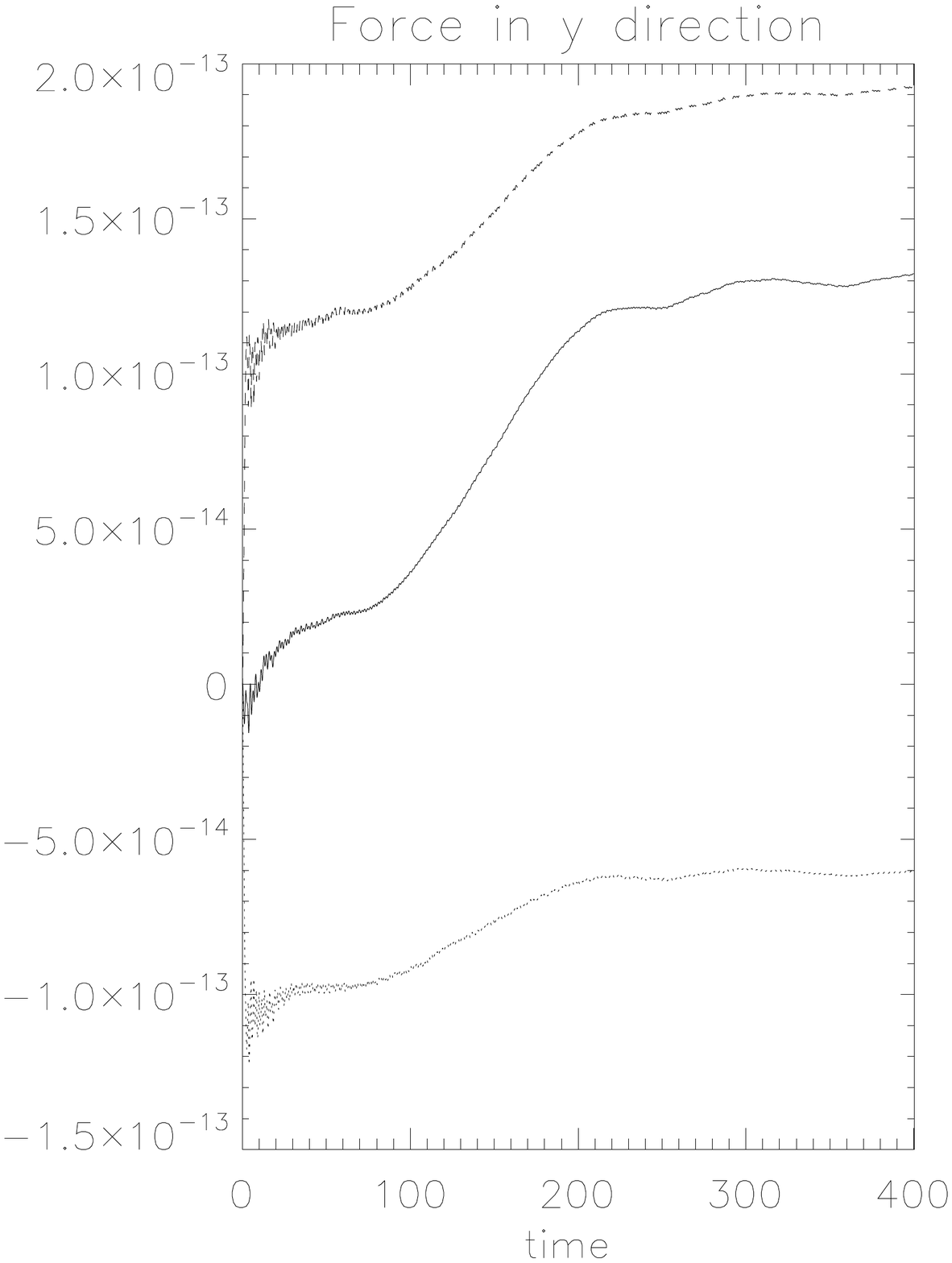}&  
\hspace{0.9cm} \includegraphics[width=90mm,height=65mm]{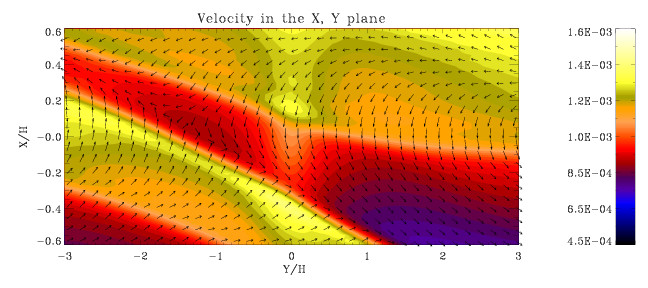}\\  
\end{array}$  
\end{minipage}  
\begin{minipage}{170mm}  
\centering  
\vspace{-0cm}  
$\begin{array}{cc}  
\hspace{-1cm}\includegraphics[width=90mm,height=65mm]{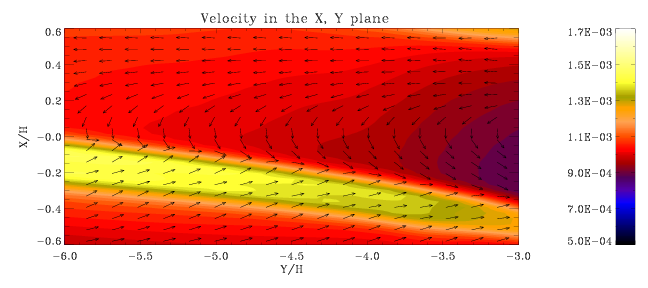}&  
\includegraphics[width=90mm,height=65mm]{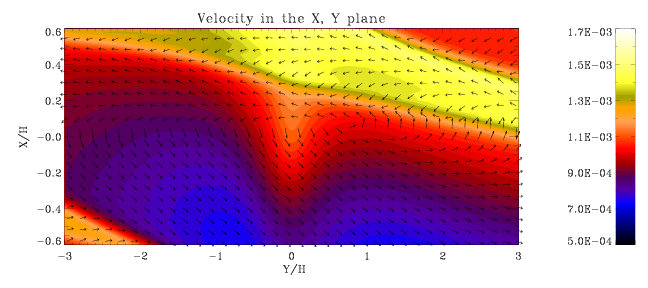}\\  
\end{array}$  
\caption{\label{inviscid} Results are shown for the case with  
$q=0.1,$ $C_0=0.2,$ and no applied viscosity$, (\nu=0).$   
The upper left panel shows surface density contours $48$  
orbits after initiation. The upper right panel shows  
the mean surface density profile averaged over the $y$ direction  
$56$ orbits after initiation. The units are such that the uniform initial value  
for all local simulations was $0.001.$ The central left panel   
shows  force components in the $y$ direction acting on the planet  
evaluated as a running time average. The units are arbitrary but the same for  
all local simulations illustrated.  
The uppermost  plot gives the contribution from $x< 0.$  
The lowermost plot  gives the contribution from $x> 0.$  
The central plot  shows  the total force component in the $y$   
direction acting on the planet  due to material in the computational domain.    
The central right hand panel indicates  streamlines  superposed on surface density  
contours for the domain $-3 < y/H < 3$ and   $-0.6 < x/H<  0.6$  
$48$ orbits after initiation.  
The lower left panel indicates   streamlines   superposed on surface density  
contours for the domain $-6 < y/H <-3$ and   $-0.6 < x/H<  0.6$ $56$  
orbits after initiation.  
The right hand panel gives the corresponding plot for    
$-3 < y/H < 3$ and   $-0.6 < x/H<  0.6.$ }  
\end{minipage}  
\end{figure*}

  
\begin{figure*}  
\begin{minipage}[!htb]{170mm}  
\centering  
\vspace{-0cm}  
$\begin{array}{cc}  
\hspace{-0cm}\includegraphics[width=90mm,height=65mm]{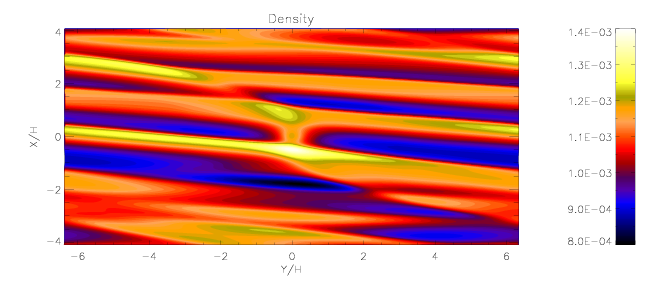}&  
\includegraphics[width=70mm,height=65mm]{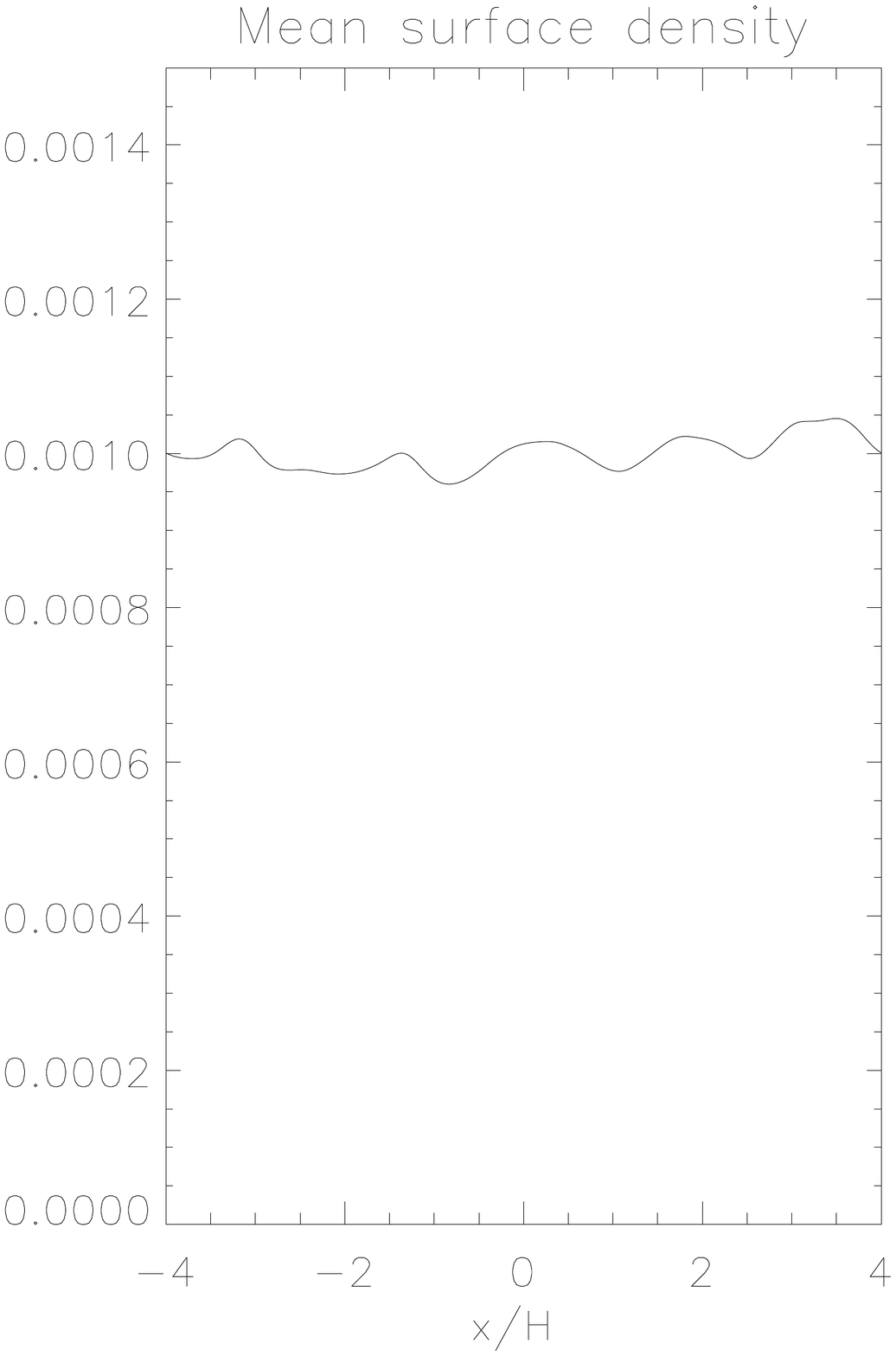}\\  
\end{array}$  
\end{minipage}  
\begin{minipage}{170mm}  
\centering  
\vspace{-0cm}  
$\begin{array}{cc}  
\hspace{-0cm}\includegraphics[width=80mm,height=65mm]{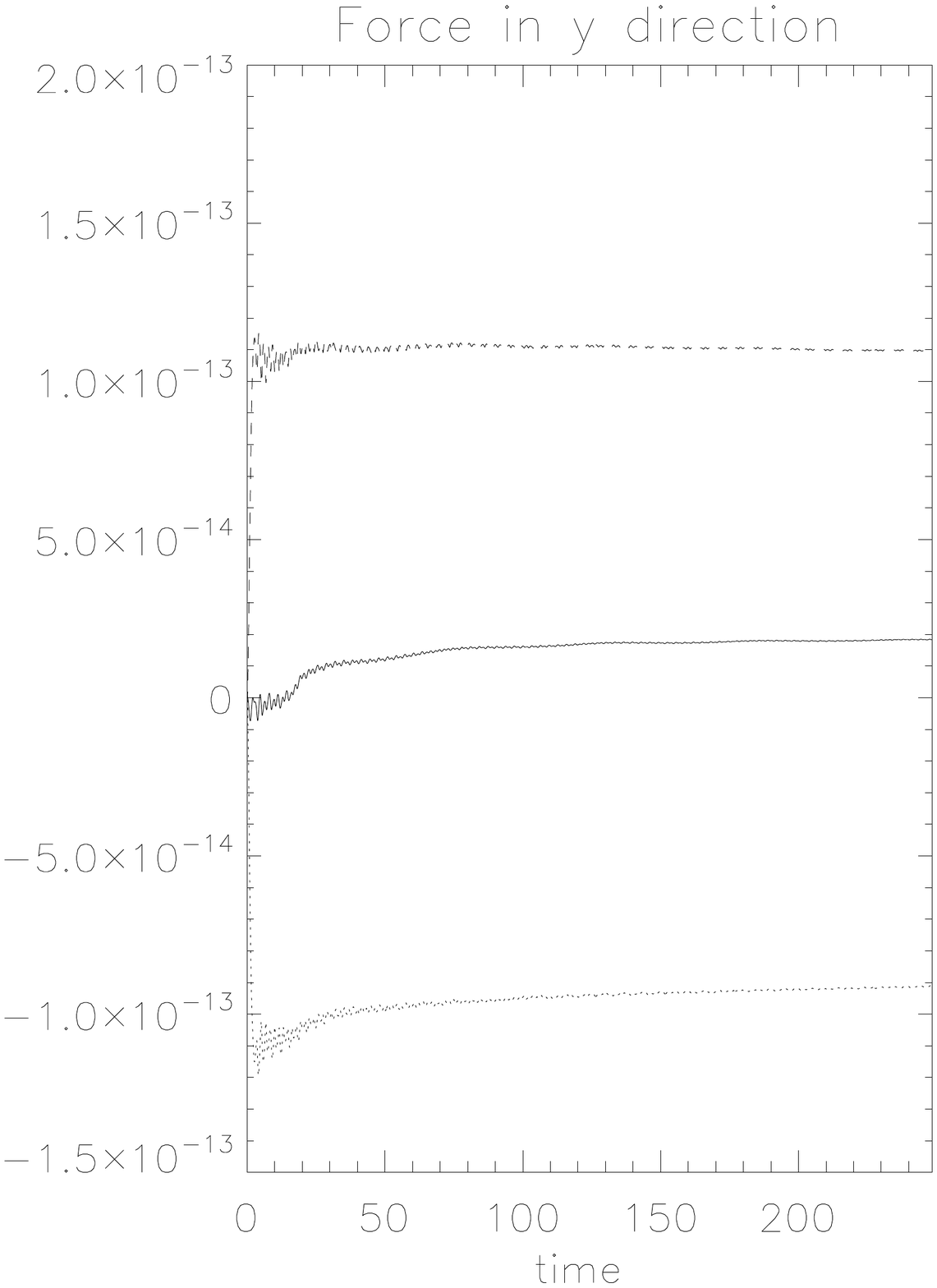}&  
\includegraphics[width=90mm,height=65mm]{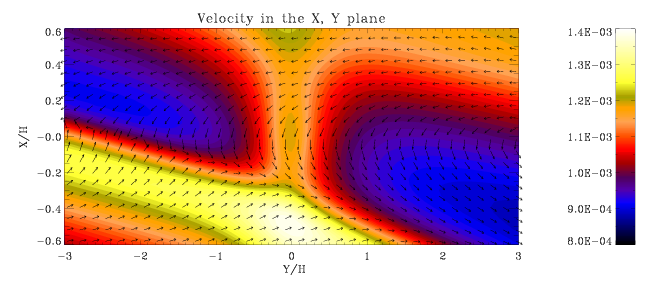}\\  
\end{array}$  
\caption{\label{lowamv}   
 Results are shown for the case with  
$q=0.1,$ $C_0=0.1,$ and  $\nu/(H^2\Omega_p)=0.02.$  
The upper left panel shows surface density contours $48$  
orbits after initiation. The upper right panel shows  
the mean surface density profile averaged over the $y$ direction  
$48$ orbits after initiation. The lower  left panel  
shows  force components in the $y$ direction acting on the planet  
evaluated as a running time average.  
The uppermost  plot gives the contribution from $x< 0.$  
The lowermost plot gives the contribution from $x> 0.$  
The central plot  shows  the total force component in the $y$  
direction acting on the planet  due to material in the computational domain.  
The lower  right hand panel indicates  streamlines  superposed on surface density  
contours for the domain $-3 < y/H < 3$ and   $-0.6 < x/H<  0.6$  
$48$ orbits after initiation.  
}  
\end{minipage}  
\end{figure*}


\begin{figure*}  
\begin{minipage}[!htb]{170mm}  
\centering  
\vspace{-0cm}  
$\begin{array}{cc}  
\hspace{-0cm}\includegraphics[width=90mm,height=65mm]{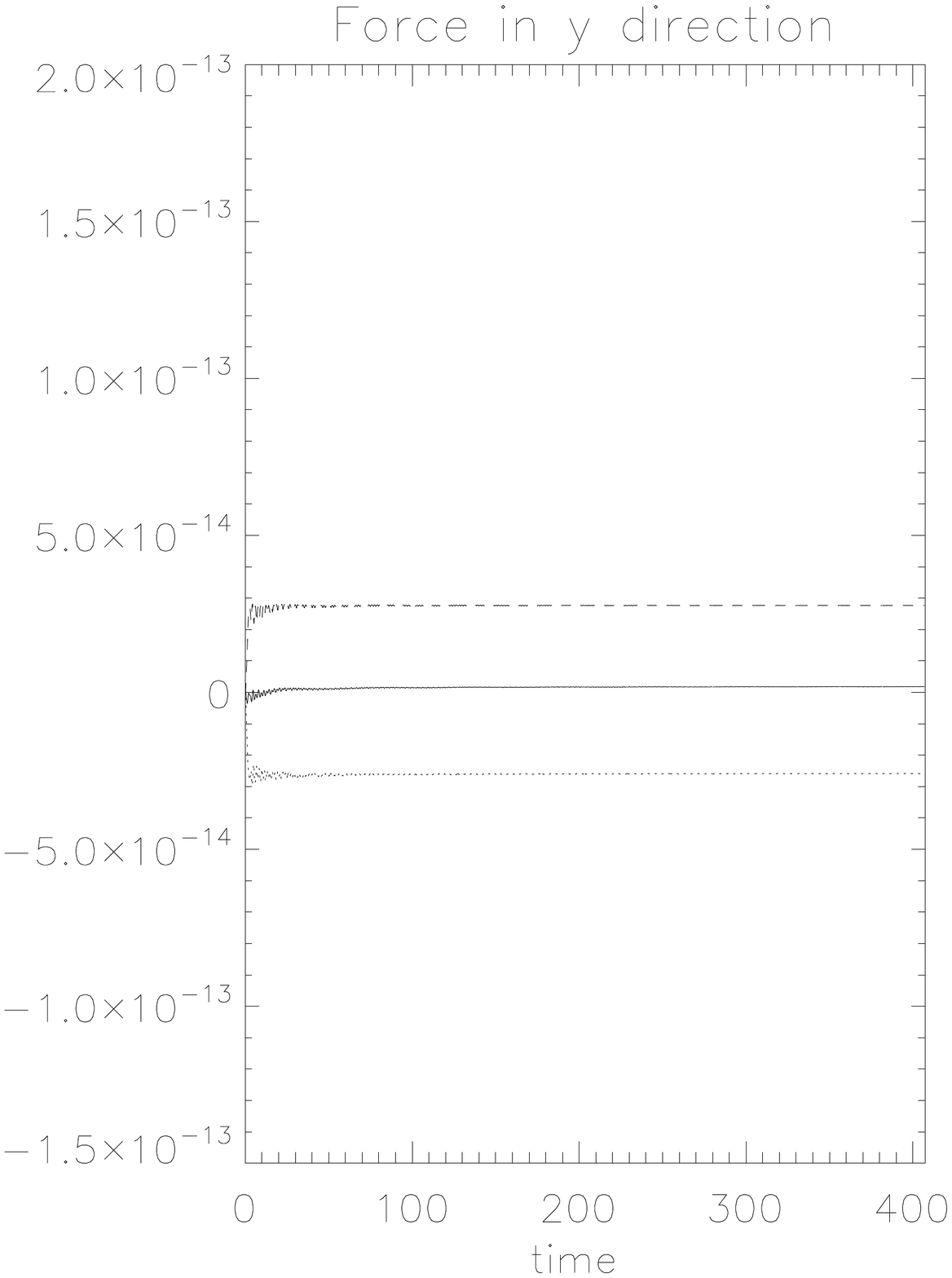}&  
\includegraphics[width=70mm,height=65mm]{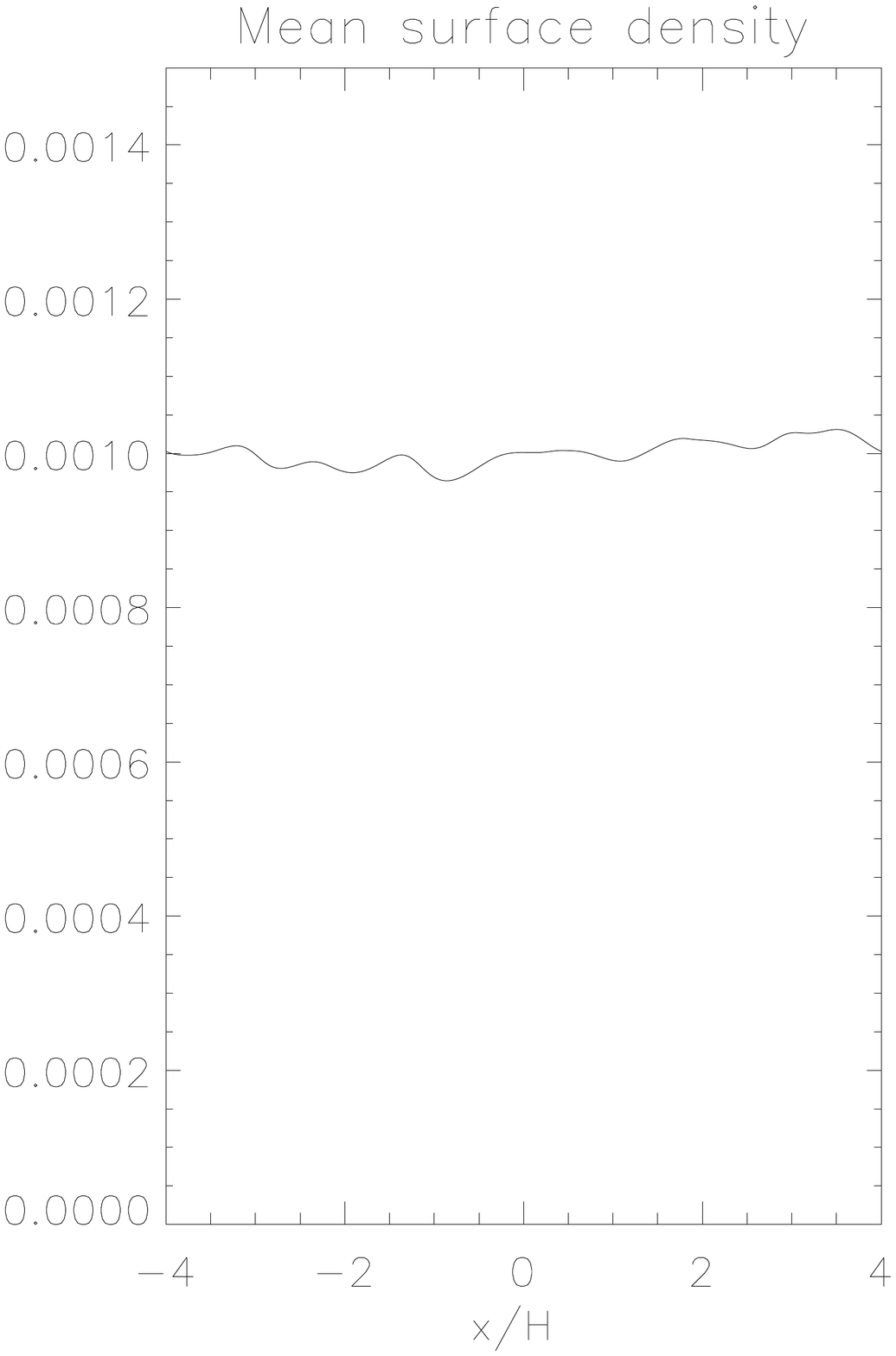}\\  
\end{array}$  
\end{minipage}  
\begin{minipage}{170mm}  
\centering  
\vspace{-0cm}  
$\begin{array}{cc}  
\hspace{-0cm}\includegraphics[width=80mm,height=65mm]{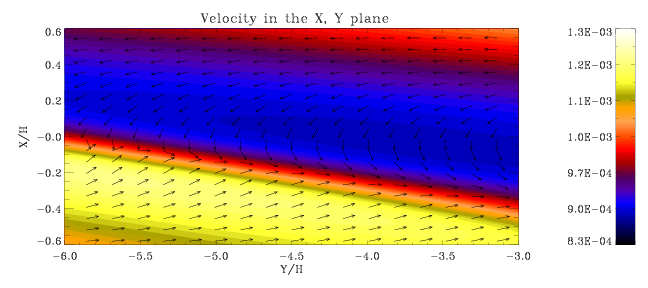}&  
\includegraphics[width=90mm,height=65mm]{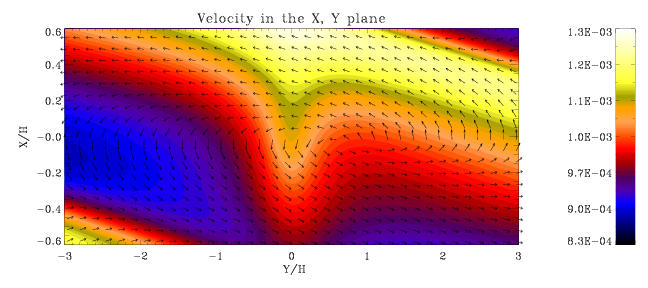}\\  
\end{array}$  
\caption{\label{lomlowamv}  
 Results are shown for the case with  
$q=0.05,$ $C_0=0.1,$ and  $\nu/(H^2\Omega_p)=0.02.$  
 The upper  left  panel  
shows  force components in the $y$ direction acting on the planet  
evaluated as a running time average.  
The uppermost  plot gives the contribution from $x< 0.$  
The lowermost  plot gives the contribution from $x> 0.$  
The central plot  shows  the total force component in the $y$  
direction acting on the planet  due to material in the computational domain.  
 The upper right panel shows  
the mean surface density profile averaged over the $y$ direction  
$56$ orbits after initiation.  
The lower left panel indicates   streamlines   superposed on surface density  
contours for the domain $-6 < y/H <-3$ and   $-0.6 < x/H<  0.6$ $56$  
orbits after initiation.  
The right hand panel gives the corresponding plot for  
$-3 < y/H < 3$ and   $-0.6 < x/H<  0.6.$ }  
\end{minipage}  
\end{figure*}  
  

  We adopt a local  Cartesian  coordinate system $(x,y,z)$  
  uniformly rotating with angular velocity $\Omega_p{\hat{\bf k}}$  
 and  with origin located at a point where the Keplerian   
angular velocity is $\Omega_p$ as seen in the inertial frame.  
The local radial coordinate is $x$  and $y$ is the orthogonal  
coordinate pointing in the direction of the shear.   
The governing hydrodynamic equations are the continuity  
equation  
\begin{align}\label{continuitybox}  
\frac{\partial \Sigma}{\partial t}=-\nabla\cdot(\Sigma \mathbf{u}),  
\end{align}  
and the equation of motion  
\begin{align}\label{momentumbox}  
\frac{\partial \mathbf{u}}{\partial t} + \mathbf{u}\cdot\nabla \mathbf{u}+2\Omega_p\hat{\mathbf{k}}\wedge\mathbf{u} =  
3\Omega^2_px {\bf i} -\frac{1}{\Sigma}\nabla P - \nabla\Phi +{\bf f}_{\nu},  
\end{align}  
where ${\bf i}$ is the unit vector in the $x$ direction.  The gravitational potential, $\Phi,$   for these simulations,  
is taken to be due to the single planet of mass $M_2,$ assumed to be fixed at the origin  
of the rotating coordinate system,  and the action of  
a specified  external forcing potential $\Phi_{ext}.$  
The potential due to the planet is given by  
\begin{align}  
\Phi_p &=   
-\frac{GM_2}{\sqrt{x^2+y^2 +  
\epsilon^2_2}} +\Phi_{ext}, \hspace{2mm}  x^2+y^2 < d_{max}^2, \hspace{1mm} {\rm and} \notag\\  
\Phi_p &= -\frac{GM_2}{\sqrt{d_{max}^2 +  
\epsilon^2_2}} +\Phi_{ext}, \hspace{2mm}  x^2+y^2 > d_{max}^2  .   
\end{align}  
  
The simulations we undertake are two dimensional  with no vertical dependence, and the gravitational potential  
is chosen to be consistent with this.  
We remark that the indirect term and self-gravity are  neglected in this model and as before  
$\epsilon_2$ is the gravitational softening length for the planet potential.  
For the simulations reported in detail here we adopted $\epsilon_2=0.3H$ and $d_{max}=3H.$  
The simulation set up is similar to that described in   
\citet{papnels2004} with the adoption of the boundary condition  
that the solution be periodic in shearing coordinates.   
The distance scale,  $d_{max},$ is used to flatten the planet   potential for $ x^2+y^2 > d_{max}^2$  
in order to make it effectively periodic in shearing coordinates when applied to the computational domain.  
The length scale is  $H,$ which as for the  
global models,  can be regarded as the putative  
local disc semi-thickness. The equation of state is thus strictly isothermal  
with $P=\Sigma H^2\Omega_p^2.$  The viscous force per unit mass,  ${\bf f}_{\nu},$ is taken to be   
the standard Navier Stokes form with kinematic viscosity $\nu$ as in the global simulations.

The planet mass  appears only through specification of the  
dimensionless ratio $ q \equiv   (M_2/M_*)/h^3$~\citep{papnels2004} which has to be specified for  
each simulation. Because all lengths are expressed in terms of $H$ and times in units of $\Omega_p^{-1},$  
no other parameters need to be specified. However,  
 when  $M_*$ is specified, the aspect ratio  $h \equiv H/R$ can be found once,   $R,$  the   
orbital radius of the origin of the coordinate system is in turn  found from  
Kepler's law in the form  $R^3= GM_*/\Omega_p^2.$

The computational domain was a  rectangle of side  $8H$ in the $x$ direction and $4\pi H$ in the $y$ direction.  
We adopt the convention that the positive $x$ direction corresponds to the outward or outer direction and the opposite  
direction to the inward  or inner direction.  
We have performed simulations with the equally spaced  grid resolutions $(N_x,N_y)= (261,300)$   
which we refer to as the standard resolution  and also $(N_x,N_y)= (522,600),$ which we refer to as higher resolution,  
in order to confirm convergence of the results. Calculations were initiated  
with a uniform surface density and velocity ${\bf u}= (0,-3\Omega_px/2,0)$ corresponding to  local  
Keplerian shear.  
The external forcing potential was applied in order to excite density waves and so for these simulations  
plays the role of the giant planet.  For $0 < x-x_{min} < 2H,$  we adopted the form  
  
\begin{align}  
\Phi_{ext} =& C_0 \Omega_p^2 H^2 \sin( \pi(x-x_{min})/(2H))\times \notag\\  
&\cos((y-y_{min} - v_{wave}t )/(2H)),   
\end{align}  
where $x_{min}$ and $y_{min}$ are respectively the minimum values of $x$ and $y$  
in the computational domain, the  propagation speed  
$ v_{wave}= (3\pi - 3/2)\Omega_p H$  
 and  $C_0$ is a scaling constant specified for each simulation.   
For $ x-x_{min} >  2H$ no forcing was applied.  
This is of the form of a harmonic forcing  with a single  wavelength  in the $y$ direction   
while being localized in the $x$ direction. It is associated with a propagation speed $v_{wave}$ in the $y$ direction  
so that there is a pattern speed as seen by the super-Earth which is at rest  in the  
centre of the  computational domain. This plays the role of the corresponding pattern speed  
of the giant planet in the global simulations.  For the case we have adopted, the putative   
material that would comove with the  
potential perturbation is located at $x= -(2\pi-1)H,$ which is a distance $1.28H$ interior to the inner  
boundary. Accordingly this mimics a perturbation from an  orbiting inner planet.  
The outer Lindblad resonance is located in the computational domain at a distance $0.21H$ from the  
inner boundary.  For the particular  choice of parameters  we have made, density waves launched from the outer Lindblad  
resonance propagate outwards. As the inner Lindblad resonance is not present  effects due to inward propagating  
waves are minimized as,  would be expected if the excitation was indeed due to an inner planet.  
As the waves propagate across the whole domain without further excitation for  
 $ x-x_{min} >  2H,$   their dissipation is optimized.  This feature also limits  
  the influence of disturbances from the periodic system of boxes that  
arise through   application of the shear periodic  boundary   conditions.


  
\subsection{Local simulation results} \label{LOCALSIM} 
We begin by considering a case with  
$q=0.1,$ $C_0=0.2,$ and zero  applied Navier-Stokes  viscosity.  
The density profile is the most strongly perturbed in this simulation which was run for  
$400\Omega_p^{-1}$ time units. Some results  
are shown in Fig. \ref{inviscid}.  
Surface density contours are shown $48$  
orbits after initiation. These typically dominated by the trailing waves  
excited by the potential $\Phi_{ext}.$ The wake associated with the embedded  
planet can scarcely be seen.   
The  surface density profile averaged over the $y$ direction  
shows that the excited waves causes  momentum and mass transport  
that results in material moving from inside the planet to outside the planet.  
Tests have shown that this effect weakens if $\nu$ is increased and/or the  
forcing amplitude $C_0$ is decreased. We also plot the running time averages of the   
force components acting on the planet  in the $y$ direction.  
As expected, these show that the contribution from exterior/interior to the planet is   
decrease/increase the $y$ momentum respectively. For these runs the exterior contribution  
is always smaller resulting in a net outward force on the planet. In the case of the  
simulation described above, the net force  is $~60\%$ of the one sided interior force  
contribution. Note that these forces convert to torques when scaled by the radius  
of the centre of the box. Tests show the net force, though always  
outwards,  decreases in magnitude  with wave forcing amplitude,   
planet mass and gravitational softening parameter. It is also important to   
remark that although the density response is dominated by the forced waves  
the response  of the planet is essential in order to produce a net mean force.  
The specific mean force is found to tend to zero with the planet mass.  
We also show the streamlines  superposed on surface density  
contours for the domain $-3 < y/H < 3$ and   $-0.6 < x/H<  0.6$  
$48$ orbits after initiation in Fig. \ref{inviscid}. In this case they indicate a structure like  
that associated with  normal horseshoes but the planet  
is not precisely centered  on a separatrix as would be the case if pressure effects  
were negligible. This type of structure was also seen in global simulations e.g. in   
the uppermost right panel of Fig. \ref{fig3def}  
We remark that as for the global simulations the streamline configuration varies with time  
and so streamlines do not precisely indicate  particle trajectories.  
However, by following fluid elements we have found that material initially with $|x|<0.25H$  
remains close to the planet such that $|x|< \sim 0.5H$ in all simulations, which  
indicates that the fluid particles do indeed turn.  
Another streamline configuration    
we find is illustrated in the domain $-6 < y/H <3$ and   $-0.6 < x/H<  0.6$ $56$  
orbits after initiation in Fig. \ref{inviscid}. In this case the flow is strongly  
affected by the density waves, being deflected  to move parallel to shocks (see lower left panel).  
This has the consequence that the flow  through $x=0$ is    
in many places reversed compared to  that expected  for  the usual   
horseshoe configuration, with separatrix located far from the planet.   
This type of configuration was also seen in the global simulations   
as in e.g. the middle  right panel of Fig. \ref{fig3abc}  
  
 In order to further illustrate the forces on an embedded planet induced  
by independently excited density waves we show results from two additional simulations.  
The first was for the same parameters as above but with an applied Navier-Stokes viscosity  
given by   $\nu/(H^2\Omega_p)=0.02.$  
Results for the simulation are presented in Fig. \ref{lowamv}.  
In this case the  presence of the wakes due to the embedded  
planet are more apparent in the surface density contours than the previous case.  The   
mean surface density profile averaged over the $y$ direction  
$48$ orbits after initiation, though  showing  some oscillatory structure,   
is generally much flatter. The increased value of $\nu$  
has resulted in lower amplitude waves with less associated angular momentum  
transport.    
The total mean  force component in the $y$  
direction acting on the planet  due to material in the computational domain is  
positive but about seven times smaller than the previous simulation again because of the smoothing action of viscosity.  
As for the previous simulation, the   streamlines    
in  the domain $-3 < y/H < 3$ and   $-0.6 < x/H<  0.6$  
$48$ orbits after initiation show indications of normal horseshoe turns  
but the streamlines are distorted  by the presence of the waves which cause the planet  
to be no longer centered on a separatrix and the flow to be parallel to shock fronts.  
  
 The second case we consider is the previous one but now we reduce the planet mass to  
$q=0.05.$  The results are shown in Fig. \ref{lomlowamv}.  
 In this case the  running time average  
of the total  force component in the $y$ direction acting on the planet  
is positive but smaller than for the other simulations we have discussed.  
 The  mean surface density profile averaged over the $y$ direction  
$56$ orbits after initiation is flatter.  
Interestingly the    streamlines  for the domain $-6 < y/H <3$ and   $-0.6 < x/H<  0.6$ $56$  
at this time are qualitatively similar to those for the case $\nu=0$ at the same time.

   We remark that   we have found, by following an appropriate set of fluid elements,
    that there is no mean mass flow through the
coorbital region of  the planet. Although   the planet  is  located on a  rising surface density profile
in  the case with $\nu =0,$ the surface density structure is  maintained by density waves rather than  by either   a mass flow,
 and/or applied  viscous stresses, 
 as needed to maintain  the surface density structure in the planet trap of \citet{masset}.
In addition we  note that,  although coorbital effects are associated with the forces/torques  
acting on the planet, it seems that the concept of saturation of corotation resonances  
  significantly affecting the positive force in the $y$ direction,
plays no role. For the simulations presented here, such effects should be manifest  
over a libration or return period which corresponds to a time of $64$ units for  a   
distance $0.25H$ from the planet. The mean force behaviour is established   
more quickly that this and remains for much longer periods of time.  
The increase of the total force magnitude that occurs on a long time scale when $\nu=0$  
is due to the secular effects associated with angular momentum transport due to the waves.


\section{The mechanism leading to outward migration of the super-Earth}  
\label{theory}  
In the previous sections we have described  the results of numerical  
experiments which  indicate that   
outward migration of  a low mass planet  may occur  in a disc in the presence of   
density waves  excited by a forcing potential,  
which in the global case was that due to  a giant planet.  
We now consider this process further  
with the aim of investigating  its nature.   
  
\subsection{Migration of the super-Earth without the source of density waves}  
\label{restore}  
  
The presence of a gas giant in the close internal  
orbit interior to  that of the   
low mass planet  affects the migration of the low-mass planet  
halting or reversing its migration.  This occurs independently   
of the inclusion of the direct gravitational   
interaction between the giant and the super-Earth.  
Accordingly it must occur through modification of the surface   
density profile through the action of density waves.    
If  the source of  density  wave excitation  
is removed,  
the low mass planet  should  undergo type I migration which in our case  
is directed  inwards.   
In order to demonstrate this,  beyond  some point  of some of   
our  simulations we  set the giant planet mass   
to zero.  The above  expectations  regarding migration are confirmed.   
We consider a simulation where the non migrating  giant planet mass was one Jupiter mass,  
the disc aspect ratio was $h=0.05,$ the  initial orbital radius of the  
super-Earth was $1.9$ and  the initial surface density profile was given   
by  Eq.(\ref{longdisc}).  The giant planet was removed at  time $t=750.$  
  
\begin{figure*}  
\begin{minipage}[!htb]{160mm}  
$\begin{array}{cc}
\includegraphics[angle=0,width=70mm]{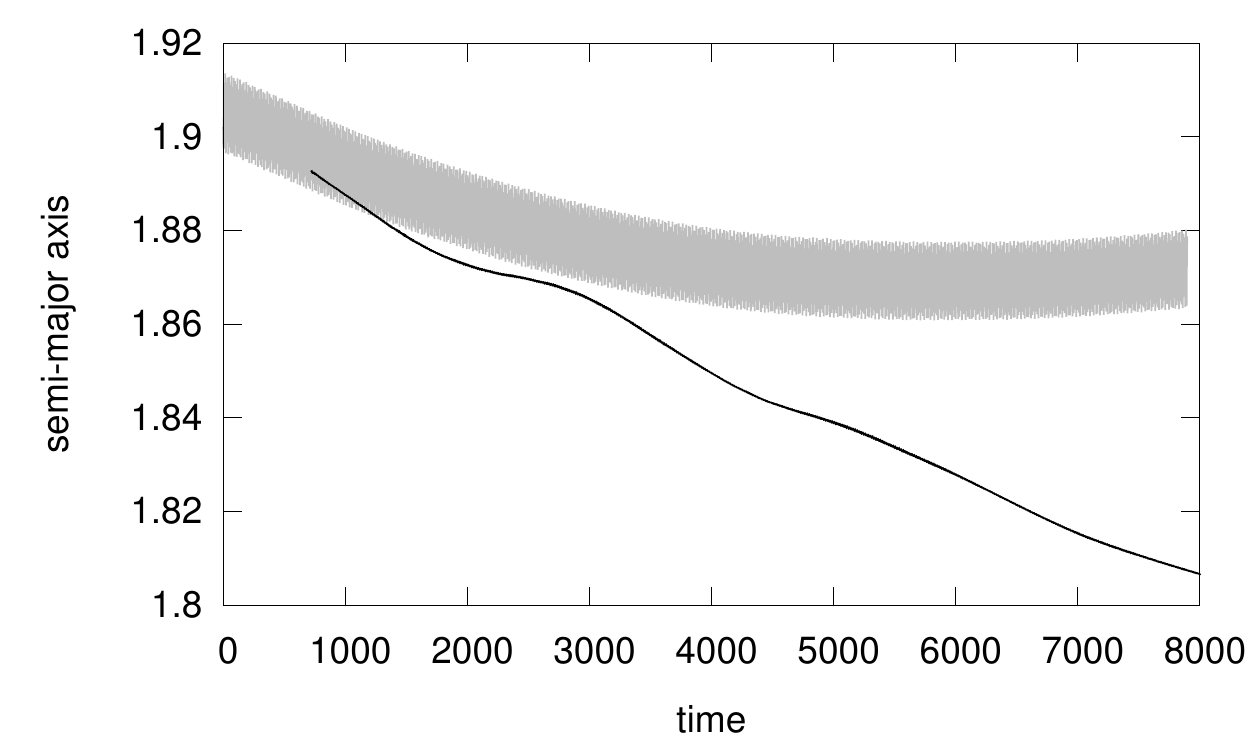}&  
\includegraphics[angle=0,width=70mm]{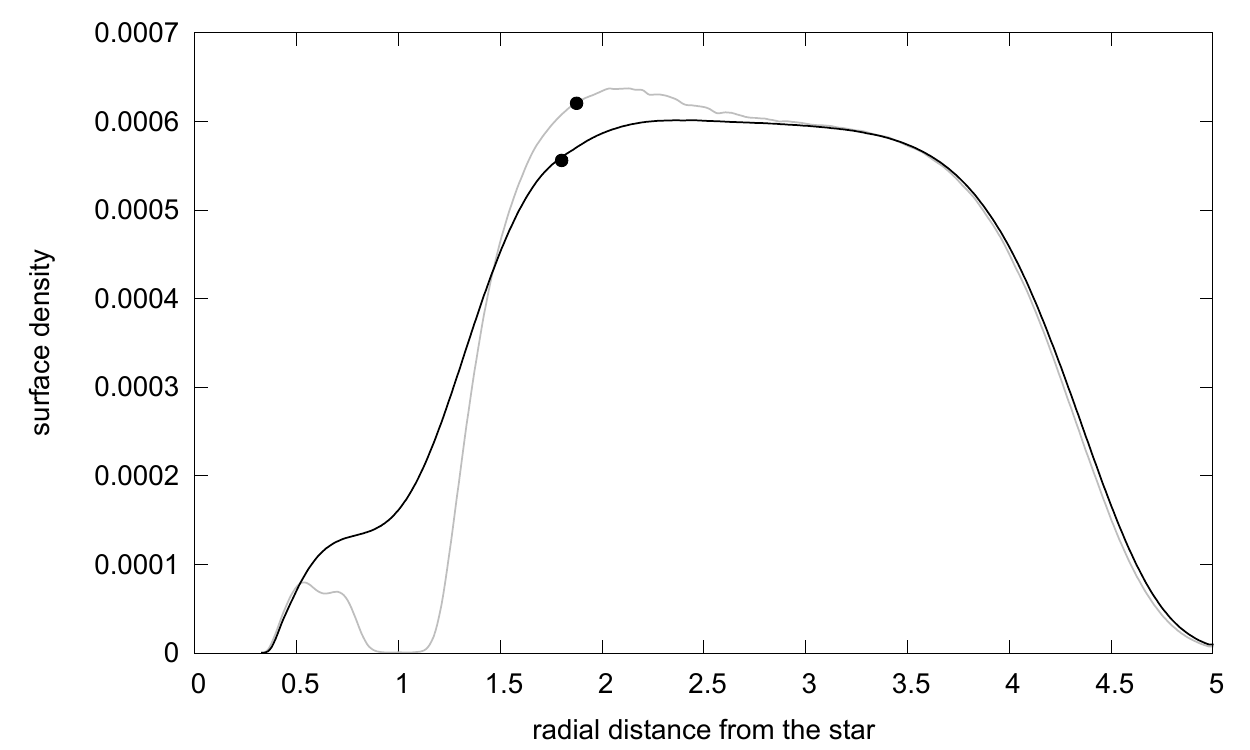}
\end{array}$
\caption{\label{1empor3}{ The left panel shows
the evolution of the semi-major axis   
of a low mass  
planet in the presence of a gas giant (grey curve), and   
after the  gas giant removal from the system   
(black curve).  
  The right panel shows the  azimuthally averaged surface density profiles
of the disc  for the cases with
(grey curve) 
and without (black curve) a gas giant 
at the end of simulations (at time $t=8000$). }}  
\end{minipage}  
\end{figure*}

The form of the   
migration, with and without the removal of the giant planet,  is shown 
in Fig. \ref{1empor3}.  
This clearly indicates that, when the giant planet is removed,  the super-Earth starts to undergo  
inward migration   at a rate similar to that it had initially, whereas when it is retained   
outward migration is ultimately attained.  The right panel 
 shows the azimuthally averaged surface density profiles   at the end of
 both simulations.

\begin{figure*}  
\begin{minipage}[!htb]{160mm}  
\centering  
\includegraphics[width=70mm]{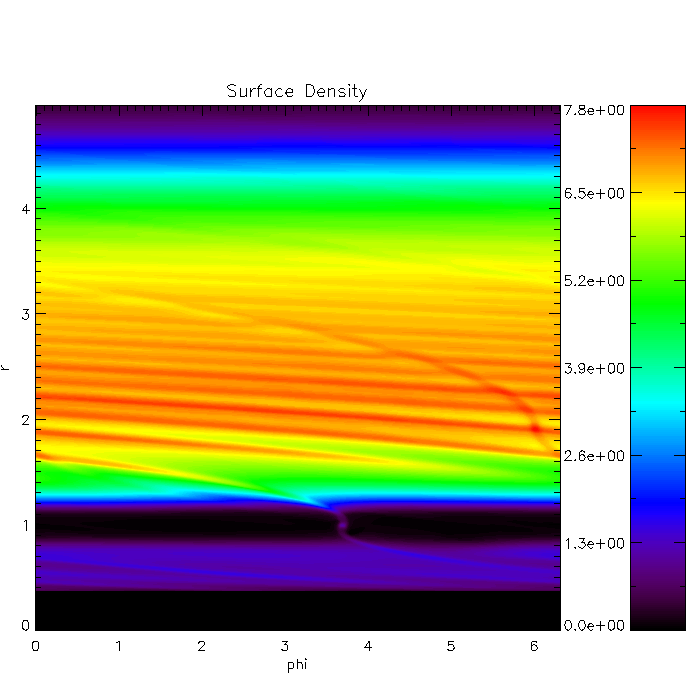}  
\includegraphics[width=70mm]{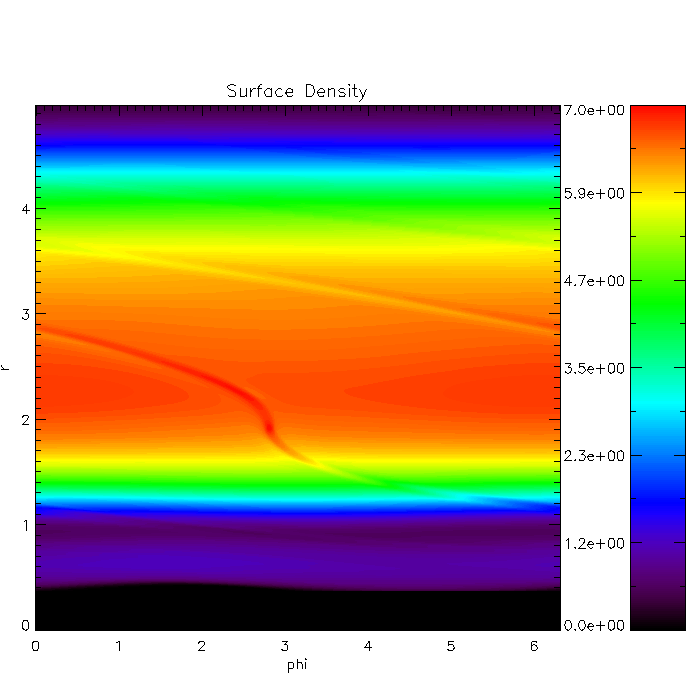}  
\caption{\label{sigma2}{The surface density contours in the disc with  
 waves excited by a gas giant and a super-Earth migrating through them   
(left) and  
 the surface density distribution in the disc after a gas giant has been   
removed with a super-Earth migrating in the resulting smooth surface    
density profile (right).  
}}  
\end{minipage}  
\end{figure*}

The disc  surface density contours in  
both these cases are  illustrated in Fig. \ref{sigma2} at time $t=795.$  
These are very smooth for the case   
where the giant planet was removed because of the absence  
of density waves. Note too that in spite of this,  the  inner  cavity  with a region of increasing 
azimuthally averaged surface density,  originally produced by the giant remains and  the presence 
 this surface density structure
 does not affect the inward migration of   
the low mass planet.
 This is in contrast to the case when the giant is allowed to 
 migrate inwards,  described in section \ref{NOTMIGG}
The edge profile in that case adjusts in a similar way while outward migration 
is maintained.

\subsection{Angular momentum transport and dissipation}  
\label{tor}  
  
To explore the nature of the positive torques  
acting on the super-Earth, we consider the relationship between angular momentum flux  
and energy dissipation due to shocks arising from the  
density waves within a coorbital annulus.  
 The conservation of angular momentum integrated over azimuth  
may be written in the form  
\begin{equation}  
\frac{\partial \rho_j}{\partial t}+\frac{\partial F_j}{\partial r}=   
 \int_0^{2\pi}\Sigma r {\cal T}d{\varphi}  
\label{consJ}  
\end{equation}  
Here the rate of  angular momentum  
flow through a circle of radius, $r,$ is  
\begin{equation}  
F_j = \int_0^{2\pi}\Sigma r^2 v_r v_{\varphi}d{\varphi},  
\label{flux}  
\end{equation}  
 the angular momentum per unit length  is  
\begin{equation}   
\rho_j = \int_0^{2\pi}\Sigma r^2 v_{\varphi}d{\varphi}  
\label{Jdensity}  
\end{equation}  
and ${\cal T}={\cal T}_{gi}+{\cal T}_{se}$ is the torque per unit mass acting on the disc which has contributions  
${\cal T}_{gi}$  and ${\cal T}_{se}$ due to   the  
giant  and super-Earth respectively.  
In the above, and also below, for simplicity  we neglect the  viscous contribution to   
the fluxes although including this would not affect our later discussion.  
  
Assuming that the forcing due to the giant and super-Earth  
has  pattern speeds $\Omega_{i,0}$ and $\Omega_p$ respectively,  
 We may also derive a conservation law   
 in the form  
\begin{equation}  
\frac{\partial \rho_{Jac}}{\partial t}+ \frac{\partial F_{Jac}}{\partial r}=   
- \epsilon_w+ (\Omega_p-\Omega_{i,0})  
\int_0^{2\pi}\Sigma r {\cal T}_{se}d{\varphi}.  
\label{consJac}  
\end{equation}  
Here    
\begin{equation}   
\rho_{Jac} = \rho_E-\Omega_{i,0}\rho_j,  
\end{equation}  
and  
\begin{equation}   
F_{Jac} = F_E-\Omega_{i,0}F_j,  
\end{equation}  
with the energy flux given by  
\begin{equation}  
F_E= \int_0^{2\pi}\Sigma r v_r \left(|{  v}|^2/2 + c_s^2\ln{\Sigma} +\Phi \right) d{\varphi},  
\label{Eflux}  
\end{equation}  
 and the energy per unit length  given by   
\begin{equation}   
\rho_E = \int_0^{2\pi}\Sigma r\left(|{\bf  v}|^2/2 + c_s^2\ln{\Sigma} +\Phi \right)  d{\varphi}.  
\label{Edensity}  
\end{equation}  
Here, for simplicity we have adopted a strictly isothermal equation of state as adopted in the local simulations. As the coorbital region is of small radial extent this should be a reasonable approximation  
for the global simulations also.  
  
The rate of energy dissipated per unit length in the radial direction  by the density waves    
is  $ \epsilon_w.$  
When  $ \epsilon_w$ and ${\cal T}_{se}$ are zero,  
on integration over the radial domain with vanishing boundary  
fluxes,  Eq.(\ref{consJac}) yields the Jacobi invariant for the flow.  
Noting that $\rho_{Jac}$ is negative,  
we see that the energy dissipation rate can be balanced either against a negative torque  
produced on the disc by the super-Earth,  or by a rate of increase in the magnitude of  $\rho_{Jac}$   
locally. The latter corresponds to torquing up the disc matter.  
Thus shock dissipation can in part be associated with balancing  
of negative torque on the disc produced by the super-Earth which implies a positive torque  
acting on  and outward migration of the super-Earth.  
Equation (\ref{consJ}) also shows that in an approximate steady state a decreasing outward angular  
momentum flux can be balanced by a negative torque acting on the disc due to the super-Earth.  
  
\begin{figure*}  
\begin{minipage}[!htb]{90mm}  
\includegraphics[angle=0,width=100mm]{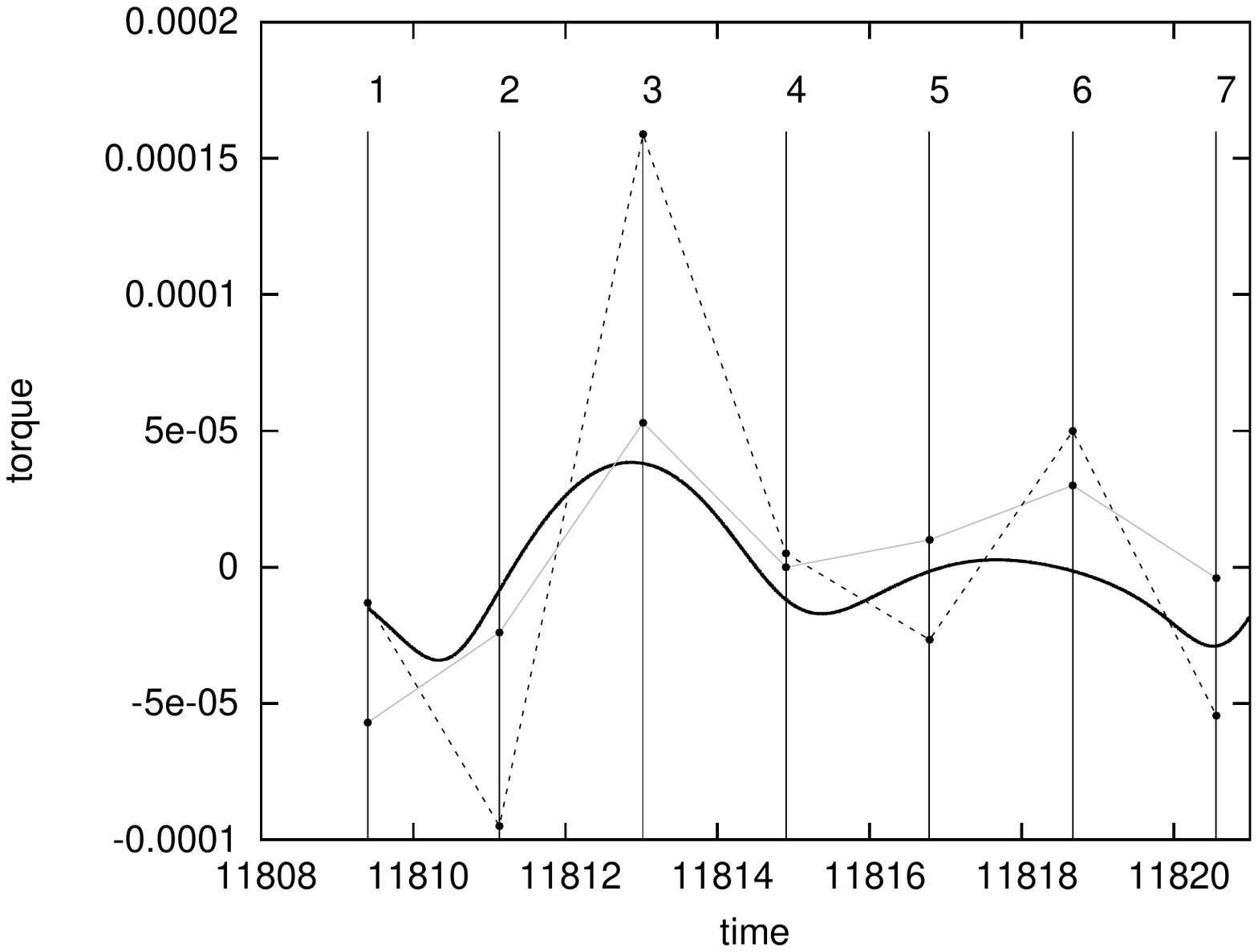}  
\caption{\label{1empor4} The torque acting on the   
super-Earth (solid black line),  
the angular momentum flow rate ($-F_j$) at the position of the super-Earth,  
across its orbital radius, averaged over  azimuth (grey)  and   
 $-F_j(r=2.3)$ averaged over  azimuth (dotted).  
 The vertical  
lines indicate  moments of time at which we show the    
surface density and the  
       azimuthally averaged angular momentum  
       flux as a function of radius below.}  
\end{minipage}  
\end{figure*}

We investigate the relationship between $F_j$ and the torque acting on the super-Earth  
during a  simulation  performed for the disc with  initial  surface density   
profile given by  Eq.(\ref{longdisc}) and  $h=0.05.$   
The giant planet is  one  Jupiter mass  and  the super-Earth is  $5.5 M_{\oplus}.$     
The giant planet is not allowed to migrate while   
the super-Earth is found to migrate slowly outwards.  
In Fig. \ref{1empor4}  we show the comparison of $-F_j$   
with the torque acting on the planet over a time span just exceeding one orbital  
period of the super-Earth after about $t=11800.$ It is seen that  
these behave similarly.  
    
\begin{figure*}  
\begin{minipage}[!htb]{160mm}  
\centering  
$\begin{array}{ccc}  
\hspace{-1.2cm}\includegraphics[width=60mm]{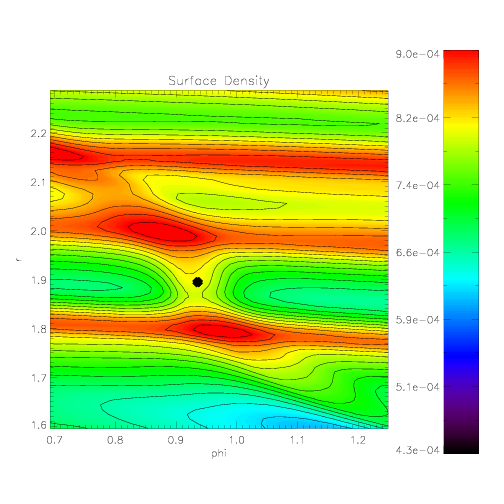}&  
\includegraphics[width=60mm]{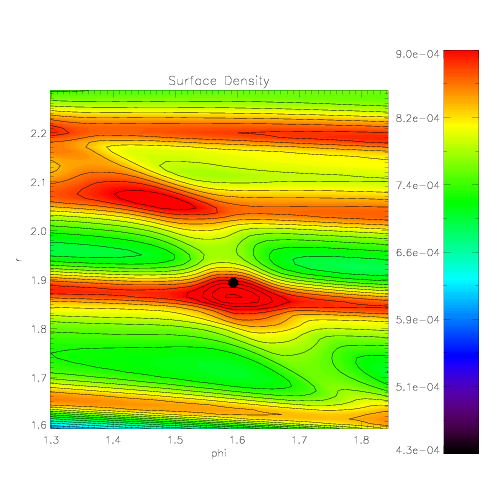}&  
\includegraphics[width=60mm]{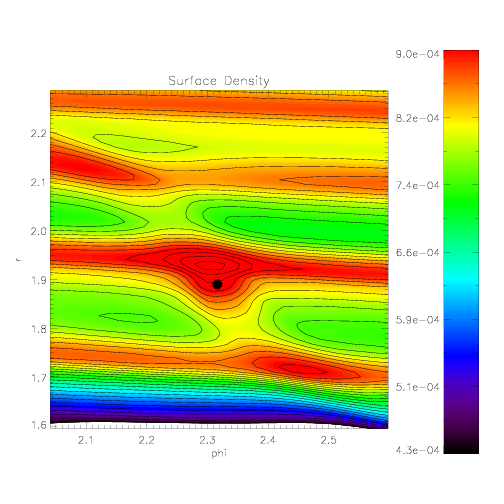}\notag \\  
\hspace{-1.2cm}\includegraphics[width=60mm]{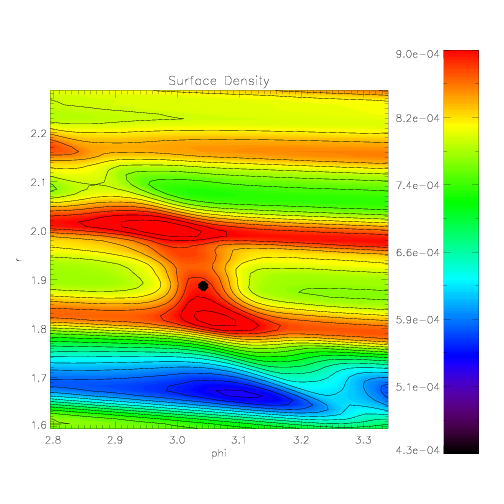}&  
\includegraphics[width=60mm]{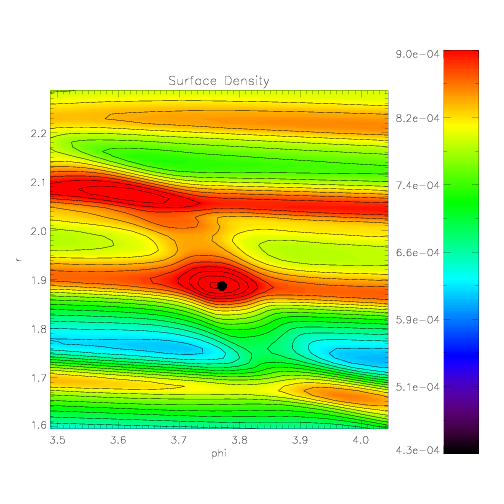}&  
\includegraphics[width=60mm]{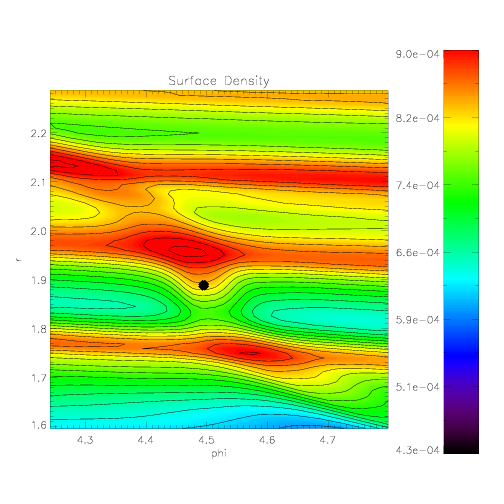}  
\end{array}$  
\caption{\label{1empor4new}The surface density contours   
for the first six  moments of time progressing from  left to  right  
,upper then lower,  marked by vertical lines in Fig. \ref{1empor4}.}  
\end{minipage}  
\end{figure*}

\begin{figure*}  
\begin{minipage}[!htb]{160mm}  
\centering  
$\begin{array}{ccc}  
\hspace{-1.2cm}\includegraphics[width=60mm]{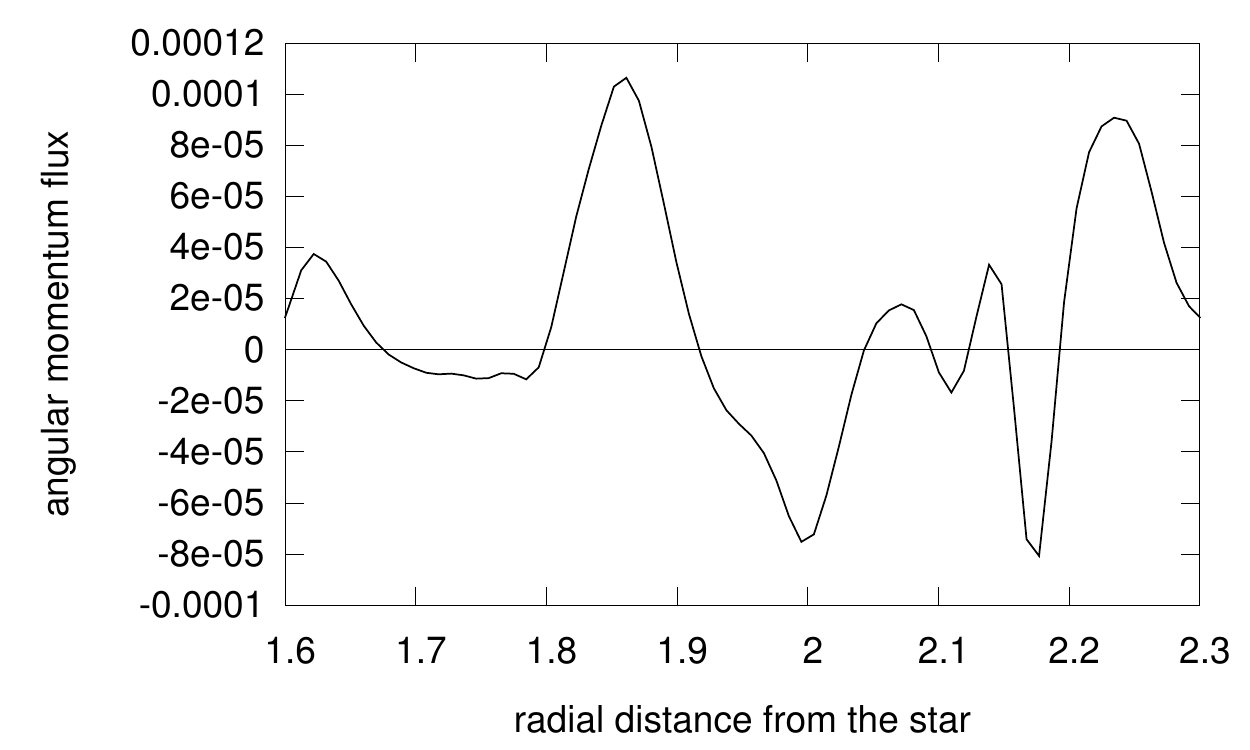}&  
\includegraphics[width=60mm]{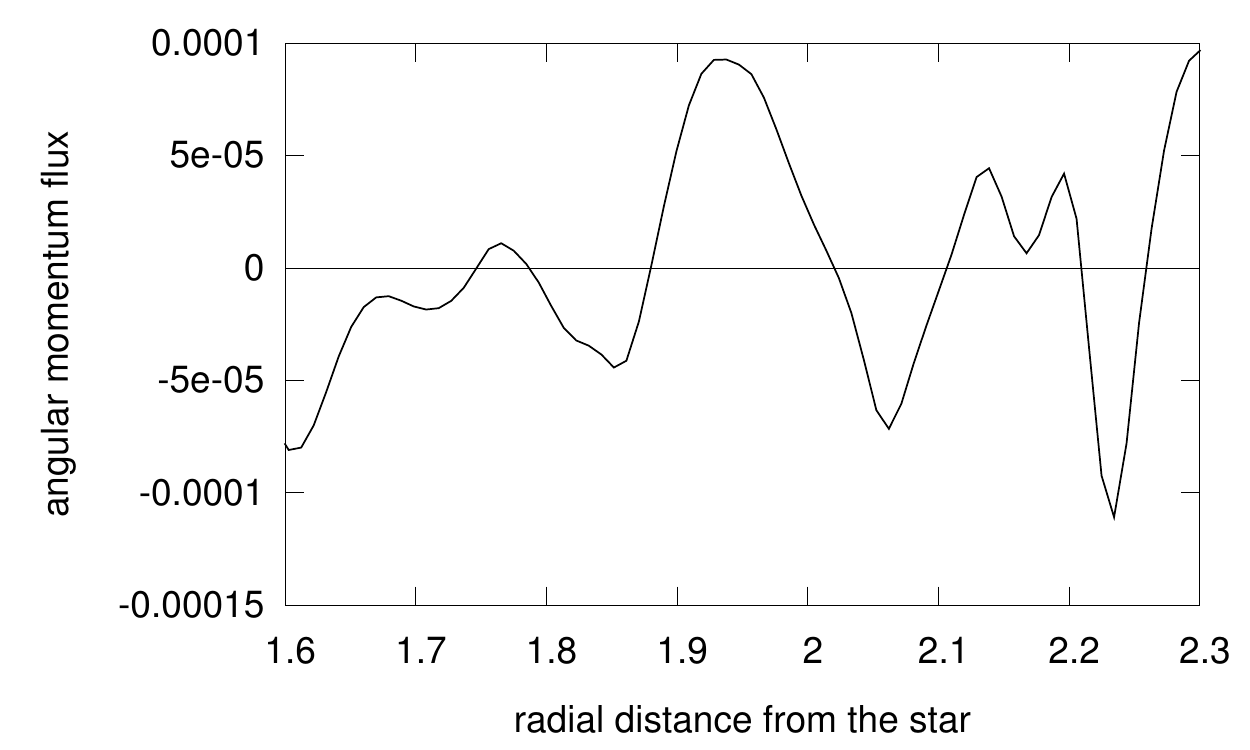}&  
\includegraphics[width=60mm]{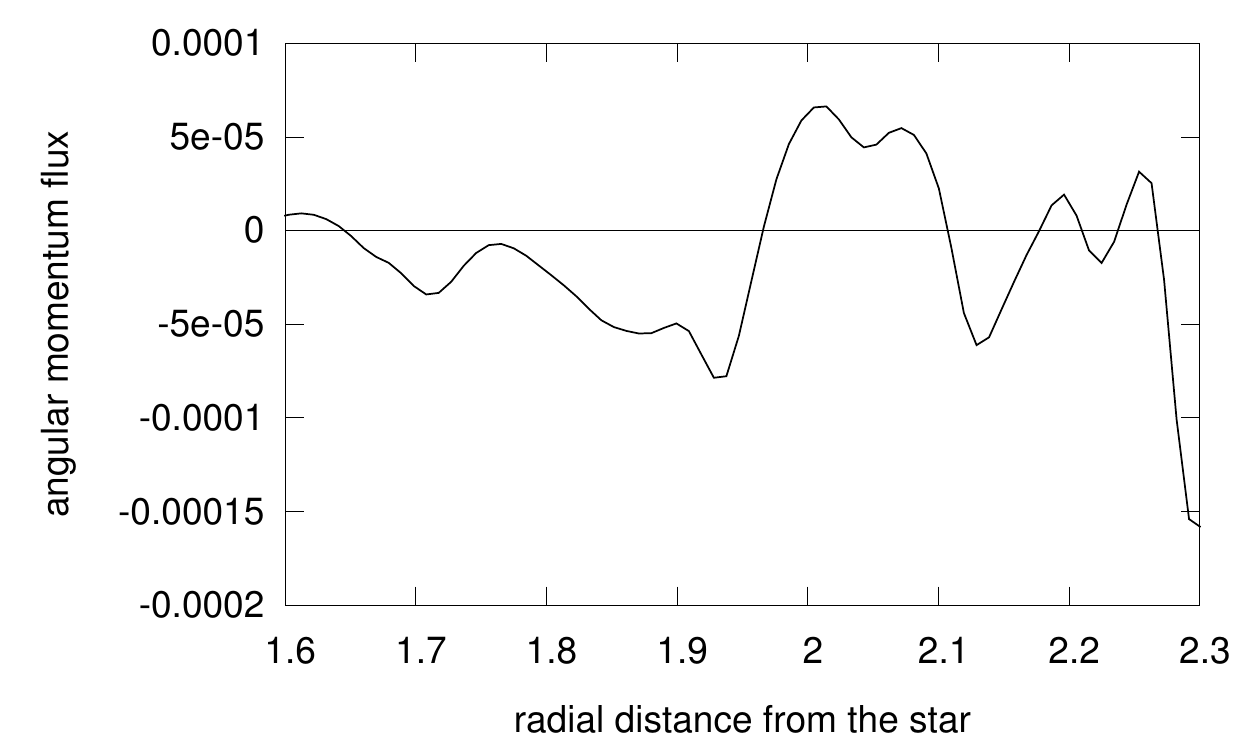}\notag \\  
\hspace{-1.2cm}\includegraphics[width=60mm]{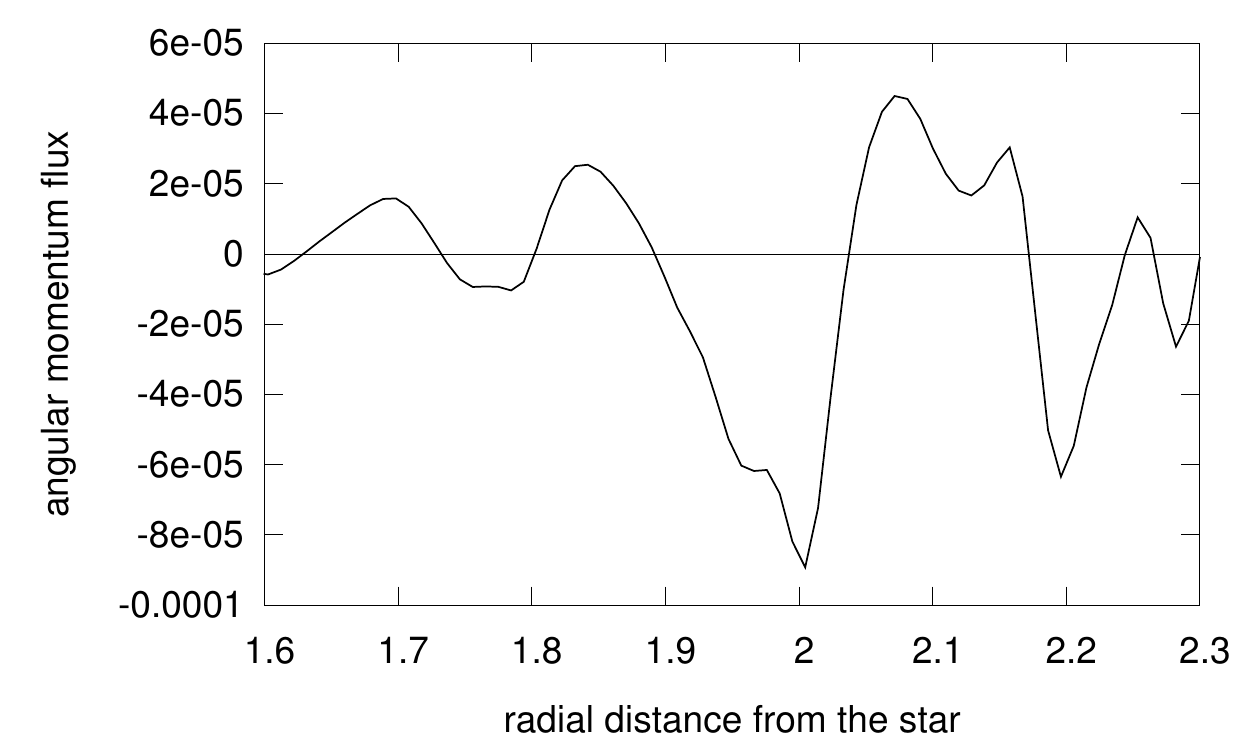}&  
\includegraphics[width=60mm]{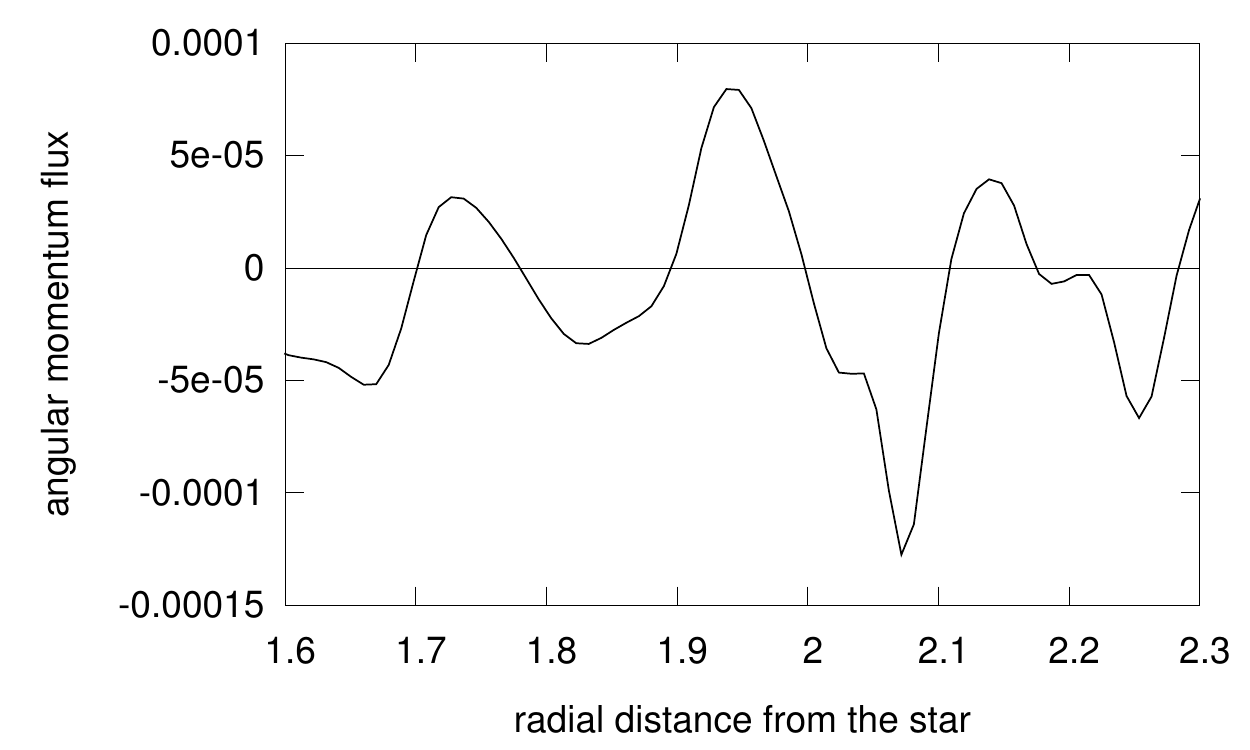}&  
\includegraphics[width=60mm]{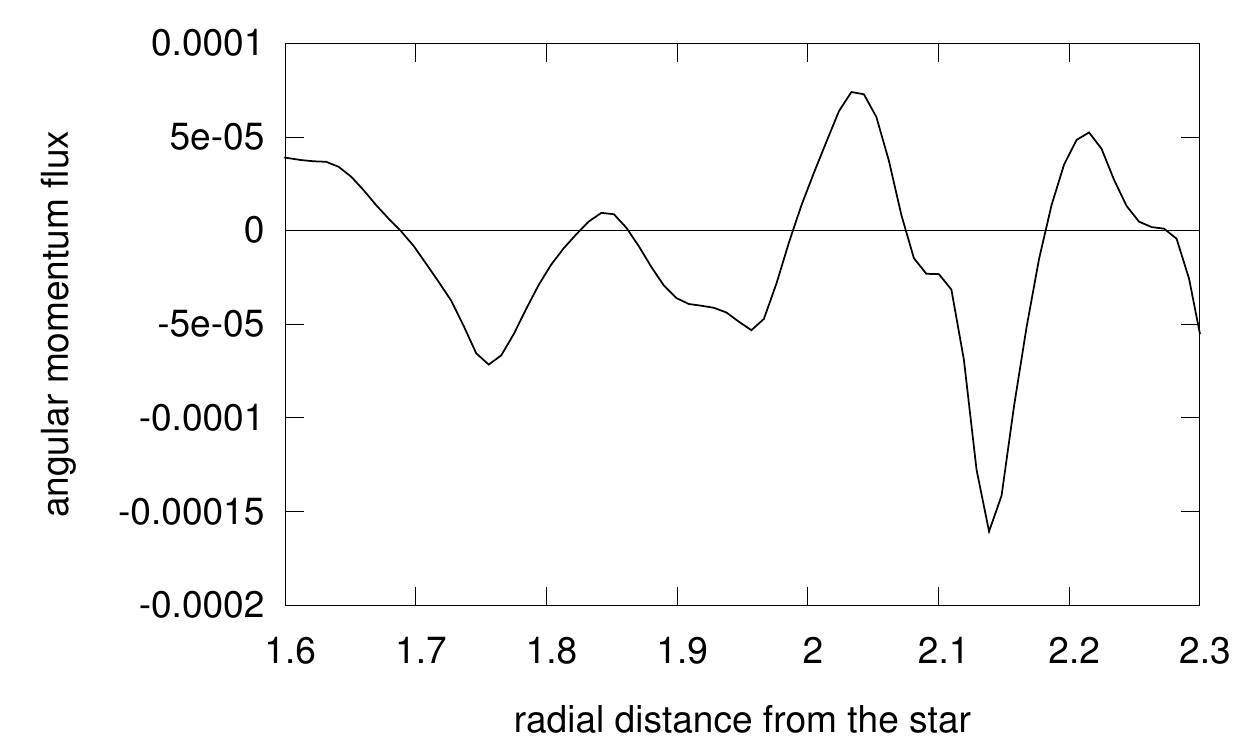}  
\end{array}$  
\caption{\label{1empor4new1}  
The azimuthally averaged angular momentum flow rate across a circle of radius, $r,$  
 against radius  
at the first six moments of time progressing  left to   
right, upper then lower,  marked by vertical lines in Fig. \ref{1empor4}.}  
\end{minipage}  
\end{figure*}

 We have   
calculated the torque acting on the planet due to the disc matter located   
in the vicinity of the planet  between $r=1.6$  and $r=2.3,$  the planet    
being located at around $r=1.89,$ at the moments of the time indicated by the   
vertical lines in Fig. \ref{1empor4}.  
In   Fig. \ref{1empor4new} we present the contours of the surface density   
and in  Fig. \ref{1empor4new1} we show  the azimuthally averaged angular momentum  
 flow rate through a circle of radius $r$  as a function of radius  
 calculated according to  Eq.(\ref{flux}) at the same times.  
Note that, although the average value is positive,  the torque values are small on account of cancellation effects.  
The largest  magnitude negative value occurs at the first time  
on account of a large surface density excess to the left of the planet.  
The largest positive value occurs at the third time on account of a positive surface  
density excess  close to the planet.  
 The azimuthally averaged angular momentum  
 flow rate through a circle of radius $r$ shows a complicated temporal and spatial dependence  
 on account of the dynamic situation and the interaction of waves excited by the giant  
 and the super-Earth. Nonetheless, we note that  Eq.(\ref{consJ})  implies that  
provided  $F_j$ vanishes in the gap region, $-F_j(r=2.3)$  
should equal the sum of the torques acting on both planets  together with the rate   
of increase of angular momentum of the disc material (note that this can actually  
be negative in some places as material does seep across the gap).  
>From Fig. \ref{1empor4}  
we see that this quantity is of a similar form as the torque  
on the super-Earth but about three times larger,   consistent with the view   
that some fraction of the  
angular momentum flow is transmitted to the super-Earth.


\begin{figure*}  
\begin{minipage}[!htb]{160mm}  
\centering  
\includegraphics[width=70mm]{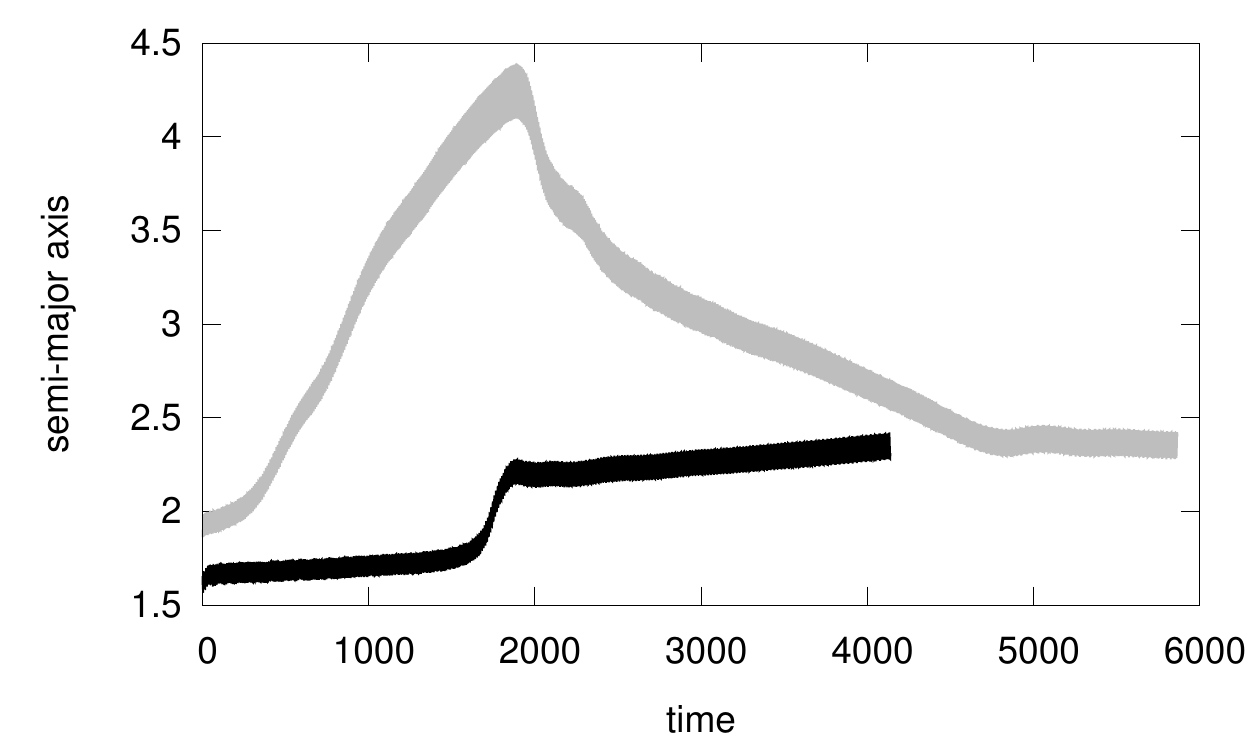}  
\includegraphics[angle=0,width=70mm]{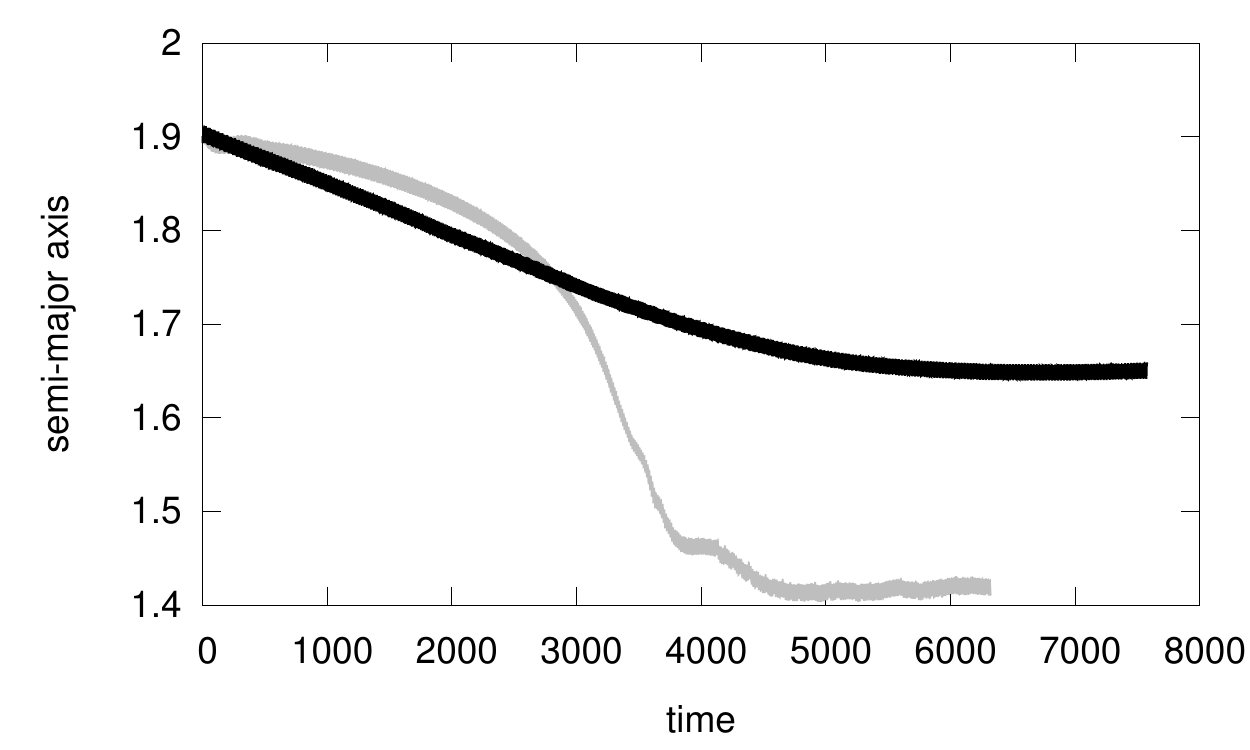}  
\caption{\label{1empor5}The left panel shows the evolution of the semi-major axis  
of a  $43M_{\oplus}$   
planet orbiting exterior to a  $7$ Jupiter mass gas giant. The  
initial orbital radius  of the super-Earth is $r= 1.9$ (grey curve) and  
$r=1.65$ (black curve), respectively.  
The right panel shows the evolution of the semi-major axis  
of a $43 M_{\oplus}$  
 (grey curve) and  
a $20 M_{\oplus}$  planet when the giant planet mass was reduced to one Jupiter mass  
(black curve).  
}  
\end{minipage}  
\end{figure*}

\begin{figure*}  
\begin{minipage}[!htb]{160mm}  
\centering  
\includegraphics[width=70mm]{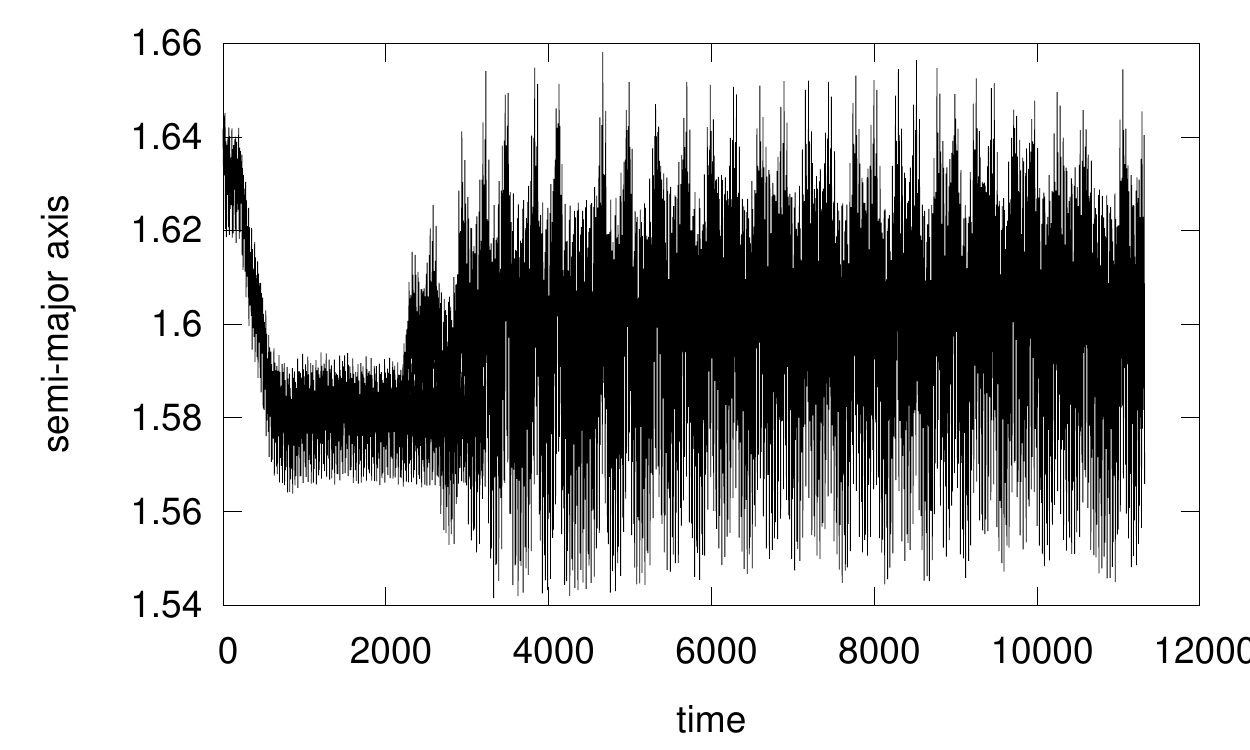}  
\includegraphics[width=70mm]{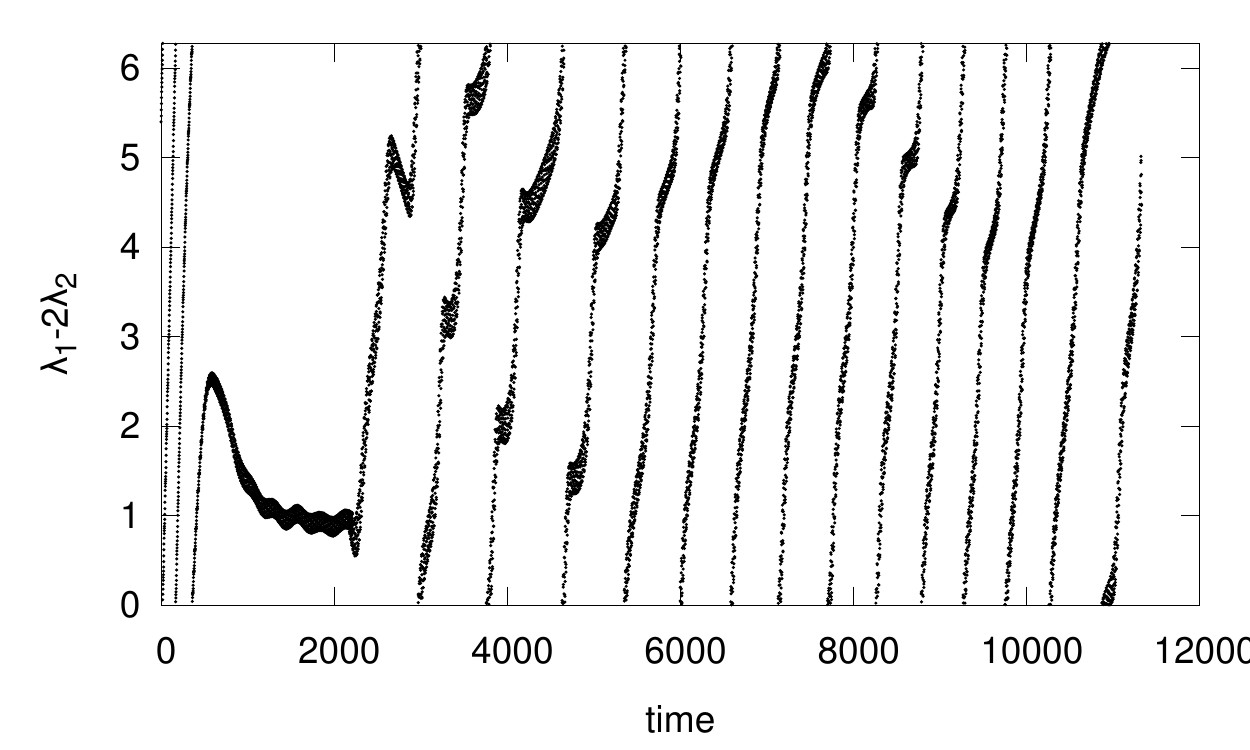}  
\includegraphics[width=70mm]{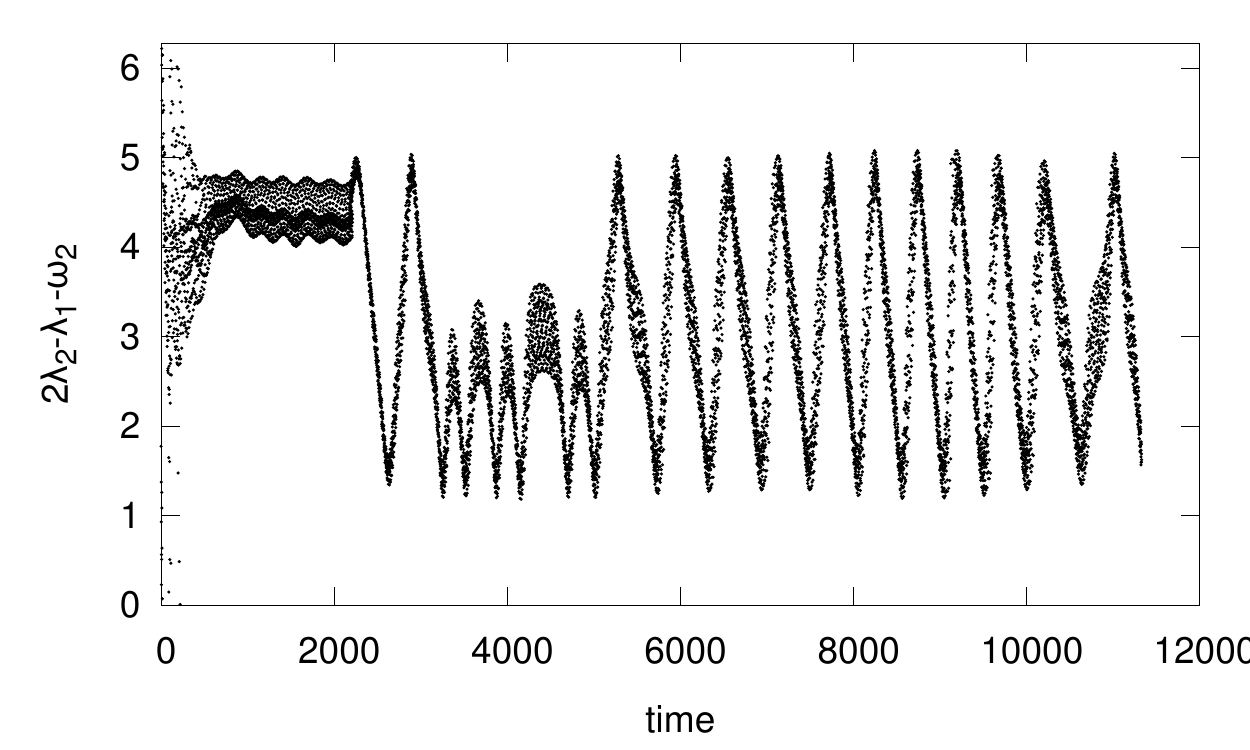}  
\includegraphics[width=70mm]{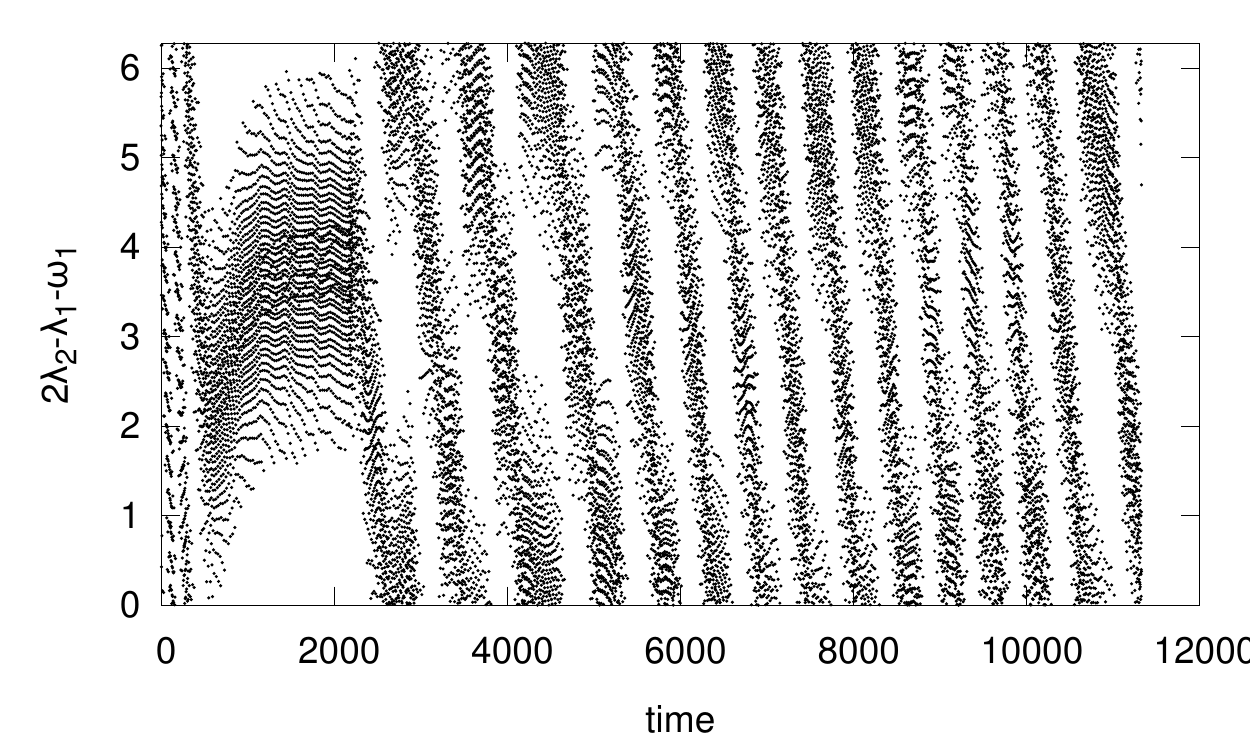}  
\caption{\label{fi24}{  
The evolution of the semi-major axis of a  $43 M_{\oplus}$  planet (upper  
left panel), both resonant angles (lower panels), and the difference in the  
mean longitudes (upper right panel) in the presence of the gas giant.  
After  2300 time units  the mass of  
the giant planet  was increased  from one  to 7 Jupiter masses.  
The oscillations of the semi-major axis  and one of the resonant angles was found to  
increase  in amplitude,  and the other angle switches from libration to circulation.}}  
\end{minipage}  
\end{figure*}


\section{ Consequences of the wave-planet interaction for the  
resonance capture  of a super-Earth by a gas giant}  
\label{observations}  
A  consequence of the results presented above is that   
outward migration induced by density waves can prevent  
an initially inwardly migrating   
 super-Earth from reaching a 2:1 commensurability with the giant.  
 The angular momentum exchange during the  
super-Earth passage through the density wave field induces its  outward migration.  
 It is thus prevented  from approaching close   
enough  to the  
gas giant's  orbit for the commensurability to be reached.   
From this we expect  that in a  system with  a  gas giant we could  
expect a low mass planet to be in an exterior  orbit, possibly  close  to,  but not exactly in   
2:1 resonance.   
 There are  two  observations  
that relate to  our studies.  The first  is  the inference from transit timing variations  
of a low  
mass companion of $15 M_{\oplus}$ that is  close to  an exterior  2:1 commensurability with  
the  
 giant planet ($ \approx 2 M_J$)  
in the Wasp-3 system \citep{maciejewski}.   
However,  further observations are needed to confirm  
this configuration.  
The other example is found in  the GJ876 system where  
the two outermost planets are close to 2:1 commensurability.  
 The planets have  
masses $2.27 M_J$ and $14.6 M_{\oplus}$ so they are within the regime where  
the mechanism described in this paper can operate.  
The situation in GJ876 is complex because apart from the configuration of the outer two   
planets,  there is another inner giant in the system with  all three planets being   
in a  Laplace resonance.   
Here we will concentrate on the outer two  planets: c and e in the  GJ876 system.  
In order to perform simulations, we scale up the masses of the outer two planets  
to be 7 Jupiter masses and 43 Earth masses so that with a central star of one solar mass,  
as adopted here, the mass ratios are the same as in the GJ876 system.

In Fig. \ref{1empor5} (left panel) we show  the evolution of the semi-major axis of the   
outermost planet in a  disc with aspect ratio $h=0.05.$  
In this particular case, the  gas giant was  not allowed to migrate   
and the  planets do not interact with each other gravitationally.    
The initial surface density profile was given by  Eq.(\ref{longdisc}).   
The outermost planet was started in an    
exterior  orbit outside the 2:1 resonance. The  evolution   
proceeds differently for different initial locations in the disc, but   
the final outcome of the  evolution is similar. The smaller planet   
always ends up  at  the same radial location, exterior to the 2:1 commensurability   
with semi-major axis exceeding twice that of the giant.  
 This  is illustrated   
in Fig. \ref{1empor5} for initial locations of the planet   
of $r= 1.65$  and $r= 1.9.$

  
In order to explore this issue further,   
in  Fig. \ref{1empor5} (right panel)  
we compare the behaviour of $43 M_{\oplus}$ and $20 M_{\oplus}$  planets in   
 the same initial   
 disc but in the presence of a lower mass  gas giant with a  mass of 1 Jupiter mass.   
 The Jupiter mass  
planet launches weaker density waves than the  7 Jupiter mass planet.   
 This enabled   the  $43 M_{\oplus}$ planet to  pass   
through the 2:1 resonance. Thus the density waves are too weak to prevent the  
attainment of the commensurability in this case.    
However the    
 $20 M_{\oplus}$   planet stopped its migration just exterior to  resonance  
which indicates that smaller mass planets do not reach the commensurability.

Finally, the above simulation for the $43 M_{\oplus}$ planet was repeated with  
the gravitational interaction between the planets included so that  
capture into  the 2:1 resonance takes place. This is illustrated   
 in Fig. \ref{fi24} where we  
show  the evolution of the semi-major axis of the low mass planet, both resonant angles  
\citep[see e.g.][for definitions]{papszusz2005} and the difference in the  
mean longitudes.   
  
Thus resonance capture is only possible for sufficiently low mass giants  
that are not efficient enough as a source of density waves to halt the inward migration.  
In order to obtain a resonant configuration for a more massive giant, we slowly increased its mass.  
However, this had to be done after removal of  significant amounts  
of disc material that could act as a carrier  
of density waves.  
For example after about 2300 time units  we started  
to increase the  mass of the gas giant as a linear function of time  
 until it reached  the mass of 7 $M_J$ at a time $t=3341.$  At  
the beginning of this procedure  we instantly  modified the surface density profile so as to  
increase the size of the  gap around the  giant by adopting the following modification:  
\begin{eqnarray}  
\label{GJ876}  
\Sigma \rightarrow  0.01\Sigma          &  {\rm for} &      r  < 2.4\notag \\  
\Sigma \rightarrow  \Sigma \times(r/2.5)^{90} & {\rm  for} &  2.4 \le r  < 2.5\notag \\  
\Sigma \rightarrow  \Sigma     &  {\rm for}  &   r \ge 2.5   
\end{eqnarray}  
After this has been carried out, the lower mass planet becomes located  
within an extensive  gap, where there is far less material, so that  density waves become  
ineffective at causing  it to  migrate outwards. Once we  started to increase  
the mass of the gas giant the  
oscillations of its  semi-major axis increased in amplitude and   one resonant angle  
moved from libration to circulation, as  can  
be seen in Fig. \ref{fi24}. However the other resonant continues to  librate,  
but with larger amplitude. These details may depend on the detailed  
procedure we followed. Nonetheless, the planet remained close to the 2:1  
resonance  
as is observed in  the Gliese 876 system \citep{rivera}.  
  
However, the above scenario requires the giant to accrete from a reservoir  
of material, possibly an interior disc,  after having removed most of the nearby outer disc.  
This would appear unlikely. It may be that the lower mass planet formed after the gas disc  
had dispersed and/or underwent planetesimal migration \citep[e.g.][]{planetesimals} to bring it into a 2:1 commensurability  
from a larger radius.   
  
\newpage

\section{Conclusions}  
\label{conclusions}  
In this paper we have studied a new mechanism that can  reverse the   
type I migration of a low mass planet  
 that would  occur in an unperturbed  locally isothermal gaseous disc.   
This mechanism operates when there is a source of   
trailing density waves that propagate through the disc.  
The most natural example of such source arises from the gravitational perturbation by    
another planet and for global simulations   
we have focused on the case when   
a gas giant is present. We have also carried out local simulations in a shearing box  
where the forcing was due to an imposed harmonically varying potential with varying   
amplitude. In all cases the density waves produce shocks which are associated with  angular  
momentum transfer to the disc material. Coorbital disc material  then transfers angular momentum  
to the low mass planet in the same direction,  
when  it is scattered by it. This is analogous to what happens with  
coorbital material undergoing a slow drag in horseshoe orbits in the case of zero pressure   
particle dynamics. However, in the cases considered here the dynamics of the coorbital material  
is not stationary in an appropriately rotating frame and generally more complex.  
In particular, we stress that although coorbital dynamics plays a role, the situation is unlike that  
pertaining to standard corotation torques in initially unperturbed discs.  
 For example,  because the dissipation  
and effective drag are produced by shocks, which are viscosity independent and the propagating  
density waves provide an external  source of angular momentum for the coorbital zone,  
issues of torque saturation become  irrelevant.  Furthermore  although there is an interior gap, the mechanism described here  
depends on the  same source of angular momentum,  rather than  applied viscous stresses or 
an accelerating inward accretion flow,
to maintain the surface  density distribution and  in this way it differs from the
trap mechanism of \citet{masset}.  
  
The  effect of the interaction with the density waves  
is  that the  migration of a low-mass planet located    
in an exterior orbit relative to the gas giant can be slowed down and  
finally reversed.   
Thus we found that a  planet  with mass in the super-Earth range cannot approach  
a Jupiter mass  planet close enough in order to form first order  
mean-motion resonances with it. The migration   
 was found to halt with semi-major axis ranging between $1.6$ and $2$  
times that of the giant. Only when the low mass planet exceeded $\sim 40 M_{\oplus}$  
was it able to attain a 2:1 commensurability.  As the giant planet mass is increased,   
even larger low mass planets would be required.  
  
Our results indicate that migrating the outermost planet in the  
GJ876 system to its observed 2:1 commensurability through planet interaction  
with the gaseous disc alone would be problematic for the reasons outlined above.  
This may indicate that migration induced by planetesimals after the clearance of the gas disc  
may have been significant in that case.  
  
\newpage  
  
\section*{Acknowledgments}  
This work has been partially supported by NSC Grant No. N N203 583740  
(2011-2012) and  MNiSW PMN grant - ASTROSIM-PL ''Computational Astrophysics.  
The formation  and evolution of structures in the universe: from planets to  
galaxies'' (2008-2011).  
We acknowledge support from the Isaac Newton Institute programme ``Dynamics  
of Discs and Planets''.  
Part of this research was performed during a stay  at the Kavli  
Institute for  
Theoretical Physics, and was supported in part by  
the National Science Foundation under Grant No.  
PHY05-51164.  
The simulations reported here were performed using  
the HAL9000  
cluster of the Faculty of Mathematics and Physics of the University of  
Szczecin.
JCBP acknowledges support through STFC grant ST/G002584/1.  
We wish also to thank Adam {\L}acny for his helpful  
comments.  
Finally, we are indebted to  
Franco Ferrari for his continuous support in the development of our  
computational  
techniques and computer facilities.

\label{lastpage}  
  
\end{document}